\documentclass[prd,aps,twocolumn,preprintnumbers, showpacs, nofootinbib,superscriptaddress,notitlepage]{revtex4-1}
\usepackage{amsmath,graphicx,epsfig,amssymb}
\usepackage{inputenc}
\usepackage[usenames]{color}
\usepackage{ulem} 
\usepackage{bigstrut}
\usepackage{slashed}
\usepackage{multirow}
\usepackage{booktabs}
\usepackage{subcaption}
\usepackage{dsfont}
\usepackage{xcolor}
\usepackage{soul} 
\usepackage{tabularx} 

\usepackage{caption}
\allowdisplaybreaks

\newcommand{\nn}{\nonumber}
\def\slash#1{#1 \hskip-0.45em /}

\begin{document}
 
\title{Determination of heavy meson light-cone distribution amplitudes: theoretical framework and lattice simulations}

\author{\includegraphics[scale=0.10]{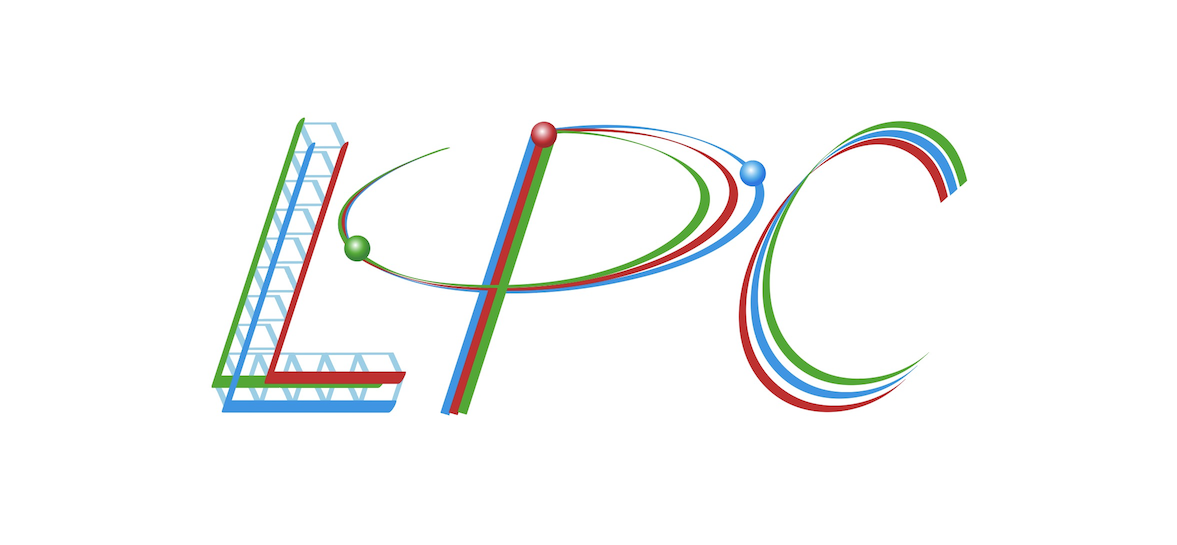}\\
Hao-Fei  Gao}
\affiliation{State Key Laboratory of Dark Matter Physics, Shanghai Key Laboratory for Particle Physics and Cosmology, Key Laboratory for Particle Astrophysics and Cosmology (MOE), School of Physics and Astronomy, Shanghai Jiao Tong University, Shanghai 200240, China}
\affiliation{Tsung-Dao Lee Institute, Shanghai Jiao Tong University, Shanghai 201210, China}

\author{Xue-Ying Han}
\affiliation{Institute of High Energy Physics, Chinese Academy of Sciences, Beijing 100049, China}
\affiliation{School of Physical Sciences, University of Chinese Academy of Sciences, Beijing 100049, China}

\author{Jun Hua}
\affiliation{Key Laboratory of Atomic and Subatomic Structure and Quantum Control (MOE), Guangdong
Basic Research Center of Excellence for Structure and Fundamental Interactions of Matter,
Institute of Quantum Matter, South China Normal University, Guangzhou 510006, China }
\affiliation{Guangdong-Hong Kong Joint Laboratory of Quantum Matter, Guangdong Provincial Key Laboratory of Nuclear Science, Southern Nuclear Science Computing Center, South China Normal University, Guangzhou 510006, China}

\author{Xiangdong Ji}
\affiliation{Maryland Center for Fundamental Physics, Department of Physics, University of Maryland, 4296 Stadium Dr., College Park, MD 20742, USA}
\affiliation{State Key Laboratory of Dark Matter Physics, Shanghai Key Laboratory for Particle Physics and Cosmology, Key Laboratory for Particle Astrophysics and Cosmology (MOE), School of Physics and Astronomy, Shanghai Jiao Tong University, Shanghai 200240, China}
\affiliation{Tsung-Dao Lee Institute, Shanghai Jiao Tong University, Shanghai 201210, China}

\author{Xiangyu Jiang}
\affiliation{CAS Key Laboratory of Theoretical Physics, Institute of Theoretical Physics, Chinese Academy of Sciences, Beijing 100190, China}

\author{Cai-Dian L\"u}
\affiliation{Institute of High Energy Physics, Chinese Academy of Sciences, Beijing 100049, China}
\affiliation{School of Physical Sciences, University of Chinese Academy of Sciences, Beijing 100049, China}

\author{Andreas Sch\"afer}
\affiliation{Institut f\"ur Theoretische Physik, Universit\"at Regensburg, D-93040 Regensburg, Germany}
\affiliation{Department of Physics, National Taiwan University, Taipei, Taiwan 106, China}

\author{Jin-Xin Tan}
\affiliation{State Key Laboratory of Dark Matter Physics, Shanghai Key Laboratory for Particle Physics and Cosmology, Key Laboratory for Particle Astrophysics and Cosmology (MOE), School of Physics and Astronomy, Shanghai Jiao Tong University, Shanghai 200240, China}
\affiliation{Tsung-Dao Lee Institute, Shanghai Jiao Tong University, Shanghai 201210, China}

\author{Ji-Hao Wang}
\affiliation{CAS Key Laboratory of Theoretical Physics, Institute of Theoretical Physics, Chinese Academy of Sciences, Beijing 100190, China}

\author{Wei Wang}
\email{Corresponding author: wei.wang@sjtu.edu.cn}
\affiliation{State Key Laboratory of Dark Matter Physics, Shanghai Key Laboratory for Particle Physics and Cosmology, Key Laboratory for Particle Astrophysics and Cosmology (MOE), School of Physics and Astronomy, Shanghai Jiao Tong University, Shanghai 200240, China}
\affiliation{Southern Center for Nuclear-Science Theory (SCNT), Institute of Modern Physics, Chinese Academy of Sciences, Huizhou 516000, Guangdong Province, China}

\author{Ji Xu}
\affiliation{School of Nuclear Science and Technology, Lanzhou University, Lanzhou 730000, China}

\author{Yi-Bo Yang}
\affiliation{CAS Key Laboratory of Theoretical Physics, Institute of Theoretical Physics, Chinese Academy of Sciences, Beijing 100190, China}
\affiliation{School of Fundamental Physics and Mathematical Sciences, Hangzhou Institute for Advanced Study, UCAS, Hangzhou 310024, China}
\affiliation{International Centre for Theoretical Physics Asia-Pacific, Beijing/Hangzhou, China}
\affiliation{School of Physical Sciences, University of Chinese Academy of Sciences, 
Beijing 100049, China}

\author{Fu-Wei Zhang}
\affiliation{School of Physical Science and Technology, Inner Mongolia University, Hohhot 010021, China}

\author{Jian-Hui Zhang}
\affiliation{School of Science and Engineering, The Chinese University of Hong Kong, Shenzhen 518172, China}

\author{Jia-Lu Zhang}
\affiliation{Tsung-Dao Lee Institute, Shanghai Jiao Tong University, Shanghai 201210, China}
\affiliation{State Key Laboratory of Dark Matter Physics, Shanghai Key Laboratory for Particle Physics and Cosmology, Key Laboratory for Particle Astrophysics and Cosmology (MOE), School of Physics and Astronomy, Shanghai Jiao Tong University, Shanghai 200240, China}

\author{Mu-Hua Zhang}
\affiliation{State Key Laboratory of Dark Matter Physics, Shanghai Key Laboratory for Particle Physics and Cosmology, Key Laboratory for Particle Astrophysics and Cosmology (MOE), School of Physics and Astronomy, Shanghai Jiao Tong University, Shanghai 200240, China}
\affiliation{Tsung-Dao Lee Institute, Shanghai Jiao Tong University, Shanghai 201210, China}

\author{Qi-An Zhang }
\email{Corresponding author: zhangqa@buaa.edu.cn}
\affiliation{School of Physics, Beihang University, Beijing 102206, China}

\author{Shuai Zhao}
 \affiliation{School of Science, Tianjin University, Tianjin 300072, China}

\collaboration{Lattice Parton Collaboration}

\begin{abstract}
We present a first-principles determination of heavy meson light-cone distribution amplitudes (LCDAs) from lattice QCD in the continuum limit, improving substantially on our previous pioneering study. 
Within the heavy-quark large-momentum effective theory (HQLaMET) framework, supplemented by lattice QCD calculations of the OPE moments, we analyze six ensembles with lattice spacings ranging from $a=0.0519-0.1053$\,fm and pion masses from $m_\pi=135.5-317.2$\,MeV, thereby enabling controlled continuum, chiral, and infinite-momentum extrapolations to the physical point.
Momentum-smeared sources, hypercubic-smeared Wilson lines, and optimized interpolating operators are adopted to significantly improved signals for the nonlocal correlators. Within a unified framework, we determine both QCD LCDAs and HQET LCDAs. Our resulting QCD LCDAs of $D$ meson peak at $y\approx 0.2-0.3$, with total uncertainties below $30\%$ for $0.1<y<0.9$. The leading-twist HQET LCDA is constructed using a peak-and-tail factorization, in which the nonperturbative peak region is obtained from lattice QCD and the perturbative tail is incorporated from HQET, with the two regions combined through a model-independent Laguerre-polynomial parametrization. At $\mu=1$\,GeV, we obtain the inverse moment of HQET LCDA $\lambda_B=0.340(20)$\,GeV and first inverse-logarithmic moment $\sigma_B^{(1)}=1.685(63)$, consistent with experimental constraints and phenomenological determinations. Direct lattice calculations based on operator product expansion provide a nontrivial cross-check of the LaMET results. Final results and phenomenological impact of these results are presented in a companion paper~\cite{HeavymesonDA_short_paper}. 
Our results remove the single-lattice-spacing limitation of the previous study, and provide a robust determinations of heavy meson LCDAs in both QCD and HQET for next-generation heavy flavor physics.

\end{abstract}
\maketitle

\section{Introduction}

In particle physics, weak decays of heavy $B$  meson  provide a powerful laboratory for precision tests of the standard model (SM) and for uncovering potential footprints of new physics beyond the SM. 
Nonleptonic decay $B\to \pi^+\pi^-$ is a cornerstone channel for studying direct CP violation and constraining the CKM unitarity triangle~\cite{Beneke:1999br,Lu:2000em}. In parallel, the rare decay $B\to K^{*}\ell\ell$, mediated by a  flavor-changing neutral-current (FCNC) transition $b\to s\ell\ell$, is loop suppressed in the SM and hence exceptionally sensitive to new physics, making it a central probe in contemporary searches for new physics~\cite{Ali:1999mm}. Achieving these goals requires efficient control over  the precision in theoretical calculations of  decay amplitudes. 
Consequently, a first-principles understanding of the decay amplitudes is indispensable for the current and next generation of high-precision heavy flavor physics phenomenology and experimental measurements.

The standard theoretical strategy for describing heavy meson weak decays is   factorization, which systematically separates short-distance physics, encoded in perturbatively calculable hard functions, from long-distance hadronic dynamics that must be treated nonperturbatively. This separation is crucial both conceptually and phenomenologically: it organizes QCD effects by scale, enables controlled resummations of large logarithms, and provides a framework in which hadronic uncertainties can be isolated, quantified, and ultimately improved with first-principles inputs. 
In heavy quark physics, several factorization formalisms are widely employed, tailored to different kinematic regimes and final states. For exclusive nonleptonic decays into light mesons, approaches such as QCD factorization~\cite{Beneke:1999br,Beneke:2000ry} and perturbative QCD approach~\cite{Keum:2000wi,Lu:2000em,Keum:2000ph} exploit the heavy-quark limit to express amplitudes in terms of convolutions of hard kernels with universal hadronic quantities. For processes involving energetic light hadrons and/or additional collinear degrees of freedom, soft-collinear effective theory (SCET)~\cite{Bauer:2000yr} provides a systematic effective field theory (EFT) organization of the relevant modes and power corrections. 
Across these formalisms, predictive power relies on a small set of universal nonperturbative inputs. As an example, in QCD factorization, decay amplitudes for processes like $B\to K^*\ell^+\ell^-$ with an energetic  $K^*$ in the final  state and $B\to \pi^+\pi^-$ can be factorized into heavy-to-light transition form factors and a hard-scattering term involving convolutions with heavy meson and light meson light-cone distribution amplitudes (LCDAs). Heavy-to-light form factors can be further computed using heavy meson LCDAs in combination with light-cone sum rules (LCSR)~\cite{DeFazio:2005dx,Khodjamirian:2006st,Wang:2015vgv,Lu:2018cfc,Gao:2019lta,Cui:2022zwm,Gao:2024vql,Huang:2025jsa,Li:2025mhq}. As a result, LCDAs play an extremely important role in the theoretical description of weak decays of $B$ meson.

At present, one of the most severe limitations facing precision phenomenology in $B$-meson weak decays is the incomplete understanding of heavy meson LCDAs. While the concept of heavy meson  LCDAs was first proposed over thirty years ago \cite{Grozin:1996pq}, their calculation was primarily only possible using model-dependent constructions~\cite{Belle:2018jqd,Beneke:2018wjp,Wang:2015vgv,Khodjamirian:2020hob,Lee:2005gza,Braun:2003wx,Grozin:1996pq}, and different model ans\"atze could yield markedly different shapes and inverse moments, translating into sizable (in many cases dominant) systematic uncertainties in factorization-based predictions.  
This issue becomes particularly transparent in light-cone sum rules (LCSRs) calculations for $B$ meson  form factors at large recoil such as $B\to K^*$ and $B\to\pi$~\cite{Gao:2019lta,Cui:2022zwm}:
\begin{eqnarray}
{\cal V}_{B\to K^*}(0) &=& 0.359^{+0.141}_{-0.085}\Big|_{\lambda_B}{}^{+0.019}_{-0.019}\Big|_{\sigma_1}{}^{+0.001}_{-0.062}\Big|_{\mu}\nonumber\\
&& {}^{+0.010}_{-0.004}\Big|_{M^2} {}^{+0.016}_{-0.017}\Big|_{s_0}{}^{+0.153}_{-0.079}\Big|_{\varphi_{\pm}(\omega)},  \label{eq:ErrorsOfBtoKFF} \nonumber\\
 f_{B\to \pi}^{0}(0)&=&0.122\times\bigg[
1\pm0.07\Big|_{S_0^\pi} {} \pm0.11\Big|_{\Lambda_q} \nonumber\\
&& {}\pm0.02
\Big|_{\lambda_E^2/\lambda_H^2}{}^{+0.05}_{-0.06}\Big|_{{M^2}}  \pm0.05\Big|_{2\lambda_E^2+\lambda_H^2}\nonumber\\
&&{}^{+0.06}_{-0.10}\Big|_{\mu_h } \pm0.04\Big|_{\mu} {}
^{+1.36}_{-0.56}\Big|_{\lambda_{B}}{}^{+0.25}_{-0.43}\Big|_{ \sigma_1, \sigma_2}\bigg].
\end{eqnarray}
In these results, the dominant portion of the uncertainty budget is tied to the heavy meson LCDA: parametric errors in the inverse moment $\lambda_B$ and logarithmic moments $\sigma_{i}$ already generate large variations in the predicted form factors, and an additional sizable uncertainty arises from the functional form (model dependence) of the leading-twist LCDAs $\varphi_\pm(\omega)$. In practice, these LCDA-induced uncertainties typically extended well beyond those from standard LCSR systematics such as the Borel parameter, or scale variation, underscoring that the absence of a precise, first-principles determination of heavy meson LCDAs has long been a primary limitation for reliable predictions of $B$-meson decay observables. 

Conventional nonperturbative methods, such as lattice QCD, are formulated after a Wick rotation and therefore do not provide direct access to real-time  correlation functions. As a result,  light-like nonlocal correlators, such as  LCDAs, are traditionally approached via operator product expansion (OPE), in which the light-like bilinear operator is expanded around the short-distance limit into a tower of local operators whose matrix elements (i.e., moments) can be computed nonperturbatively.  However, this standard route encounters a fundamental obstruction for heavy meson LCDAs defined in heavy quark effective theory (HQET): the defining operator involves   effective heavy quark field $h_v$ and a light-like Wilson line, and the cusp at their junction induces a cusp divergence and removes a well-defined local limit~\cite{Braun:2003wx}. Consequently, the OPE breaks down in this case, leading to intrinsic ambiguities in defining the non-negative moments of heavy meson LCDAs, which is one of the principal reasons why first-principles nonperturbative determinations were historically so difficult.

Large-momentum effective theory (LaMET) \cite{Ji:2013dva,Ji:2014gla,Ji:2020ect,Cichy:2018mum} opened a new avenue for computing light-like partonic structures from lattice QCD. Rather than relying on a local OPE in the light-cone separation, LaMET provides a systematic way to connect the Euclidean equal-time correlators to their Minkowski (light-like) counterparts in the large-momentum limit, through factorization and perturbative matching. In this approach, one computes on the lattice a quasi observable, which is defined with a spatial separation at equal Euclidean time, at large hadron momentum $P^z$, and then reconstructs the full distribution of the desired light-cone quantity. This conceptual breakthrough is particularly appealing for heavy meson LCDAs, as it offers a potential path to bypass the cusp-related obstruction inherent to the HQET light-like operator, and thereby enable first-principles access to the LCDA beyond a finite set of moments. The successful LaMET implementations for light meson LCDAs as well as PDFs \cite{Ji:2020ect} further suggest a viable strategy to pursue the heavy meson cases. 

Motivated by merits of LaMET, Refs.~\cite{Kawamura:2018gqz, Wang:2019msf,Zhao:2020bsx,Xu:2022krn,Xu:2022guw,Hu:2023bba,Hu:2024ebp} developed a LaMET program formulated directly in HQET. In the heavy-quark limit, they established a factorization relation between a highly boosted heavy-light quark bilinear equal-time correlator and the corresponding light-like correlation, and proposed to extract the heavy meson LCDAs via an HQET quasi-distribution amplitude (quasi-DAs) computable on the lattice. In this setup, HQET effectively sits ``above” LaMET in the hierarchy of scales, with $m_Q$ parametrically larger than the large hadron momentum $P^z$ used for the LaMET expansion; accordingly, the lattice observable involves an HQET heavy quark field $h_v$ with large velocity $v$. However, realizing an HQET propagator at large velocity on the lattice has long been recognized as technically difficult \cite{Mandula:1990fit,Mandula:1993sj,Meinel:2010uji}, severe signal-to-noise degradation makes the numerical implementation highly challenging, and this has so far prevented the HQET-based LaMET proposal from becoming a fully practical lattice calculation.

An important step toward a practically implementable lattice program was made in 2024, when the first feasible strategy for the lattice determination of heavy meson LCDAs was proposed in Refs.~\cite{Han:2024fkr,LatticeParton:2024zko}. The central obstacle of the earlier HQET-based LaMET constructions can be traced to the assumed “nesting” of effective theories. 
If one insists on integrating out the heavy quark mass first $m_Q\gg P^z$ to arrive at HQET, then the subsequent LaMET step inevitably requires matrix elements of highly boosted HQET operators, i.e., an effective quark field $h_v$ with large velocity. 
The new approach resolves this by reordering the scale separation in a way that is compatible with what lattice QCD can compute directly. One starts from equal-time correlation functions built from full QCD fields in the regime with a clean hierarchy $P^z \gg m_Q \gg \Lambda_{\rm QCD}$, and then sequentially integrates out the two short-distance scales. First, the large momentum $P^z$ is integrated out via LaMET, and only afterward the heavy quark mass $m_Q$ is integrated out via HQET. This ordered two-step EFT, which combining three distinct scales with two effective theories, is referred to as heavy-quark LaMET (HQLaMET), and it enables a genuine first-principles lattice determination of the heavy meson LCDAs as a full distribution rather than only model-dependent inverse moments \cite{LatticeParton:2024zko}.

In this work, we advance the pioneering framework from our earlier study by addressing key limitations and introducing critical improvements to achieve a more rigorous, precise, and comprehensive determination of heavy meson LCDAs. 
\begin{itemize}
\item 
Building on the sequential effective theory methodology, we extend the analysis to multiple lattice ensembles with varying lattice spacings (ranging from $0.0519$ fm to $0.1053$ fm) and pion masses, enabling controlled continuum extrapolation and chiral extrapolation to the physical light-quark mass point. This resolves the single-lattice-spacing constraint of our previous work, allowing us to quantify and reduce discretization artifacts that could impact the reliability of the LCDA predictions. 
\item We refine the lattice simulation setup by incorporating momentum-smeared sources, Hypercubic (HYP) smearing for gauge links in Wilson lines, and optimized interpolating operators, which substantially enhance the signal-to-noise ratio of nonlocal correlators, especially at large spatial separations and high boost momenta $P^z$ (up to 3.5 GeV on the finest ensemble).

\item 
We further strengthen the theoretical consistency of the approach by introducing a cross-validation benchmark using lattice OPE moments. By directly computing the lowest moments of the QCD LCDA from local twist-two operators,  we quantitatively assess power corrections  in LaMET, ensuring these systematic effects to be well-controlled. 

\item We improve the perturbative matching procedure by implementing renormalon resummation to stabilize endpoint behavior, refine the hybrid renormalization scheme by a more rigorous treatment of scale separation $z_s$, and adopt a model-independent Laguerre polynomial parametrization to smoothly merge the lattice QCD results in the peak region  and  the perturbative  QCD calculation in the tail region, yielding a continuous HQET LCDA distribution over the full $\omega$ range. 

\item We simultaneously derive the results for QCD LCDAs (full QCD description) and HQET LCDAs (heavy-quark limit description), each tailored to distinct phenomenological contexts:  QCD LCDAs for processes with momentum transfer much larger than the heavy quark mass 
and HQET LCDAs for heavy meson decays  where the heavy quark behaves as a static source.

\end{itemize}
These advancements collectively reduce theoretical uncertainties, enhance the robustness of our results, and extend the phenomenological applicability of heavy meson LCDAs to precision studies of processes. More details of the phenomenonlogical discussions can be found in Ref.~\cite{HeavymesonDA_short_paper}.


The remainder of this paper is organized as follows. In Sec. II, we review the theoretical framework for heavy meson LCDAs, including the definitions and basic properties of the QCD and HQET LCDAs, as well as the HQLaMET formalism for extracting them from lattice QCD.  In Sec. III, we describe the numerical setup of the lattice calculation, including the gauge ensembles, quark propagators, smearing procedures, and momentum choices used in the quasi-DA calculation. In Sec. IV, we present the lattice determination of the QCD LCDAs, including the extraction of bare quasi-DA matrix elements, nonperturbative renormalization in the hybrid scheme, $\lambda$-extrapolation, Fourier transformation, LaMET matching with endpoint-logarithm resummation, and the continuum, chiral, and infinite-momentum extrapolations to the physical limit. In Sec. V, we benchmark the lowest moments of the QCD LCDAs obtained from LaMET against direct lattice calculations based on the OPE of local twist-two operators, providing a nontrivial check on the control of power corrections. In Sec. VI, we derive the HQET LCDAs by separating the peak and tail regions, combining them through a model-independent Laguerre-polynomial parametrization, and determining the inverse and inverse-logarithmic moments. Finally, Sec. VII contains our conclusions and an outlook. Additional details on the nonperturbative renormalization of OPE moments in the RI/SMOM scheme, together with supplementary fit results, are collected in the Appendix.

\section{Recipes for Heavy Meson LCDAs from Lattice QCD}

\subsection{Definitions of Heavy Meson LCDAs}

LCDAs are universal nonperturbative functions that characterize the longitudinal-momentum structure of a hadron in hard exclusive reactions. By definition, an LCDA is defined through a gauge-invariant, light-like nonlocal quark bilinear correlator between the vacuum and a hadron state, and it describes the probability amplitude for finding the hadron in a minimal Fock configuration (e.g., a $\bar{q}$ pair for a meson) with a given partition of light-ray momentum. 
In QCD factorization theorems, LCDAs enter as the long-distance building blocks in convolution with perturbatively calculable hard kernels, thereby encoding the hadronization dynamics associated with energetic, nearly collinear partons. For heavy mesons, where heavy quark mass $m_Q\gg \Lambda_{\rm QCD}$ (the QCD scale), two distinct classes of LCDAs are physically relevant, each tailored to specific kinematic regimes: QCD LCDAs (relevant for processes with momentum transfer much larger than heavy quark mass) and HQET LCDAs (optimized for the heavy-quark limit, e.g., heavy meson decays).

\subsubsection{QCD LCDAs}
In hard exclusive production of heavy mesons at a large momentum transfer $Q^2$, such like $W\to B\gamma$ and  $\gamma^*/Z^*\to B\bar{B}$ in the kinematic region where the final mesons are energetic and collinear, the relevant hierarchy in factorization is $Q^2\gg m_H^2$. In this case, the heavy meson behaves as a highly boosted collinear hadron and is most conveniently described using full QCD fields. The appropriate long-distance input is therefore a QCD LCDA, defined in close analogy with the standard light meson LCDAs through a gauge-invariant light-like quark bilinear,
\begin{align}
   if_H&\phi(x,\mu) = \int \frac{d\tau}{2\pi} e^{iy\tau n_+ \cdot P}       \nonumber\\
	& \quad\times\langle 0|\bar{q}(\tau n_+)\slash{n}_+\gamma_5 W_c(\tau n_+,0)Q(0)|H(P)\rangle,
\label{eq:definition_of_QCDLCDA}
\end{align}
where $q$ and $Q$ denote the light and heavy quark fields, respectively, and $x\in[0,1]$ represents the light-ray momentum fraction carried by the light quark. $f_H$ is the decay constant of the heavy meson. The gauge invariance of the bilocal operator is ensured by a Wilson line along the light-like direction,
\begin{align}
	W_c(tn_+,0)=P\exp\left[ig\int_0^t ds n_+\cdot A(sn_+) \right],
\end{align}
where $t$ denotes the separation along the light-cone between $q$ and $Q$.

Both the decay constant $f_H$ in full QCD and $\tilde{f}_H$ in HQET are defined from local matrix elements, 
\begin{align}
&\langle 0|\bar q(0)\gamma^\mu\gamma_5Q(0)|H(P)\rangle
= i f_H P^\mu , \\
&\langle 0|\bar q(0)\gamma^\mu\gamma_5h_v(0)|H(v)\rangle
= i\tilde{f}_H(\mu)v^\mu. 
\end{align}
The difference between them is purely short-distance at leading power of $1/m_Q$, and their relation is given by \cite{Eichten:1989zv}
\begin{align}
	f_H=\tilde{f}_H(\mu) \left[1-\frac{\alpha_sC_F}{4\pi}\left(\frac{3}{2}\ln\frac{\mu^2}{m_Q^2}+2 \right) +\mathcal{O}(\alpha_s^2) \right].
\end{align}

For QCD LCDAs, it is common to parameterize them in terms of a finite set of moments or partial-wave (conformal) coefficients. In nonperturbative determinations, such as QCD sum rules \cite{Grozin:1996pq,Braun:2003wx} and traditional lattice QCD calculations\cite{Blum:2001sr,Braun:2006dg,Arthur:2010xf,Bali:2017ude,Bali:2020isn}, a standard strategy is to relate the light-like bilocal operator in Eq.~\eqref{eq:definition_of_QCDLCDA} to a set of local operators through a short-distance OPE. Expanding the bilocal operator around $\tau=0$, one obtains
\begin{align}
&\bar{q}\left(\frac{\tau n_{+}}{2}\right)
\slash{n}_{+}\gamma_{5}
W_c\left(\frac{\tau n_{+}}{2},-\frac{\tau n_{+}}{2}\right)
Q\left(-\frac{\tau n_{+}}{2}\right)
\nonumber\\
&\quad =
\sum_{n=0}^{\infty}
\frac{1}{n!}
\left(\frac{i\tau}{2}\right)^n
\bar{q}(0)\slash{n}_{+}\gamma_{5}
\left(i n_{+}\cdot\overleftrightarrow D\right)^n
Q(0)
+\cdots ,
\label{eq:OPE_bilocal}
\end{align}
where $\overleftrightarrow D\equiv \overrightarrow D-\overleftarrow D$, and $D_{\mu}$ denotes the covariant derivative. The ellipsis denotes higher-twist contributions. Sandwiching Eq.~\eqref{eq:OPE_bilocal} between the vacuum and the meson state, one obtains the standard moment relations, 
\begin{align}
&\langle \xi^{n}\rangle(\mu)
\equiv \int_{0}^{1} dx\, (2x-1)^{n}\phi(x,\mu)
\nonumber\\
&=\frac{1}{i f_H (n_+\cdot P)^{n+1}}
\langle 0|
\bar{q}(0)\slash{n}_+\gamma_5
\left(i n_+\cdot\overleftrightarrow D\right)^{n}
Q(0)
|H(P)\rangle ,
\label{eq:moments_def}
\end{align}
with $\xi\equiv 2x-1$.

An alternative and widely used parametrization is to expand $\phi(x,\mu)$ in Gegenbauer polynomials, which are the eigenfunctions of the evolution kernel~\cite{Efremov:1979qk,Lepage:1980fj}: 
\begin{align}
\phi(x,\mu)
= 6x(1-x)\left[1+\sum_{n=1}^{\infty} a_n(\mu)C_n^{(3/2)}(2x-1)\right],
\label{eq:gegenbauer_expansion}
\end{align}
where $C_n^{(3/2)}$ are Gegenbauer polynomials and the coefficients $a_n(\mu)$ are Gegenbauer moments encoding the nonperturbative shape of the LCDAs. The moments $a_n(\mu)$ can be defined using the orthogonality of $C_n^{(3/2)}$
\begin{align}
a_n(\mu)
= \frac{2(2n+3)}{3(n+1)(n+2)}
\int_0^1 dx C_n^{(3/2)}(2x-1)\phi(x,\mu).
\label{eq:gegenbauer_coeff_def}
\end{align}

With the Gegenbauer representation in Eq.~\eqref{eq:gegenbauer_expansion} and the normalization $\int_0^1dx\phi(x,\mu)=1$, the local OPE moments and the Gegenbauer moments are related by a simple linear transformation:
\begin{align}
	\left(\begin{array}{c}
a_1 \\
a_2 \\
a_3 \\
a_4 \\
\ldots
\end{array}\right)=\left(\begin{array}{cccc}
\frac{5}{3} & 0 & 0 & 0 \\
0 & \frac{35}{12} & 0 & 0 \\
-\frac{9}{4} & 0 & \frac{21}{4} & 0 \\
0 & -\frac{77}{12} & 0 & \frac{77}{8} \\
\cdots &
\end{array}\right)\left(\begin{array}{c}
\langle\xi\rangle \\
\left\langle\xi^2\right\rangle \\
\left\langle\xi^3\right\rangle \\
\left\langle\xi^4\right\rangle \\
\cdots
\end{array}\right)+\left(\begin{array}{c}
0 \\
-\frac{7}{12} \\
0 \\
\frac{11}{24} \\
\cdots
\end{array}\right).
\label{eq:GandLmoments}
\end{align}

\subsubsection{HQET LCDAs}
In heavy-quark decays such as $B\to\pi\pi$, the $B$ meson typically enters factorization theorems as a soft hadron characterized by a heavy-quark velocity $v^{\mu}$ (with $p_B^{\mu}\simeq m_B v^{\mu}$) and a soft spectator quark. The appropriate description is provided by HQET, and the relevant nonperturbative input is the HQET LCDA, whose natural argument is the light-cone projection of the spectator momentum. 

The leading twist HQET LCDA of a heavy pseudoscalar meson $H$ with quark content $Q\bar{q}$ is defined through a vacuum-to-meson matrix element of a gauge-invariant bilocal heavy-light operator, built from the HQET heavy-quark field $h_v$ and the light antiquark field, separated by a light-like distance \cite{Lange:2003ff,Braun:2003wx},
\begin{align}
	i\tilde{f}_H m_H & \varphi ^+(\omega,\mu) = \int\frac{dt}{2\pi}e^{it\omega n_+\cdot v}\nonumber\\
	& \times\langle0\left|\bar{q}_s(tn_+)\slash{n}_+\gamma_5 W_c(tn_+,0)h_v(0) \right|H(v)\rangle, 
\label{eq:definition_of_HQETLCDA}
\end{align}
where $\omega$ is the momentum carried by the light quark, $\tilde{f}_H$ is the HQET decay constant. $h_v$ is the effective heavy-quark field describing a quark moving with four-velocity $v^{\mu}$, and $q_s$ represents the light quark field with soft momentum. The gauge invariance of the bilocal operator is ensured by a Wilson line along the light-like direction $W_c(tn_+,0) $
where $t$ denotes the separation along the light-cone between $q_s$ and $h_v$.

For the HQET LCDA neither the OPE nor the Gegenbauer expansion is well defined. The underlying reason is the cusp divergence \cite{Korchemskaya:1992je} of the heavy-to-light light-like operator. In HQET definition in Eq.~\eqref{eq:definition_of_HQETLCDA}, the effective heavy quark field $h_v$ can be treated as being accompanied by a semi-infinite Wilson line along the time-like direction $v^{\mu}$,  while gauge invariance of the bilocal operator requires an additional Wilson line along the light-like direction $n_+^{\mu}$. Their vertex forms a time-like-to-light-like cusp, whose renormalization is governed by the cusp anomalous dimension and generates UV singularities that are not captured by a naive local limit $t\to0$ \cite{Korchemskaya:1992je,Braun:2003wx,Lange:2003ff}. Consequently, the HQET bilocal operator does not admit a smooth short-distance expansion into the local HQET operators with finite matrix elements. In other words, the non-negative moments for HQET LCDA are ill-defined: radiative corrections generate a hard UV tail of the renormalized distribution at large $\omega$, such that the integral $\int_0^{\infty}d\omega\,\omega^n\varphi^+(\omega,\mu)$ diverges for $n\geq0$ \cite{Braun:2003wx,Lange:2003ff}.

The cusp divergence also prevents  a straightforward conformal (Gegenbauer) expansion. For light mesons, the Gegenbauer basis is singled out by collinear conformal symmetry of light-like operators and by the fact that Gegenbauer polynomials diagonalize the leading-order ERBL evolution kernel \cite{Efremov:1979qk,Lepage:1980fj}. In contrast, the heavy-to-light HQET operator is intrinsically non-conformal: the presence of $h_v$ breaks the conformal structure underlying the ERBL eigenfunctions, and the renormalization-group evolution (RGE) of $\varphi^+(\omega,\mu)$ is governed by a different evolution kernel containing cusp-driven logarithms rather than an ERBL kernel diagonal in Gegenbauer polynomials \cite{Braun:2003wx,Lange:2003ff}.

As a result, the non-negative moments of HQET LCDAs are not available. Instead, phenomenological analyses are typically organized in terms of the inverse moment $\lambda_B^{-1}$ and inverse-logarithmic moments $\sigma_B^{(n)}$, defined by \cite{Braun:2003wx,Lange:2003ff}
\begin{align}
  \lambda_B^{-1}(\mu)=&\int_0^{\infty} \frac{d\omega}{\omega}  \varphi^+(\omega,\mu) , \\
  \sigma^{(n)}_B(\mu) =& \lambda_B(\mu)\int_0^\infty \frac{d\omega}{\omega} \ln \left(\frac{\mu}{\omega}\right)^{(n)}\varphi^+(\omega,\mu), 
 \label{eq:HQETmoments}
\end{align} 
which enter leading-power factorization formulae and therefore play a pivotal role in setting the theoretical precision for many exclusive $B$ decay observables. However, even these inverse (and inverse-logarithmic) moments are not directly accessible in present lattice QCD, since their definitions still rely on the same light-like HQET correlator and inherit the practical obstacles discussed above. 

In short,  while the HQET LCDA is arguably the more ubiquitous nonperturbative input for heavy flavor phenomenology than its QCD counterpart, the cusp-induced breakdown of the local expansion has prevented a nonperturbative determination from lattice QCD. Therefore, since the early development of HQET, no first-principles computation had been achieved for decades. 

\subsection{Theoretical Framework of HQLaMET}

To overcome the long-standing difficulty that neither the full distribution of HQET LCDA nor its inverse moments are directly accessible in lattice QCD, Refs.~\cite{Han:2024fkr,LatticeParton:2024zko,Wang:2025uap} proposed and implemented a new effective theory framework that combines LaMET with a boosted heavy-quark expansion. We will refer to this approach as heavy-quark LaMET (HQLaMET). The basic idea is to start from equal-time Euclidean correlation functions of a highly boosted heavy meson that are directly computable on the lattice, and to construct a quasi-DA whose defining matrix element features a controlled hierarchy of three characteristic scales,
\begin{align}
	\Lambda_{\mathrm{QCD}}\ll m_H (\sim m_Q) \ll P^z,
\end{align}
where $m_Q$ and $m_H$ denote the masses of heavy quark and heavy meson, respectively, and their difference is parametrically soft, $m_H-m_Q\sim\mathcal{O}(\Lambda_{\mathrm{QCD}})$. In this setup, both $m_H$ and $P^z$ are taken to be perturbative scales, so that one can systematically integrate them out in sequence. 
The quasi-DA is defined from an equal-time spatial correlator of a fast-moving heavy pseudoscalar meson $H$ boosted along the $z$-direction,
\begin{align}
	\tilde{\phi}(x,P^z) =& \int\frac{dz}{2\pi}e^{-ixP^zz} \nonumber\\
	&\times \frac{\langle0|\bar{q}(z\hat{n}_z)\Gamma W_c(z\hat{n}_z,0)Q(0)|H(P^z)\rangle}{\langle0|\bar{q}(0)\Gamma Q(0)|H(P^z)\rangle}_R,
\label{eq:def_of_quasiDA}
\end{align}
where $\hat{n}_z=(0,0,0,1)$. The subscript $R$ indicates that the nonlocal operator has been properly renormalized. The matrix $\Gamma$ specifies the Dirac structure of the bilocal quark operator. For a pseudoscalar meson, both $\Gamma=\gamma^t\gamma_5$ and $\gamma^z\gamma_5$ match onto the leading-twist LCDA in the large-momentum limit. 
In practice, $\Gamma=\gamma^z\gamma_5$ is highly preferred because its Dirac structure aligns with the spatial Wilson line. It simplifies the operator's transformation under residual discrete lattice symmetries. Consequently, it rigorously suppresses operator-mixing effects \cite{Liu:2018tox} and ensures a multiplicative renormalization structure, while allowing residual subleading effects to be systematically absorbed into power corrections.

As $P^z$ is increased and eventually taken to the infinite-momentum limit, the quasi-DA $\tilde{\phi}(x,P^z)$ progressively approaches the corresponding QCD LCDA $\phi(y,\mu)$. In the large-$P^z$ limit, quasi and light-cone correlators share the same infrared physics, while their difference is purely short-distance and can be absorbed into a perturbative matching kernel \cite{Ji:2013dva,Ji:2014gla,Ji:2024oka}. One thus obtains the LaMET factorization formula \cite{Ji:2024oka}
\begin{align}
	\phi(y,\mu)=&\int_{-\infty}^{+\infty}dx~ \mathcal{C}\left(x,y,\frac{\mu}{P^z}\right)\tilde{\phi}(x,P^z) \nonumber\\
	&\quad +\mathcal{O}\left(\frac{m_H^2}{(P^z)^2}, \frac{\Lambda_{\mathrm{QCD}}^2}{(yP^z, \bar{y}P^z)^2} \right),
\label{eq:LaMET_match}
\end{align}
where $y$ denotes the light-quark momentum fraction in the QCD LCDA and $\bar{y}\equiv1-y$. In this step, the large boost $P^z$ plays the role of the ultraviolet scale, where $P^z\gg m_H, \Lambda_{\mathrm{QCD}}$, and is integrated out into the perturbative kernel $\mathcal{C}$. 
The matching kernels relating (heavy) meson quasi-DAs to QCD LCDAs have been derived in Refs.~\cite{Liu:2018tox,Xu:2018mpf,Liu:2019urm,Ji:2020brr,LatticeParton:2024zko}.Up to one-loop accuracy, the matching kernel can be written as
\begin{align}
  C\left(x, y, \frac{\mu}{P^z}\right) 
  =~& \delta(x-y)+ C_{B}^{(1)}\left(x, y, \frac{\mu}{P^z}\right)   \nonumber\\
  &  - C_{CT}^{(1)}\left(x, y\right) +\mathcal{O}(\alpha_s^2), 
 \label{eq:C_1loop}
\end{align}
where $C_B^{(1)}$ is the bare one-loop kernel and $C_{CT}^{(1)}$ denotes the counterterm associated with the chosen renormalization prescription for the nonlocal operator defining $\tilde{\phi}$. 

The bare matching kernel $C_B^{(1)}$ is determined entirely by short-distance physics and is therefore insensitive to the hadronic state (i.e., whether the external meson is light or heavy) at leading power. Consequently, the bare kernel coincides with the light-meson result, and we adopt the expression from Ref.~\cite{Liu:2018tox}, 
\begin{align}
 & C_{B}^{(1)}\left(x, y, \frac{\mu}{P^z}\right)\nonumber\\
  =& \frac{\alpha_s C_F}{2 \pi} \begin{cases}{\left[H_1(x, y)\right]_{+}} & x<0<y \\ {\left[H_2\left(x, y, P^z / \mu\right)\right]_{+}} & 0<x<y \\ {\left[H_2\left(1-x, 1-y, \frac{P^z}{ \mu}\right)\right]_{+}} & y<x<1 \\ {\left[H_1(1-x, 1-y)\right]_{+}} & y<1<x \end{cases} \,, 
 \end{align}
where
\begin{align}
  H_1(x, y) &=\frac{1+x-y}{y-x} \frac{1-x}{1-y} \ln \frac{y-x}{1-x} \nonumber\\
 &+\frac{1+y-x}{y-x} \frac{x}{y} \ln \frac{y-x}{-x} \,,\nonumber\\
 H_2\left(x, y, P^z / \mu\right) &=\frac{1+y-x}{y-x} \frac{x}{y} \ln \frac{4 x(y-x)\left(P^z\right)^2}{\mu^2} \nonumber\\
 & +\frac{1+x-y}{y-x}\left(\frac{1-x}{1-y} \ln \frac{y-x}{1-x}-\frac{x}{y}\right) \nonumber\,.
\end{align}

The counterterm $C_{CT}^{(1)}$ depends on the renormalization scheme for the nonlocal operator. In this work we employ the hybrid renormalization scheme \cite{Ji:2020brr}, in which a perturbatively controlled short-distance correction is introduced to remove the $z^2\to0$ singularity and to consistently connect perturbation theory with lattice-renormalized matrix elements. In momentum space, the one-loop correction generates a counterterm contribution $\tilde{\phi}_{CT}$ of the form \cite{Ji:2020brr,LatticeParton:2024zko}
\begin{align}
&\tilde{\phi}_{CT}^{(1)}(x, P^z; x_0) \nonumber\\
&= \int \frac{d \lambda}{2\pi} e^{-i \lambda (x-x_0)}\frac{\alpha_s C_F}{2\pi} \frac{3}{2} \ln\left(\frac{\lambda^2}{z_s^2 (P^z)^2}\right) \theta\left(z_s-\left|\frac{\lambda}{P^z}\right|\right) \nonumber\\
&+ \int \frac{d \lambda}{2\pi} e^{-i \lambda (x-x_0)}\frac{\alpha_s C_F}{2\pi} \frac{3}{2}\ln(z_s^2 (P^z)^2)\\
&= - \frac{3\alpha_s C_F}{4\pi} \frac{2 \, \text{Si}[(x-x_0) z_s P^z] }{\pi (x-x_0)} \ ,
\end{align}
where $\text{Si}$ is the sine integral function. Accordingly, the counterterm kernel entering Eq.~\eqref{eq:C_1loop} can be written as
\begin{align}
C_{CT}^{(1)} = - \frac{3\alpha_s C_F}{4\pi} \left[\frac{2 \, \text{Si}[(x-y) z_s P^z] }{\pi (x-y)}\right]_{+} \ ,
\label{eq:LaMETkernel_ct}
\end{align}
where the plus prescription ensures that the matching preserves the correct normalization of the distribution. Actually, the convolution with the sine integral can be implemented efficiently using a simple Fourier transform, while requiring fewer computational resources.

It is worth noting that the factorization used in this work is formulated at leading power. Accordingly, the residual corrections are organized as an expansion in $m_H^2/(P^z)^2$ and $\Lambda_{\mathrm{QCD}}^2/(yP^z, \bar{y}P^z)^2$ \cite{Ji:2024oka}. The latter terms are enhanced in the endpoint regions $y\to0, 1$, where one of the partonic longitudinal momenta $yP^z$ or $\bar{y}P^z$ ceases to be hard. 
As pointed out in Ref.~\cite{Su:2022fiu}, the perturbative series for the matching kernel contains renormalon ambiguities correlated with these $\mathcal{O}(\Lambda_{\mathrm{QCD}}^2/(yP^z)^2)$-type contributions. By exploiting renormalon resummation, one can systematically reduce the associated uncertainties and stabilize the endpoint behavior.

By contrast, the heavy hadron mass correction $m_H^2/(P^z)^2$ unambiguously dominates the power corrections in the present factorization. This is because $m_H\gg\Lambda_{\mathrm{QCD}}$, making this power expansion parametrically larger than the ones proportional to $\Lambda_{\mathrm{QCD}}$ at a fixed $P^z$. Moreover, unlike the endpoint-enhanced higher-twist corrections, the hadron mass effect contributes over the entire $y$ range. This feature is especially relevant for our subsequent extraction of the HQET LCDA from the peak region, where we will suppress the endpoint domain to minimize the $\Lambda_{\mathrm{QCD}}^2/(yP^z, \bar{y}P^z)^2$ corrections, but must still quantify the $m_H^2/(P^z)^2$ effect. A practical and systematically improvable strategy to estimate this leading correction is to use constraints from an OPE analysis of the first few moments of heavy meson QCD LCDA, which we detail in the next section.

After matching the quasi-DA to the QCD LCDA at large $P^z$, the next step is to disentangle the two remaining scales in $\phi(y,\mu)$: the perturbative heavy mass scale $m_H\sim m_Q$ and the nonperturbative scale $\Lambda_{\mathrm{QCD}}$. At $\mu\sim m_H$, the QCD LCDA is strongly asymmetric. Because the light spectator quark carries only a parametrically small momentum fraction,
\begin{align}
	y\sim\frac{\Lambda_{\mathrm{QCD}}}{m_H}\ll1, 
\end{align} 
so that $\phi(y,\mu)$ develops a narrow peak near the endpoint. An expansion-by-regions analysis \cite{Beneke:2023nmj} shows that the structure of the QCD LCDA at $\mu\sim m_H$ is most transparently organized by separating two parametric domains: the peak region at $y\sim \Lambda_{\mathrm{QCD}}/m_H$ and the tail regionat $y\sim \mathcal{O}(1)$.

In the peak region, the light antiquark carries a soft light-ray momentum $\omega \equiv y m_H \sim \mathcal{O}(\Lambda_{\mathrm{QCD}})$, so the nonperturbative shape is governed by the universal HQET LCDA $\varphi^+(\omega,\mu)$. The dependence on the heavy scale $m_H\sim m_Q\gg\Lambda_{\mathrm{QCD}}$ is short-distance and can be factorized into a perturbative matching coefficient, often referred to as a jet function in the boosted-HQET formulation. This scale separation yields the following factorization relation at leading power in $\Lambda_{\mathrm{QCD}}/m_H$ \cite{Ishaq:2019dst,Zhao:2019elu,Beneke:2023nmj}
\begin{align}
	\varphi^+_{\mathrm{peak}}(\omega,\mu) = \frac{1}{m_H}\frac{f_H}{\tilde{f}_H} \frac{1}{\mathcal{J}_{\mathrm{peak}}} \phi(y,\mu),
  \label{eq:HQET_matching_peak}
\end{align}
where $\mathcal{J}_{\mathrm{peak}}$ collects the hard and hard-collinear contributions associated with the scale $m_H$, while $\varphi^+$ encodes the soft dynamics at $\omega\sim\mathcal{O}(\Lambda_{\mathrm{QCD}})$. At one-loop accuracy, this perturbative jet function reads \cite{Beneke:2023nmj},
\begin{align}
	\mathcal{J}_{\mathrm{peak}} =& 1+\frac{\alpha_sC_F}{4\pi} \left( \frac{1}{2}\ln^2\frac{\mu^2}{m_H^2} + \frac{1}{2}\ln\frac{\mu^2}{m_H^2} + \frac{\pi^2}{12} + 2  \right) \nonumber\\
	& \quad + \mathcal{O}\left(\alpha_s^2\right).
\end{align}

By contrast, in the tail region $y\sim\mathcal{O}(1)$, or equivalently $\omega\sim m_H$, the light spectator quark carries a hard-collinear light-ray momentum on the order of the heavy scale. In this region, the HQET LCDA is generated purely by short-distance radiation and is therefore perturbatively calculable. Starting at $\mathcal{O}(\alpha_s)$ \cite{Lee:2005gza}, the tail distribution is given by
\begin{align}
	\varphi^+_{\mathrm{tail}}(\omega, \mu) = \frac{\alpha_sC_F}{\pi\omega} \left[\left(\frac{1}{2}-\ln\frac{\omega}{\mu} \right) + \frac{4\bar{\Lambda}}{3\omega} \left(2- \ln\frac{\omega}{\mu} \right) \right],
  \label{eq:HQET_matching_tail}
\end{align}
where $\bar{\Lambda}=m_H-m_Q$ parameterizes the power suppressed effects in   heavy-quark expansion.

In the following calculations, we will exploit this ``peak-and-tail" structure by extracting $\varphi^+$ in the peak region from the lattice-determined QCD LCDA, while treating the tail $\omega\sim m_H$ with perturbative QCD. In the intermediate window, $\Lambda_{\rm QCD} \ll \omega \ll m_H$, neither a fixed-order perturbative expansion nor a direct nonperturbative determination is strictly reliable in practice. We therefore adopt a model-independent parametrization strategy that smoothly interpolates between these two domains, rigorously preserving the known constraints from both the nonperturbative peak and the perturbative tail. The corresponding numerical implementations will be presented in Sec.~\ref{sec:HQET_and_Pheno}.

\subsection{Moments of QCD LCDAs from  OPE }

The LaMET factorization in Eq.~\eqref{eq:LaMET_match} is formulated at leading power in the large-momentum limit, and its dominant systematic uncertainty for heavy mesons is expected to come from the hadron mass correction $\sim m_H^2/(P^z)^2$. In this subsection we introduce a moment-based cross-check to quantify whether this leading finite-$P^z$ effect is under control in our LaMET determination of the QCD LCDA $\phi(y,\mu)$. The key observation is that two complementary nonperturbative strategies access the same QCD LCDA but involve different hard scales.


On the one hand, LaMET reconstructs the full distribution $\phi(y,\mu)$ from equal-time correlators of a highly boosted heavy meson, where the largest scale is $P^z$ and residual finite-$P^z$ effects enter as power corrections \cite{Ji:2013dva,Ji:2014gla,Liu:2018tox}. On the other hand, the traditional lattice OPE approach does not aim to reconstruct the full light-ray correlator, instead it computes vacuum-to-meson matrix elements of local twist-two operators and thus directly yields the first few moments of $\phi(y,\mu)$ \cite{Efremov:1979qk,Lepage:1980fj}. In this approach the relevant short-distance scale is set by the heavy mass $m_H\sim m_Q$, and no additional scale larger than $m_H$ is introduced.


To define the lowest moments, it is convenient to use $\langle \xi^n\rangle(\mu)$ defined in Eq.~\eqref{eq:moments_def}  and 
\begin{align}
\langle \mathbf{1}\rangle &\equiv \int_0^1 dy \phi(y,\mu)=1.
\label{eq:xi_mom_def}
\end{align}
These moments can be obtained from the short-distance expansion of light-ray operator and are represented by matrix elements of the following (bare) local operators:
\begin{align}
\mathcal{O}_\rho & =\bar{q} \gamma_\rho \gamma_5 Q, \label{eq:localoperformoments1}\\
\mathcal{O}_{\rho \mu}^{-} & =\bar{q}\,\gamma_{(\rho} \gamma_5\left[\overleftarrow{D}_{\mu)}-\overrightarrow{D}_{\mu)}\right]  Q, \\
\mathcal{O}_{\rho \mu \nu}^{-} & =\bar{q}\,\gamma_{(\rho} \gamma_5\left[\overleftarrow{D}_{\mu} \overleftarrow{D}_{\nu)}-2 \overleftarrow{D}_{\mu} \overrightarrow{D}_{\nu)}+\overrightarrow{D}_{\mu} \overrightarrow{D}_{\nu)}\right]  Q, \\
\mathcal{O}_{\rho \mu \nu}^{+} & =\bar{q}\, \gamma_{(\rho} \gamma_5 \left[\overleftarrow{D}_{\mu} \overleftarrow{D}_{\nu)} +2 \overleftarrow{D}_{\mu} \overrightarrow{D}_{\nu)}+\overrightarrow{D}_{\mu} \overrightarrow{D}_{\nu)}\right] Q, \label{eq:localoperformoments2}
\end{align}
where parentheses denote symmetrization over Lorentz indices and subtraction of traces is implied to project onto the leading-twist (symmetric-traceless) component. On the lattice, the covariant derivatives $D_{\mu}$ are implemented with a symmetric discretization. To avoid operator mixing and preserve rotational symmetry on lattice, in the calculation we adopt the off-diagonal components with all Lorentz indices distinct, that is $\mathcal{O}_{4i}^{-}\,(i=1,2,3)$ for the first moment and $\mathcal{O}_{4ij}^{\pm}\, (i,j=1,2,3,\, i\neq j)$ for the second moment. These operators transform in an appropriate irreducible representation of $H(4)$ and therefore do not mix with lower-dimensional operators on the lattice \cite{Braun:2015axa,Bali:2017ude,RQCD:2019osh}.

The corresponding vacuum-to-meson matrix elements define the lowest moments through
\begin{align}
\langle 0| \mathcal{O}_\rho|H(P)\rangle & =i f_H P_\rho\langle\mathbf{1}\rangle, \label{eq:OPEmoments1}\\
\langle 0| \mathcal{O}_{\rho \mu}^{-}|H(P)\rangle & = f_H P_\rho P_\mu\langle\xi\rangle, \\
\langle 0| \mathcal{O}_{\rho \mu \nu}^{+}|H(P)\rangle & =-i f_H P_\rho P_\mu P_\nu\langle\mathbf{1}\rangle, \\
\langle 0| \mathcal{O}_{\rho \mu \nu}^{-}|H(P)\rangle & =-i f_H P_\rho P_\mu P_\nu\left\langle\xi^2\right\rangle, \label{eq:OPEmoments2}
\end{align}
so that the zeroth, first and second moments $\langle\mathbf{1}\rangle$, $\langle\xi\rangle$ and $\langle\xi^2\rangle$ are accessible from lattice calculations of these local-current matrix elements. 
In this work, these local operators are renormalized nonperturbatively in the RI/SMOM scheme \cite{Martinelli:1994ty,Sturm:2009kb}. The details of the lattice extraction of the bare matrix elements and their nonperturbative renormalization are presented in Sec.~\ref{sec:LaMET}.

\begin{table*}[htbp]
\centering
\renewcommand{\arraystretch}{1.8}
  \setlength{\tabcolsep}{2.5mm}
\begin{tabular}{l l l}
\hline
 & LaMET & OPE \\
\hline
Output
&\parbox[t]{7.1cm}{\raggedright
$x$-dependent partonic distributions 
of LCDAs}
&
A finite set of moments of LCDAs
\\
Pros
&\parbox[t]{7.1cm}{\raggedright
  1) Direct access to the full $x$-dependence of partonic distributions \\
2) Direct comparison with global fits and phenomenology
}

&\parbox[t]{7.1cm}{\raggedright
1) No requirement for large hadron momentum \\
2) Precise determination of lowest-order moments \\
3) QCD sum-rule constraints for consistency checks
}
\\
Cons
&\parbox[t]{7.1cm}{\raggedright
1) Requirement for large hadron momentum $P^z$ \\
2) Difficulties with discretization effects, signal-to-noise degradation, and excited-state contamination
}
&\parbox[t]{7.1cm}{\raggedright
1)Difficulties with higher moments (larger noise, operator mixing, complex renormalization)\\
2)  Inability to directly determine full partonic distributions \\
}
\\
\noalign{\vskip 10pt}
\hline
\end{tabular}
\caption{Comparison between the LaMET and Lattice OPE approaches. The two methods are complementary: LaMET provides access to the $x$-dependent structure, while Lattice OPE yields precise low-moment constraints.}
\label{tab:lamet_ope_comparison}
\end{table*}

A comparison of the OPE calculation with LaMET is given in Tab.~\ref{tab:lamet_ope_comparison}. These two methods are complementary: LaMET provides access to the $x$-dependent structure, while Lattice OPE yields precise low-moment constraints.

\section{Numerical Simulation Setup}
\label{sec:latticesetup}

\begin{table*}[htbp]
  \centering
  \renewcommand{\arraystretch}{1.8}
  \setlength{\tabcolsep}{2.5mm}
  \begin{tabular}{l c c c c c c c}
  \hline
    \multirow{2}{*}{Ensemble} &
    \multirow{2}{*}{$a$ (fm)} &
    \multirow{2}{*}{$L^3\times T$} &
    \multirow{2}{*}{$m_\pi$ (MeV)} &
    \multirow{2}{*}{$m_D$ (MeV)} &
    \multicolumn{2}{c}{$n_{\rm cfg}\times n_{\rm meas}$} \\
    \cmidrule(lr){6-7}
     & & & & & LaMET & OPE moment $\langle\xi\rangle$ & OPE moment $\langle\xi^2\rangle$ \\
    \hline
    C24P29 & \multirow{2}{*}{0.1053} & $24^3\times 72$  & 292.7(1.2) & 1885.7(3.6)  & ---      &   \multirow{2}{*}{$50\times16$}  & $440\times 16$ \\
    C48P14 &                          & $48^3\times 96$  & 135.5(1.6) & 1864.8(3.2) & ---      &     & $304\times 48$ \\
    \hline
    F32P30 & \multirow{2}{*}{0.0775}  & $32^3\times 96$  & 303.2(1.3) & 1887.9(1.3) & $900\times102$ & $50\times16$   & $231\times 16$ \\
    F32P21 &                          & $32^3\times 64$  & 210.9(2.2) & 1869.3(3.1)  & $459\times128$ & --- & ---           \\
    
    \hline
    G36P29 & 0.0683 & $36^3\times 108$ & 295.1(1.2) & 1873.1(1.0) & $656\times86$ & $50\times16$   & $117\times 16$ \\
    \hline
    H48P32 & 0.0519 & $48^3\times 144$ & 317.2(0.9) & 1882.4(0.8) & $550\times108$ & $50\times16$   & $111\times 16$ \\
    \hline
  \end{tabular}
  \caption{Gauge ensembles used in this work. We list the lattice spacing $a$, volume $L^3\times T$, pion mass $m_\pi$, and the tuned $D$-meson mass $m_D$ for each ensemble. The last two columns show the statistics used in the LaMET quasi-DA calculation and in the lattice-OPE determination of the lowest LCDA moments, quoted as $n_{\rm cfg}\times n_{\rm meas}$.}
  \label{tab:ensembles}
\end{table*}

The lattice-QCD calculations in this work are performed on $N_f=2+1$ gauge ensembles generated by  CLQCD collaboration with a tree-level Symanzik-improved gauge action and stout-smeared clover Wilson fermions \cite{Zhang:2021oja,Hu:2023jet,CLQCD:2023sdb,CLQCD:2024yyn}. A single step of stout-link smearing is applied to the gauge links entering the clover action, which improves the numerical stability of the simulations at fixed bare quark masses and helps reduce discretization artifacts in hadron observables \cite{Hu:2023jet}. The ensemble parameters used in this work are summarized in Table~\ref{tab:ensembles}, including the lattice spacing $a$, lattice volume $L^3\times T$, pion mass $m_{\pi}$, etc.

A key practical challenge in the lattice calculation is the hierarchy $\Lambda_{\mathrm{QCD}}\ll m_H\ll P^z$ underlying HQLaMET. Achieving parametrically large boosts $P^z$ is difficult to achieve on current lattices, particularly at finer lattice spacings where discretization effects at large momentum are better controlled. For this reason, we choose the $D$ meson as our representative heavy meson: its mass is sufficiently large for the heavy-quark expansion to be applicable, yet moderate enough that the regime $P^z\gg m_H$ can be approached with momenta that remain feasible on the available ensembles. 
The charm quark masses used in this work are taken from the interpolated bare charm-mass parameters provided in Ref.~\cite{CLQCD:2024yyn}. With these inputs, the $D$ meson masses on each ensemble are determined from standard two-point spectroscopy and are listed in Table~\ref{tab:ensembles}. Extrapolating to the continuum limit and the physical light quark masses yields $m_D=1.862(12)~\mathrm{GeV}$, consistent with its physical values \cite{ParticleDataGroup:2024cfk}.

To determine the QCD LCDA of the $D$ meson, we compute the full $y$-dependent distribution using LaMET and use the lattice-OPE moments as a quantitative benchmark. Since the two approaches probe different hard scales, they impose different numerical requirements. The LaMET calculation requires a large hadron boost $P^z$, and hence benefits from relatively fine lattices to better control discretization effects at high momentum. In practice, our LaMET analysis is carried out on the four finest ensembles F32P30, F32P21, G36P29, H48P32, which span multiple lattice spacings and pion masses, enabling controlled continuum and chiral extrapolations. For these ensembles, the lattice spacings range from $a\simeq0.0775$ fm down to $a\simeq0.0519$ fm, allowing us to reach momenta in the few-GeV regime. Concretely, we take $P^z=2\pi n_z/(La)$, with $n_z=5$ on F32P30, F32P21, and G36P29, corresponding to $P^z\simeq2.5$ GeV. On the finest ensemble H48P32, we further compute at $n_z=\{5,6,7\}$, i.e., $P^z\simeq\{2.5,3,3.5\}$ GeV, which allows an explicit large-$P^z$ extrapolation. The statistics used on each ensemble, quoted as $n_{\rm cfg}\times n_{\rm meas}$, are summarized in Table~\ref{tab:ensembles}.

For the lattice OPE determination of the lowest moments, we additionally include two coarser ensembles, C24P29 and C48P14, to strengthen control over discretization effects in the continuum extrapolation of the local-operator matrix elements. In extracting the zeroth-, first-, and second-moment matrix elements in Eq.~\eqref{eq:OPEmoments1}, we evaluate correlators at several small spatial momenta, $\vec{P}=\left\{(0,0,0),~(0,0,1),~(0,1,1),~(1,1,1)\right\}\times(2\pi)/(La)$, where the largest momentum is $\sim0.86$ GeV.  This provides an internal consistency check of the moment extraction across different $\vec{P}$. For the final quoted moments, we take the $\vec{P}=0$ results, which have the smallest statistical uncertainties, and use them as the benchmark for validating the LaMET determination. The LaMET analysis and the OPE-moment calculation are detailed in Secs.~\ref{sec:LaMET} and \ref{sec:OPE_moments}, respectively.

In the calculations we employ quark propagators with momentum-smeared sources and point sink. The momentum smearing procedure \cite{Bali:2016lva} is proposed to optimize the overlap of the interpolating field with a hadron carrying a prescribed spatial momentum, which significantly improves the signal quality for boosted hadron correlators and simultaneously helps to suppress the excited-state contaminations at practical source-sink separations. 
In addition, to enhance the signal of the nonlocal operator matrix elements at large spatial separations $z$, we apply one step of Hypercubic (HYP) smearing \cite{Hasenfratz:2001hp} to the gauge links entering the straight Wilson line in the quasi-DA operator. This link smearing reduces ultraviolet fluctuations on the gauge connection and substantially improves the signal-to-noise behavior of the spatially extended correlators that define $\tilde{\phi}(x,P^z)$, especially in the large-$z$ region relevant for resolving the momentum dependence. A detailed discussion of HYP smearing for nonlocal operators in the quasi-distributions can be found in Ref.~\cite{Tan:2025ofx}.

\begin{figure}[http]
\centering
\includegraphics[width=1.00\linewidth]{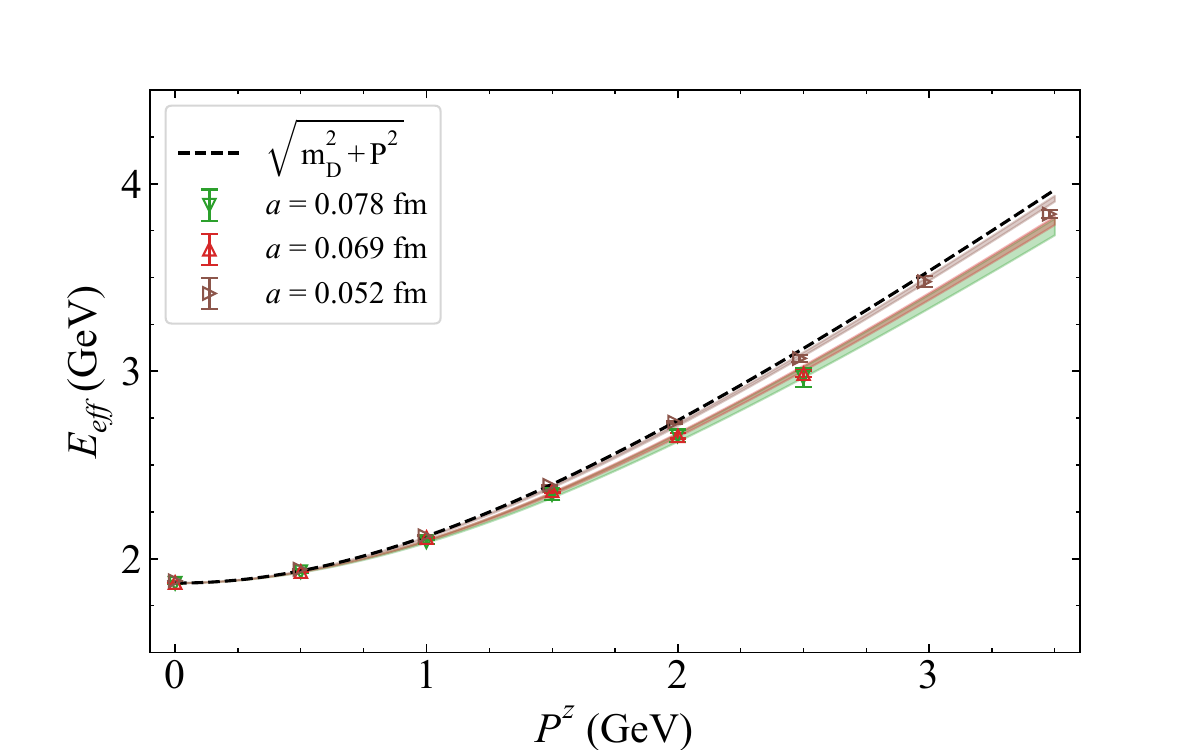}
\caption{ Dispersion relation of the boosted $D$ meson on the ensembles used in this work. The effective ground-state energy $E_{\rm eff}$ extracted from two-point correlators is shown as a function of the momentum magnitude $P$ for C24P29, C48P14, F32P30, F32P21, G36P29, and H48P32. The dashed curve denotes the continuum relativistic dispersion relation $E^2={m_{D}}^2+P^2$, where $m_{D}=1.87\,{\rm GeV}$.}
\label{fig:disprelation}
\end{figure}

To demonstrate that the discretization effects are under control for the boosted $D$ mesons, we examine the dispersion relation on each ensemble. As shown in Fig.~\ref{fig:disprelation}, we determine the ground-state $D$ meson energies at the same momentum choices employed in  LaMET and OPE-moment calculations, and compare the result with the continuum relativistic form $E^2=m_D^2+\vec{P}^2$. We observe that, for the momenta considered in this work, the lattice results are globally   consistent with the continuum dispersion relation within uncertainties. Moreover, the agreement systematically improves as the lattice spacing is reduced, as expected if the leading cutoff effects scale with powers of $aP$. This behavior suggests that the subsequent continuum extrapolation is reliable.

\section{LaMET Determination of the QCD LCDA}
\label{sec:LaMET}
\subsection{Lattice QCD Calculation of the Bare Quasi-DA Matrix Elements}

Building on the ensemble choices and momentum setup summarized in Sec.~III, we now compute the bare quasi-DA matrix elements entering Eq.~\eqref{eq:def_of_quasiDA} in lattice QCD. The basic lattice observable is the vacuum-to-meson matrix element of a quark bilinear formed by a heavy and a light quark operator, connected by a  straight Wilson   line along the $z$ direction. We access it through boosted Euclidean two-point correlation function with an insertion of the nonlocal operator at spatial separation $z$,
\begin{align}
	C_2(z,P^z,t) =&~ a^3\sum_{x^3}e^{iP^zx^3} \left\langle G_q(x^3+z,t;\vec{0},0)\gamma^z\gamma_5 \right. \nonumber\\
	& \times \left. W_c(x^3+z,x^3)\gamma_5 G_Q^{\dagger}(x^3,t;\vec{0},0)\gamma_5\Gamma_{\mathrm{src}}\right\rangle,
\end{align}
where $z$ is the spatial separation between the light and heavy fields in the nonlocal operator, $G_f(\vec{x},t;\vec{0},0)$ denotes the $f$-flavored smeared-source-to-point-sink quark propagator from $(\vec{0},0)$ to $\vec{x},t$, and $W_c(x^3+z,x^3)$ is the straight Wilson line along the $z$ direction. As mentioned above, we use HYP smeared links for $W_c$ to suppress UV fluctuations of the gauge connection and improve the signal quality of the nonlocal correlator at large separations $z$, which are essential for resolving the momentum dependence of the quasi distributions \cite{Tan:2025ofx}.
In addition, we use kinematically optimized (momentum-enhanced) interpolating operators for the boosted pseudoscalar meson. In particular, we use the meson interpolator containing $\Gamma=\gamma^t\gamma_5$ as advocated in Ref.~\cite{Zhang:2025hyo}, which significantly improves the stability of boosted correlators at large $P^z$.

The bare quasi-DA matrix element associated with Eq.~\eqref{eq:def_of_quasiDA} is
\begin{align}
	\tilde{M}^{(0)}(z,P^z)\equiv \langle 0|\bar{q}(z)\gamma^z\gamma_5 W_c(z,0)Q(0)|H(P^z)\rangle,  
\label{eq:bare_ME}
\end{align}
which we extract from the ratio of the nonlocal to local correlation functions,
\begin{align}
\frac{C_2(z,P^z,t)}{C_2(z=0,P^z,t)}=
\tilde{M}^{(0)}(z,P^z)\left[
1+\sum_{n} A_ne^{-(E_n-E_0)t}
\right],
\label{eq:param_of_ratio}
\end{align}
where $E_0$ and $E_n$ denote the ground-state and excited-state energies of the pseudoscalar heavy meson, and the amplitudes $A_n$ parameterize excited-state contamination. We determine $\tilde{M}^{(0)}(z,P^z)$ from correlated multi-state fits according to the $t$-dependence of the ratio, following the systematic strategy developed in Ref.~\cite{LatticeParton:2024zko}. While Ref.~\cite{LatticeParton:2024zko} presented a detailed analysis for the $D$ meson quasi-DA on the ensemble H48P32, in the present work we carry out the same procedure  to perform the fits on a broader set of ensembles listed in Table~\ref{tab:ensembles}. Representative fit examples on these ensembles are provided in the Appendix.~\ref{ax:MEs_fits}.

\subsection{ Nonperturbative Renormalization}

\begin{figure}[http]
	\centering
	\includegraphics[width=1.00\linewidth]{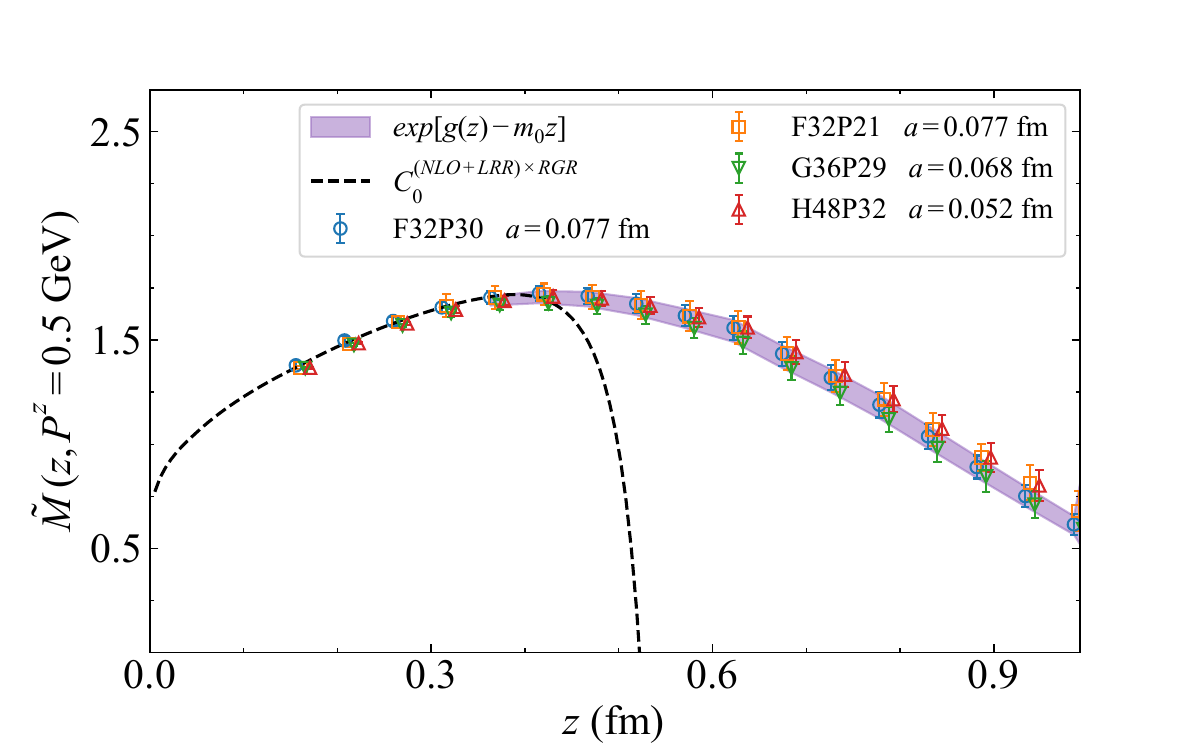}
\caption{Hybrid-scheme renormalization in coordinate space. The reference-momentum matrix element at $P_{\rm ref}^z\simeq0.5~{\rm GeV}$ is shown as a function of the spatial separation $z$ (points), together with the perturbative short-distance curve \cite{Izubuchi:2018srq} and the long-distance fit used to determine $Z_R$.}
\label{fig:smallPz_ME}
\end{figure}

The bare nonlocal matrix elements $\tilde{M}^{(0)}(z,P^z)$ defined in Eq.~\eqref{eq:bare_ME} contain both a linear divergence associated with the self-energy of the straight Wilson line and the usual logarithmic UV divergences of the composite operator. These divergences are multiplicative and can be removed nonperturbatively. In this work we adopt the hybrid renormalization scheme proposed in Ref.~\cite{Ji:2020brr,LatticePartonLPC:2021gpi}, which combines a ratio renormalization at short distances with an explicit subtraction at long distances, thereby maintaining perturbative control at small $|z|$ while providing a practical treatment of the Wilson-line power divergence at large $|z|$.

Concretely, we define the renormalized matrix element in coordinate space as
\begin{align}
	\tilde{M}^R&(z,P^z)=\frac{\tilde{M}^{(0)}(z,P^z,1/a)}{\tilde{M}^{(0)}(z, P^z_{\mathrm{ref}},1/a)}\theta(z_s-|z|) \nn\\
	& + \frac{\tilde{M}^{(0)}(z,P^z,1/a)}{Z_R(z,1/a)}\frac{Z_R(z_s,1/a)}{\tilde{M}^{(0)}(z_s,P^z_{\mathrm{ref}},1/a)}\theta(|z|-z_s),
\end{align}
where $z_s$ separates a short-distance region in which perturbation theory is reliable from a long-distance region in which we determine an explicit renormalization factor $Z_R$.

In the short distance region $|z|\leq z_s$, where the perturbation theory works well, the renormalization can be carried out by dividing by the same hadronic matrix element at small reference momentum $P^z_{\mathrm{ref}}$ \cite{Ji:2020brr}. In many applications one can choose $P^z_{\mathrm{ref}}=0$, however, our operator choice $\Gamma=\gamma^z\gamma_5$ implies that the corresponding matrix element from Eq.~(\ref{eq:bare_ME}) vanishes at $P^z=0$. We therefore take the smallest nonzero lattice momentum as reference, $P^z_{\mathrm{ref}}=0.5\,\mathrm{GeV}$ and verify that the matrix element at $P^z_{\mathrm{ref}}$ agrees well with the perturbative expectation \cite{Izubuchi:2018srq} for $|z|\lesssim0.2\,\mathrm{fm}$, as shown in Fig.~\ref{fig:smallPz_ME}. Accordingly, we set the nominal short-distance cutoff to $z_{s}=0.156\,\mathrm{fm}$ and vary it within $z_s\in[0.104,\,0.208]\,\mathrm{fm}$ to estimate the associated systematic uncertainty.

At long distances $|z|> z_s$, we remove the UV divergences through an explicit renormalization factor $Z_R(z,1/a)$. The parametrization of the renormalization factor in the nonperturbative region can be expressed as \cite{Ji:2020brr,Ji:2020brr}
\begin{align}
 	Z_R&(z,1/a)=\exp\left\{ \frac{kz}{a\ln[a\Lambda_{\mathrm{QCD}}]} +m_0z + f(z)a^2  \right. \nn\\
 	& \left.+ \frac{3C_F}{b_0} \ln\left[\frac{\ln[1/(a\Lambda_{\mathrm{QCD}})]}{\ln[\mu/\Lambda_{\mathrm{QCD}}^{\mathrm{\overline{MS}}}]} \right] + \ln\left[1+\frac{d}{\ln(a\Lambda_{\mathrm{QCD}})} \right] \right\},
 \end{align}
where $k$ encodes the coefficient of the linear divergence, $m_0$ parameterizes the scheme-dependent finite linear-$z$ contribution associated with renormalization ambiguity, $f(z)a^2$ captures residual discretization effects, and the last two terms account for leading and subleading logarithmic dependence as in Ref.~\cite{Ji:2020brr}.
The parameters are determined from a global fit to lattice data in the region $|z|> z_s$. For $z_{s}=0.156\,\mathrm{fm}$, we obtain 
\begin{align}
k=1.03(36),\qquad
m_0=0.13(15)\,{\rm GeV},\nn\\
\Lambda_{\rm QCD}=0.05 (11)\,{\rm GeV},\qquad
d=0.9(1.2),
\end{align}
with $\chi^2/\mathrm{d.o.f}=0.42$. The resulting fit curve in Fig.~\ref{fig:smallPz_ME} describes the long-distance behavior well, indicating that the UV divergences in the bare matrix elements have been properly subtracted.

\begin{figure}[http]
	\centering
	\includegraphics[width=1.00\linewidth]{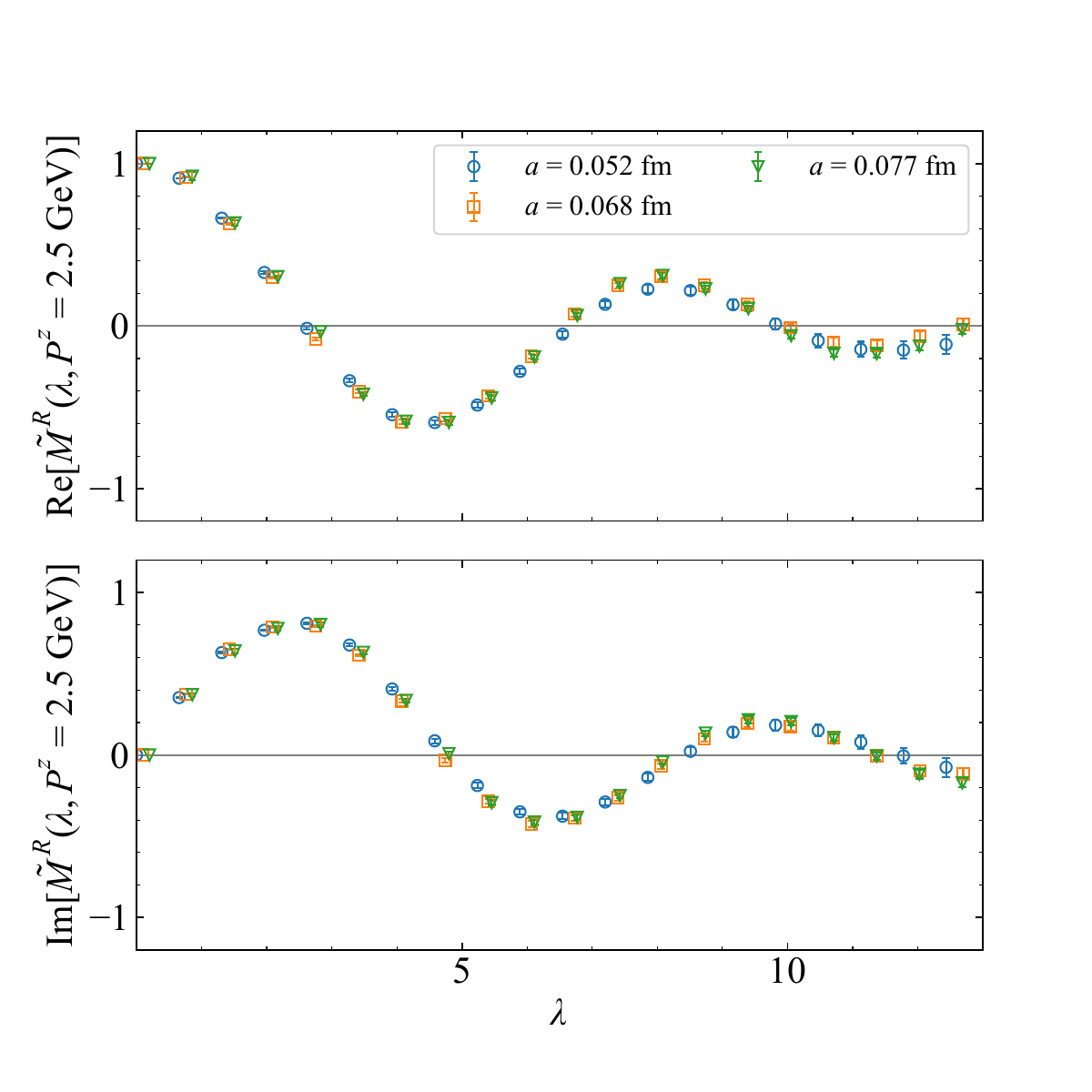}
\caption{Renormalized matrix elements $\tilde{M}^R(\lambda,P^z)$ at $P^z\simeq2.5~{\rm GeV}$ as functions of $\lambda=zP^z$ on the F32P30, G36P29 and H48P32 ensembles (upper: real part; lower: imaginary part).}
\label{fig:renormME_with_a}
\end{figure}

\begin{figure}[http]
	\centering
	\includegraphics[width=1.00\linewidth]{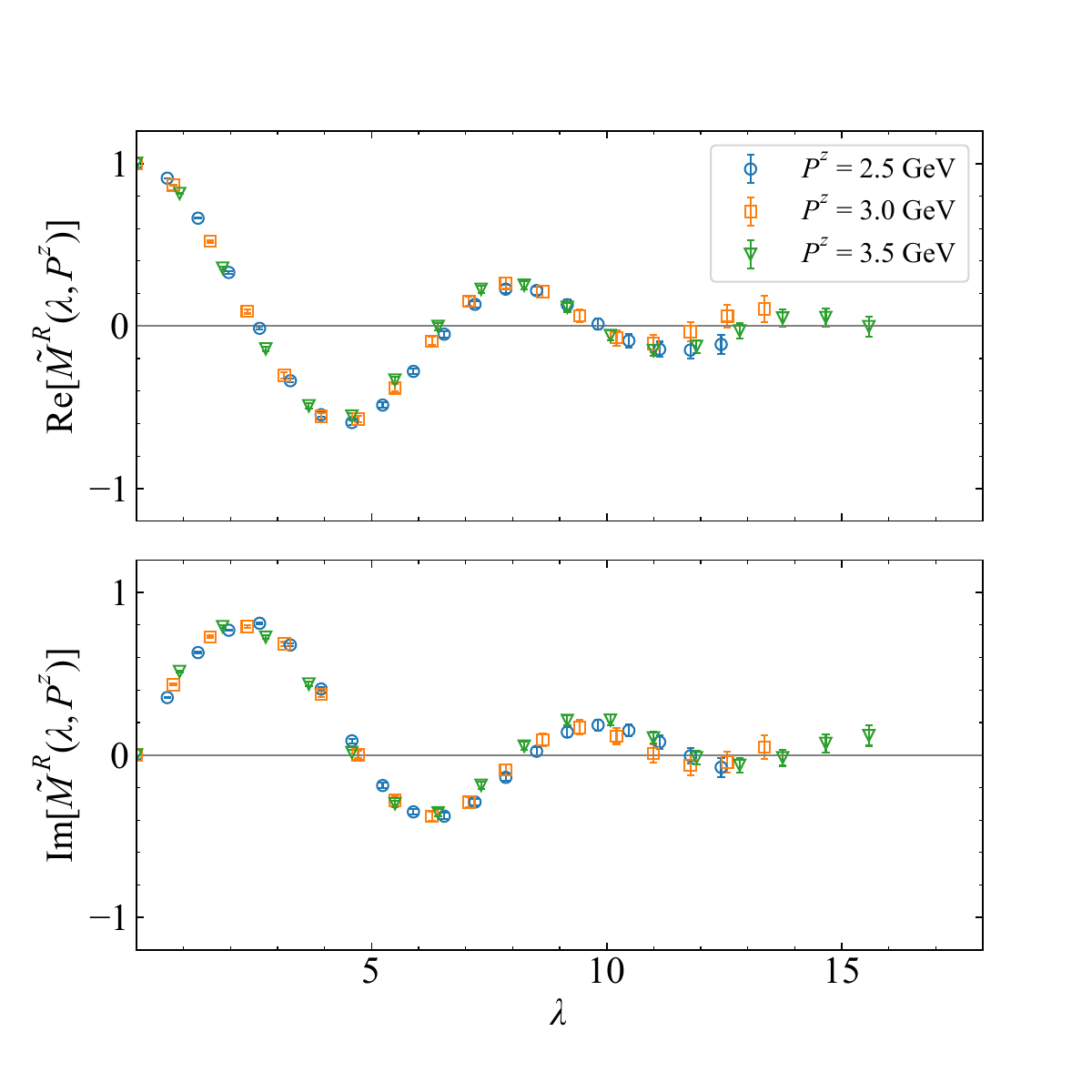}
\caption{Momentum dependence of the renormalized matrix elements on the H48P32 ensemble. Results for $P^z\simeq2.5$, $3.0$ and $3.5~{\rm GeV}$ are shown as functions of $\lambda=zP^z$ (upper: real part; lower: imaginary part).}
\label{fig:renormME_with_Pz}
\end{figure}

\begin{figure}[http]
	\centering
	\includegraphics[width=1.0\linewidth]{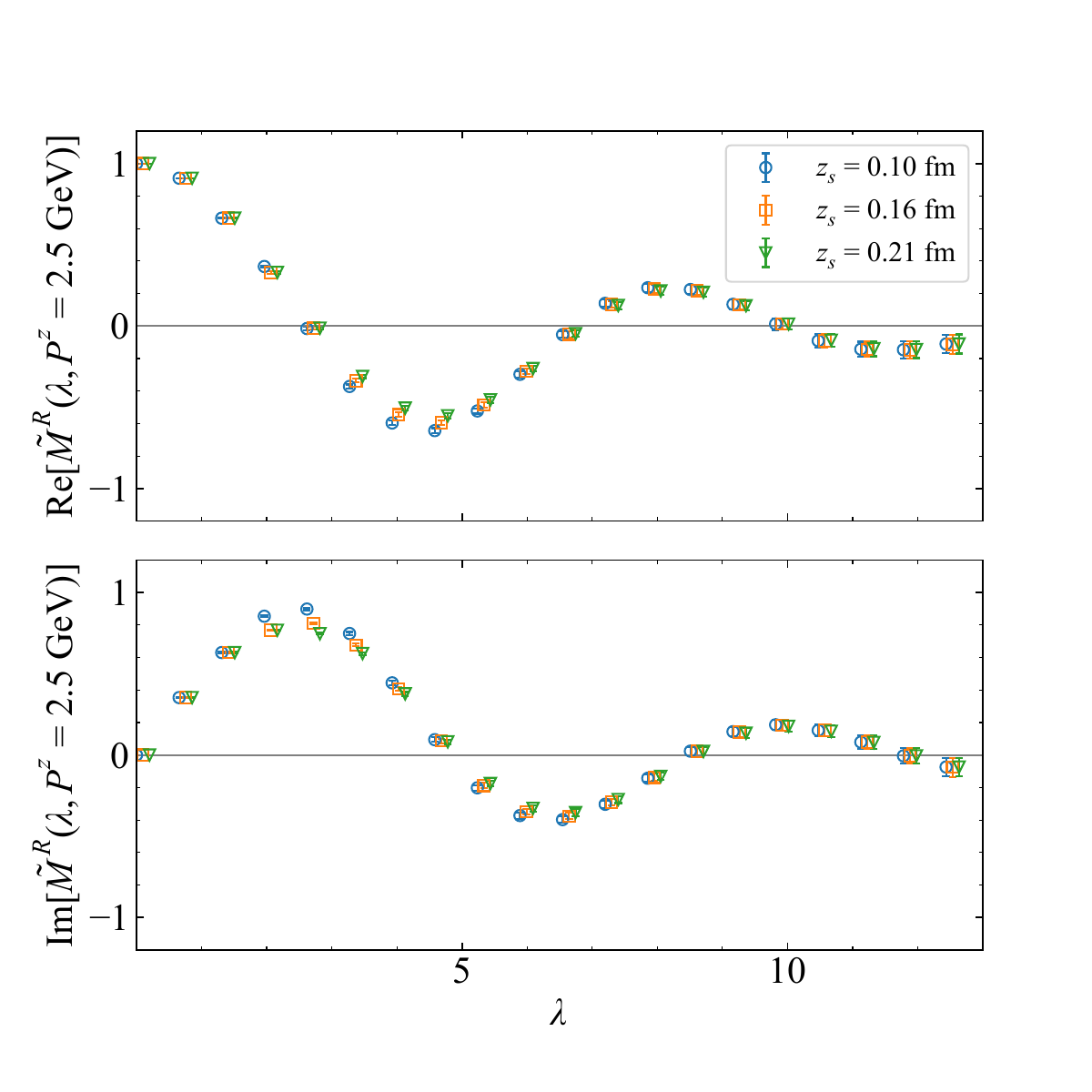}
\caption{Dependence of $\tilde{M}^R(\lambda,P^z)$ on the hybrid separation scale $z_s$ on H48P32 at $P^z\simeq3.5~{\rm GeV}$, comparing two representative choices of $z_s$ (upper: real part; lower: imaginary part).}
\label{fig:zs_dependence}
\end{figure}

We further assess the robustness of the hybrid prescription by repeating the renormalization procedure for different choices of $z_s$ within the window quoted above. We find that, once $z_s$ is chosen inside a reasonable range where short-distance perturbation theory remains valid and long-distance fits are stable, the resulting $\tilde{M}^{R}(z,P^z)$ are essentially insensitive to $z_s$ within statistical uncertainties. A representative comparison is shown in Fig.~\ref{fig:zs_dependence}, where the real and imaginary parts of $\tilde{M}^{R}(z,P^z)$ at $P^z\simeq3.5\,{\rm GeV}$ nearly overlap for $z_s=0.10\,{\rm fm}$, $0.16\,{\rm fm}$ and $0.21\,{\rm fm}$. We take the residual spread under $z_s$ variation as a systematic uncertainty in the renormalized matrix elements.

The renormalized matrix elements $\tilde{M}^{R}(z,P^z)$ as functions of $\lambda=zP^z$ are shown in Fig.~\ref{fig:renormME_with_a} and Fig.~\ref{fig:renormME_with_Pz}. 
Figure~\ref{fig:renormME_with_a} compares results at $P^z\simeq2.5\,{\rm GeV}$ and $m_{\pi}\simeq210\,{\rm MeV}$, $300\,{\rm MeV}$ across the ensembles F32P30, G36P29, and H48P32, illustrating a mild lattice-spacing dependence over the renormalized quasi-DAs at coordinate space.  
Figure~\ref{fig:renormME_with_Pz} displays the $P^z$-dependence on H48P32 for $P^z\simeq\{2.5,3,3.5\}\,{\rm GeV}$. The consistency at larger $P^z$ provides an important validation that residual finite-$P^z$ effects are approaching the expected power-suppressed behavior.

\subsection{$\lambda$ Extrapolation and Fourier Transformation}

Owing to the finite lattice extent and the exponentially deteriorating signal-to-noise ratio at large spatial separations, the renormalized coordinate-space matrix elements $\tilde{M}^{R}(z,P^z)$ are accessible only up to a finite correlation length in $\lambda=zP^z$. A Fourier transform to momentum space quasi-DAs therefore requires a controlled reconstruction of $\tilde{M}^{R}$ beyond the largest $\lambda$ directly reachable in the lattice simulation. In the long tail region of the coordinate-space matrix elements, the oscillatory decaying behavior is dominated by the vicinity of the endpoints of momentum space quasi-DAs, which vanishes as $\tilde{\phi}(x)\sim x^{\alpha}(1-x)^{\beta}$ at the momentum fraction $x\to0$ or $1$, then  $\tilde{M}^{R}$ exhibits an endpoint-controlled power suppression at large $\lambda$ with exponents determined by $\alpha$ and $\beta$ \cite{Ji:2020brr}. Retaining the leading contributions associated with the two endpoints leads to the commonly used large-$\lambda$ extrapolation ansatz
\begin{align}
	\tilde{M}^R_{\rm tail}(\lambda)=\left[\frac{c_1}{\left(-i \lambda\right)^{\alpha+1}}+e^{i \lambda} \frac{c_2}{\left(i \lambda\right)^{\beta+1}}\right] e^{-\lambda / \lambda_0}, 
	\label{eq:lambda_extra}
\end{align}
where the two terms in brackets parameterize the algebraic falloff dictated by the endpoint powers, with the phase $e^{i\lambda}$ reflecting the contribution from the $x\to1$ endpoint. The exponential factor accounts for the observed damping of the long-distance tail and provides a numerically stable representation of the asymptotic behavior in practical lattice reconstructions \cite{Ji:2020brr,Gao:2021dbh,Chen:2025cxr,Xiong:2025obq,Ling:2025olz}. Here $\lambda_0$ is an effective correlation length that controls the onset of the exponential suppression in the renormalized quasi distributions. 
For heavy-to-light systems, where $x$ denotes the light-quark momentum fraction, the distribution is strongly asymmetric. As a result, the $x\to0$ endpoint typically provides the numerically dominant contribution, whereas the $x\to1$ term only induces a subleading oscillatory component. This expectation is consistent with what we observe in our fits.

\begin{figure}[http]
  \centering
  \includegraphics[width=1.00\linewidth]{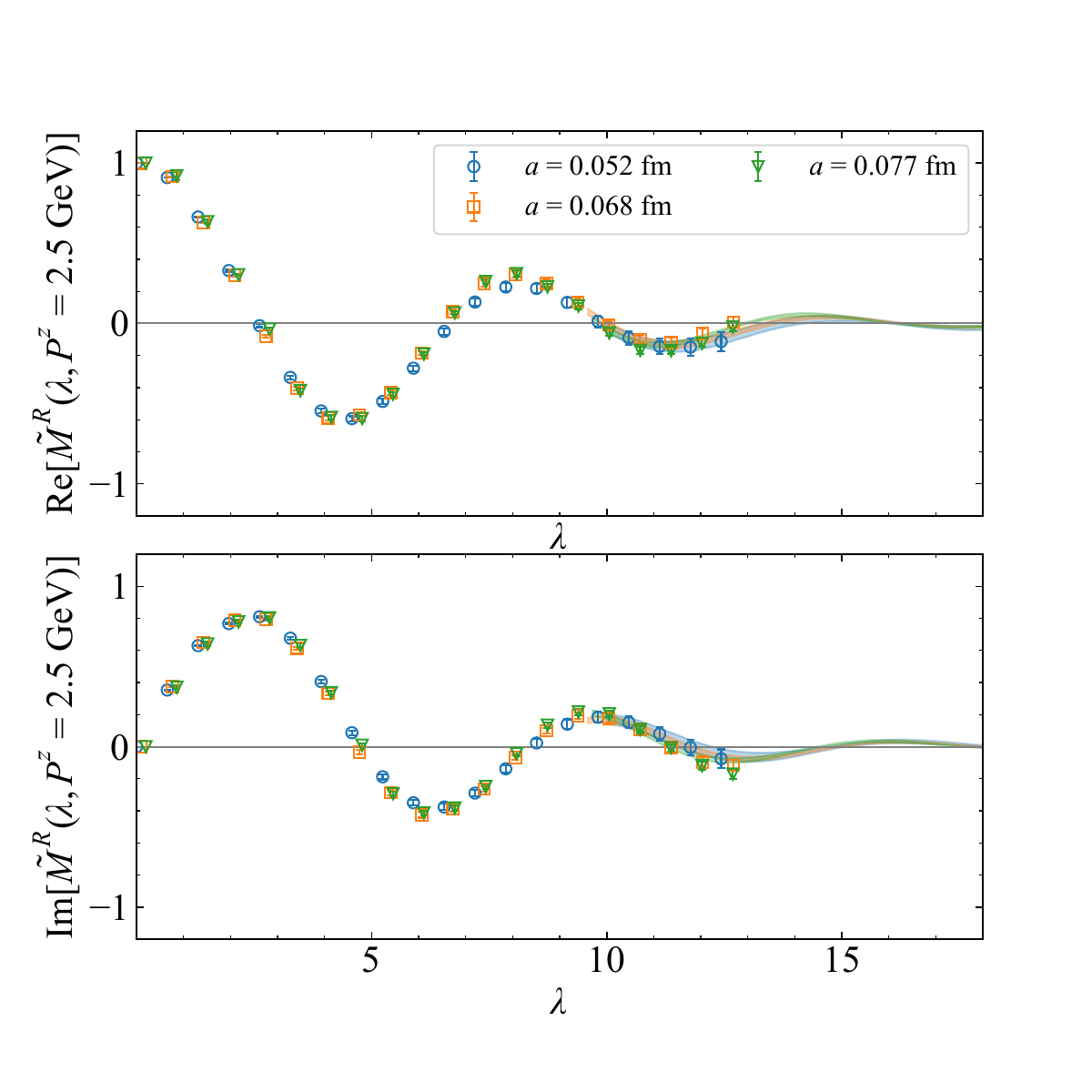}
  \caption{Comparison of the original (data points) and extrapolated results (colored bands) at $P^z\simeq2.5\,{\rm GeV}$ with different lattice spacings. More cases can be found in the Appendix.}
  \label{fig:extra_renormME_with_a}
\end{figure}

In practice, we perform the extrapolation using lattice data in a moderate-$\lambda$ window. Concretely, we fit the parameters in Eq.~\eqref{eq:lambda_extra} to data with $\lambda \gtrsim 9.5$, and we further vary the fit window to check the stability of the extrapolated tail. We then use the fitted ansatz to reconstruct the long-distance tail of the correlator, effectively extending it to $\lambda\to\infty$. Further numerical details of the $\lambda$-extrapolation procedure can be found in Refs.~\cite{LatticeParton:2024zko}. The resulting extrapolated quasi-DAs are shown in Fig.~\ref{fig:extra_renormME_with_a}, where the points denote the lattice data and the shaded bands represent the extrapolated curves. Additional examples of the $\lambda$-extrapolation are provided in the Appendix.~\ref{ax:more_lambda_extrapolation}

With the reconstructed $\tilde{M}^{R}(z,P^z)$ over the full $\lambda$ range, we obtain the momentum space quasi-DA via the Fourier transform,
 \begin{align}
\tilde\phi(x,P^{z})= \int_{-\infty}^{+\infty}\frac{dz}{2\pi} e^{-ix\lambda}\tilde M^{R}(\lambda,P^{z}),
\label{eq:FT_quasiDA}
\end{align}
which is presented as the blue band in Fig.~\ref{fig:lamet_matching_example} and then used as the nonperturbative input for the perturbative LaMET matching to the QCD LCDA in the next subsection.

\subsection{Heavy meson QCD LCDA from LaMET}

\begin{figure}[http]
  \centering
  \includegraphics[width=1.00\linewidth]{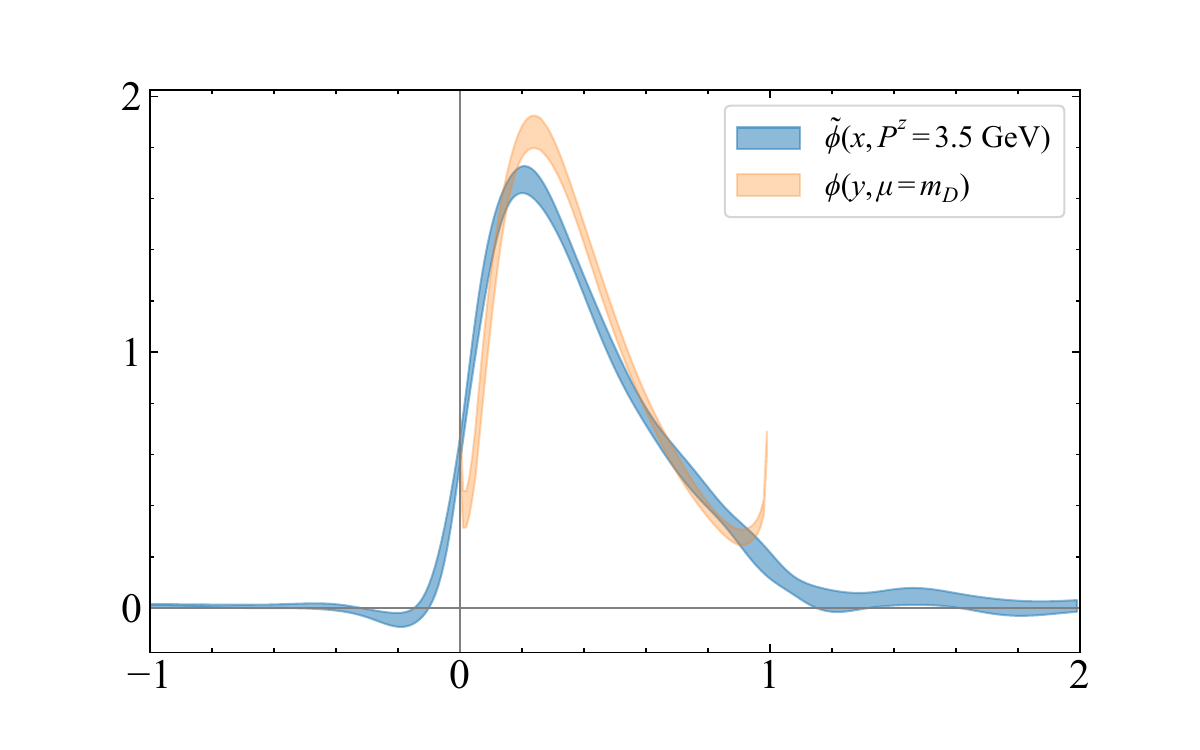}
  \caption{Comparison between the renormalized quasi-DA $\tilde{\phi}(x,P^z)$ on the H48P32 ensemble with $P^z\simeq3.5~{\rm GeV}$, and the matched QCD LCDA $\phi(y,\mu)$ at $\mu=m_D$.}
  \label{fig:lamet_matching_example}
\end{figure}

Applying the LaMET factorization formula in Eq.~\eqref{eq:LaMET_match}, one can match the renormalized quasi-DA of the $D$ meson with the QCD LCDA $\phi(y,\mu)$ in the $\overline{\rm MS}$ scheme. The matching kernel at next-to-leading order in $\alpha_s$ has been provided in Eqs.~\eqref{eq:C_1loop}--\eqref{eq:LaMETkernel_ct}, and it should be noted that, the matching with fixed-order kernel will suffer large logarithms in the endpoint regiont. Following the procedures in Ref.~\cite{LatticeParton:2024zko}, we resum these endpoint logarithms by introducing a $y$-dependent initial scale $\mu_0$ in the two endpoint regions: for $y\to0$ we take $\mu_0=2yP^z$ to resum terms of the form $\ln[(2yP^z)^2/\mu^2]$, while for $y\to1$ we take $\mu_0=2\bar y P^z$ to resum $\ln[(2\bar y P^z)^2/\mu^2]$. The evolution from $\mu_0$ to the common scale $\mu=m_D$ is carried out with Efremov--Radyushkin--Brodsky--Lepage (ERBL) evolution, equivalently implemented as the $\mu$ evolution of the matching coefficient \cite{LatticeParton:2024zko},
\begin{align}
	\frac{d}{d \ln \mu^2} \mathcal{C}\left(x, y, \mu, P^z\right)=\int_0^1 d \zeta \, V\left[\zeta, y, \alpha_s(\mu)\right] \mathcal{C}\left(x, \zeta, \mu, P^z\right),
\label{eq:ERBLevolution}
\end{align}
where $V(y,\zeta;\alpha_s)$ is the ERBL kernel of the QCD LCDAs \cite{Efremov:1979qk,Lepage:1980fj}.

In this work we choose the $\overline{\rm MS}$ scale as $\mu=m_D$, so that the LaMET-matched QCD LCDA is obtained at the same short-distance scale that enters the subsequent heavy-quark factorization in HQLaMET. The dominant perturbative uncertainty in the numerical matching arises from the choice of the initial resummation scale. We estimate this uncertainty by varying $\mu_0\to r\,\mu_0$ with $r\in[1/\sqrt2,\sqrt2]$ and taking the envelope of the resulting $\phi(y,\mu=m_D)$ as the corresponding systematic error. This variation corresponds to an $\mathcal{O}(20\%)$ change of $\alpha_s(\mu_0)$ at a representative momentum fraction $y\sim0.2$, which corresponds to the peak of the $D$ meson QCD LCDA.

\begin{figure}[http]
  \centering
  \includegraphics[width=1.00\linewidth]{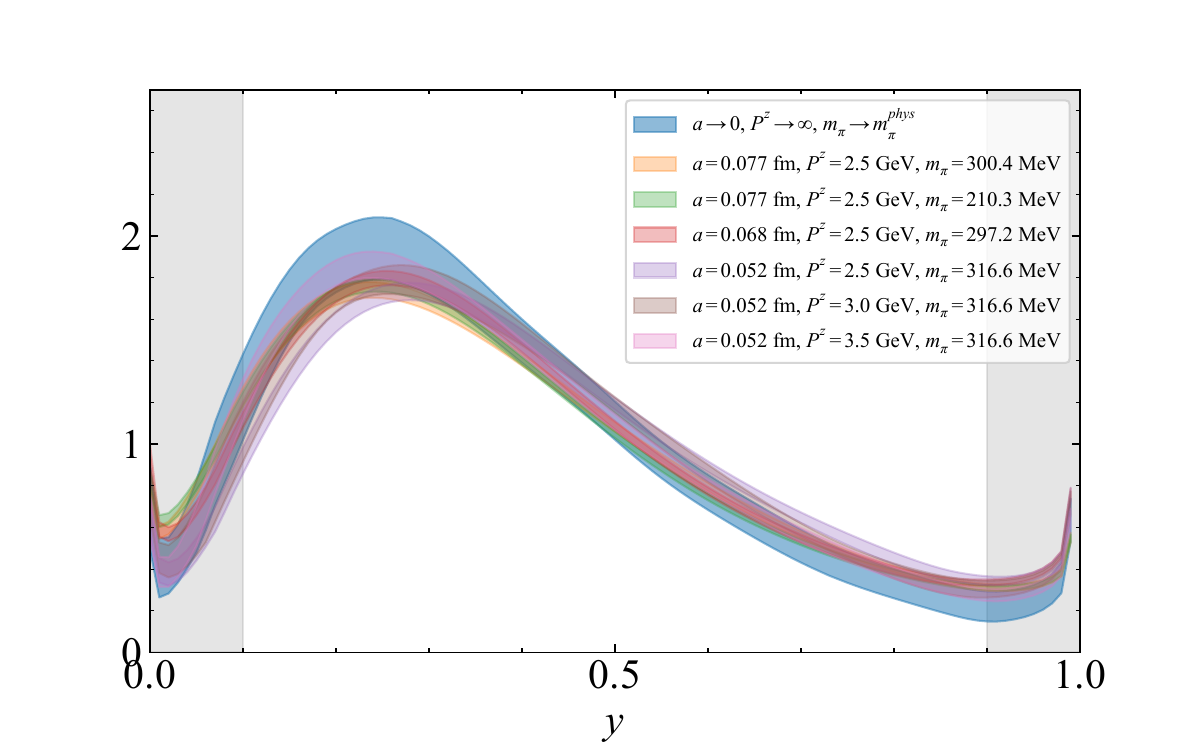}
  \caption{
  $D$ meson QCD LCDA $\phi(y,\mu=m_D)$ obtained from LaMET matching for different
  ensembles and momenta. The gray band shows the final result of $\phi(y,\mu=m_D)$ after extrapolating to $a\to0$, $P^z\to\infty$, and
  $m_\pi\to m_\pi^{\rm phy}$. Only statistic errors are included in the resulting bands. }
  \label{fig:qcd_lcda_ensembles}
\end{figure}

Applying the RG-improved matching, we obtain the QCD LCDAs of the $D$ meson on each ensemble and for each available boost momentum $P^z$. A representative example is shown in Fig.~\ref{fig:lamet_matching_example}, where we compare the renormalized quasi-DA computed on the H48P32 ensemble at $P^z\simeq3.5~{\rm GeV}$ and the matched QCD LCDA at $\mu=m_D$.
And in Fig.~\ref{fig:qcd_lcda_ensembles}, we collect the matched $D$ meson QCD LCDAs obtained on all ensembles and at the available boosts listed in Table~\ref{tab:ensembles}. One can observe a consistent qualitative pattern across all curves, including a peak around $y\simeq0.2$--$0.3$ and a slowly falling tail towards $y\to1$, as expected from the general discussion in Sec.~II. The residual spread among the curves reflects deviations from the physical continuum and infinite-momentum limits: variations with $P^z$ predominantly encode finite-momentum power corrections, while the mild dependence on $a$ and $m_\pi$ quantifies discretization and unphysical light-quark-mass effects, respectively. 

\begin{figure}[http]
  \centering
  \includegraphics[width=1.00\linewidth]{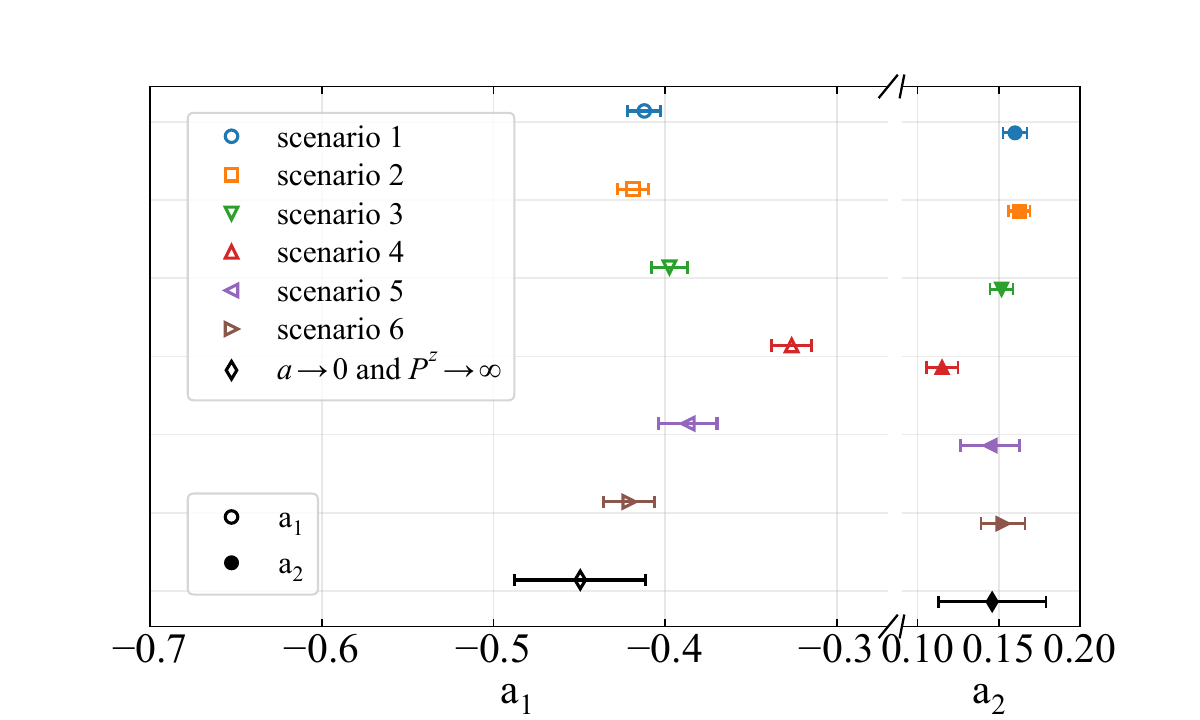}
  \caption{Comparison of the first two Gegenbauer moments, $a_1$ and $a_2$, extracted from the fitted QCD LCDAs on individual ensembles and from the final extrapolated QCD LCDA at the physical point. The filled and open markers represent $a_1$ and $a_2$, respectively.}
  \label{fig:gegenbauer_moments_qcd_lcda}
\end{figure}

To reach the continuum limit, the infinite-momentum limit, and the physical pion mass, we perform a point-by-point extrapolation in $y$ using the following ansatz
\begin{align}
\phi(y, \mu; a, m_{\pi}, P^z) = \,&\phi(y,\mu) + d_1\left(m_{\pi}^2-(m_{\pi}^{\rm phy})^2\right) \nonumber\\
& + d_2 a^2 + \frac{d_3}{(P^z)^2}\,,
\label{eq:global_extrapolation_phi}
\end{align}
where the $a^2$ term captures the leading discretization effects for our lattice actions and the $1/(P^z)^2$ term parameterizes the finite-$P^z$ corrections in the LaMET expansion. The resulting extrapolated QCD LCDA is shown as the gray band in Fig.~\ref{fig:qcd_lcda_ensembles}. To further characterize the shape of the fitted QCD LCDAs, we extract the first two Gegenbauer moments, $a_1$ and $a_2$, for the same six representative cases used in the LaMET analysis, namely: scenario 1, $a=0.077\,\mathrm{fm}$, $P^z=2.5\,\mathrm{GeV}$, $m_\pi=300.4\,\mathrm{MeV}$; scenario 2, $a=0.077\,\mathrm{fm}$, $P^z=2.5\,\mathrm{GeV}$, $m_\pi=210.3\,\mathrm{MeV}$; scenario 3, $a=0.068\,\mathrm{fm}$, $P^z=2.5\,\mathrm{GeV}$, $m_\pi=297.2\,\mathrm{MeV}$; scenario 4, $a=0.052\,\mathrm{fm}$, $P^z=2.5\,\mathrm{GeV}$, $m_\pi=316.6\,\mathrm{MeV}$; scenario 5, $a=0.052\,\mathrm{fm}$, $P^z=3.0\,\mathrm{GeV}$, $m_\pi=316.6\,\mathrm{MeV}$; and scenario 6, $a=0.052\,\mathrm{fm}$, $P^z=3.5\,\mathrm{GeV}$, $m_\pi=316.6\,\mathrm{MeV}$. The corresponding results are shown in Fig.~\ref{fig:gegenbauer_moments_qcd_lcda}.

In addition to statistical errors, our determination of the $D$ meson QCD LCDA has several systematic errors several sources of systematic uncertainty associated with renormalization, $\lambda$-extrapolation, perturbative matching, and the physical extrapolations. We assess these systematics by carrying out a set of controlled variations of the analysis procedure and taking the induced spread as the corresponding uncertainty. The main sources include:
\begin{itemize}
  \item { Hybrid renormalization (choice of $z_s$):}
  The parameter $z_s$ separating the short-distance ratio renormalization from the long-distance subtraction in the hybrid scheme introduces a residual scheme dependence. We vary $z_s$ within the window discussed in Sec.~IV.B and quantify the resulting variation of $\phi(y,\mu)$. The envelope is taken as the associated systematic uncertainty.

  \item {$\lambda$ extrapolation:}
  Since lattice data for $\tilde{M}^R(\lambda)$ are available only up to a finite correlation length, the reconstruction of the long-$\lambda$ tail required for the Fourier transform introduces a modeling uncertainty \cite{Ji:2020brr}. We estimate it by shifting the starting point of the $\lambda$ extrapolation by one forward and backward, and propagating the resulting spread to the matched LCDA. The extrapolation ranges used for the $\lambda$ extrapolation are summarized in Table~\ref{tab:lambda_extrapolation_ranges}.

\begin{table}[http]
  \renewcommand{\arraystretch}{1.8}
  \setlength{\tabcolsep}{2.5mm}
  \centering
  \begin{tabular}{lcccc}
    \hline\hline
    Ensemble &$P_z(\text{GeV})$ &$\lambda_L^{-1}$ &$\lambda_L^0$ &$\lambda_L^1$\\
    \hline
    F32P30 &2.50 &9.20 &9.86 &10.51\\
    F32P21 &2.50 &9.20 &9.86 &10.51\\
    G36P29 &2.52 &9.28 &9.95 &10.61\\
    H48P32 &2.49 &9.16 &10.47 &11.13\\
    H48P32 &2.99 &9.42 &10.21 &11.00\\
    H48P32 &3.49 &10.08 &11.00 &11.91\\
    \hline\hline
  \end{tabular}
  \caption{The starting points used for the $\lambda$ extrapolation at different configurations and momenta are summarized here. The quantity $\lambda_L^{0}$ denotes the central choice of the extrapolation starting point, while $\lambda_L^{1}$ and $\lambda_L^{-1}$ represent the alternative $\lambda_L$ values adopted to estimate the associated systematic uncertainty.}
  \label{tab:lambda_extrapolation_ranges}
\end{table}

  \item {Scale uncertainty in LaMET matching:}
  The numerical matching is stabilized by resumming endpoint logarithms with a $y$-dependent initial scale $\mu_0$. The dominant perturbative systematic uncertainty is associated with this choice. We estimate it by varying $\mu_0\to r\,\mu_0$ with $r\in[1/\sqrt2,\sqrt2]$ and taking the envelope of the resulting $\phi(y,\mu=m_D)$ as the perturbative matching uncertainty.

  \item {Physical mass, continuum and infinite-momentum extrapolations:}
  We take the difference between the extrapolated result and the data point closest to the extrapolation region as an estimate of the systematic uncertainty.


\end{itemize}

\begin{figure}[tbp]
  \centering
  \includegraphics[width=1.0\linewidth]{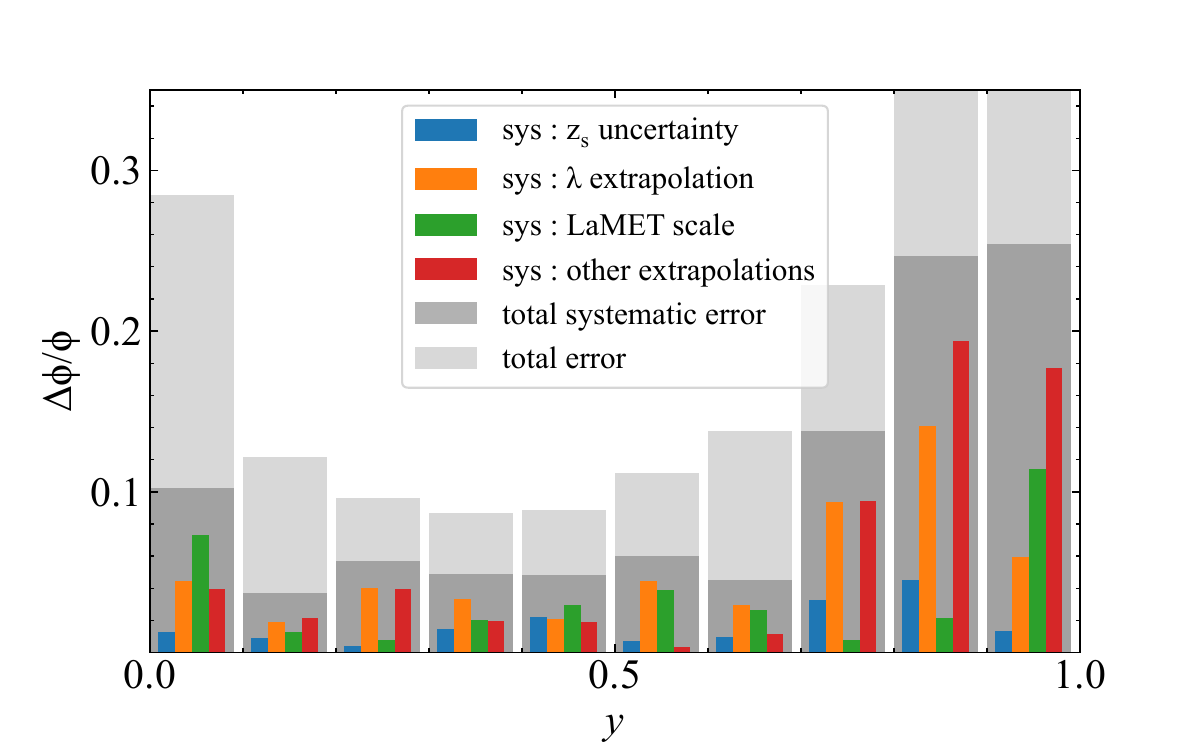}
  \caption{Error budget for the final result of $D$ meson QCD LCDA, showing statistical uncertainties and systematic ones from different sources. Total uncertainties are obtained by adding these contributions in quadrature.}
  \label{fig:error_budget_phi}
\end{figure}

The statistical uncertainty and the individual systematic components of the final $D$ meson QCD LCDA are summarized in Fig.~\ref{fig:error_budget_phi}. In the moderate-$y$ region ($0.1<y<0.9$) where the LaMET prediction is reliable, all error components are well controlled. The statistical uncertainty stays below $20\%$, and each systematic contribution is typically of comparable size or smaller.
As a consequence, the total uncertainty in $\phi(y,\mu)$ at $\mu=m_D$ does not exceed about $30\%$ across $0.1<y<0.9$, yielding a quantitatively precise lattice determination in the kinematic window most relevant for the subsequent HQLaMET matching and for phenomenological applications in exclusive $D$-meson processes.

\section{Moments From OPE}
\label{sec:OPE_moments}

In the following subsections, we present the lattice OPE determination of the first and second moments and use it as a quantitative consistency check on the LaMET extraction before proceeding to the HQLaMET matching to the HQET LCDA.

\subsection{Lattice QCD Calculation of the OPE Moments}

To determine the local matrix elements related to the OPE moments of heavy meson QCD LCDA, we compute the two-point correlation functions with operators $\mathcal{O}_{\rho}$, $\mathcal{O}_{\rho\mu}^-$ and $\mathcal{O}_{\rho\mu\nu}^{\pm}$ at the sinks, and pseudoscalar interpolating field at the source. For a heavy-to-light pseudoscalar meson, we employ the source interpolator as $\bar{q}(0)\gamma_5 Q(0)$ and consider the following correlators
\begin{align}
	C_{2,\rho}(\vec{p}, t)=&~ a^3 \sum_{\vec{x}} \left\langle G_q(\vec{x},t; \vec{0},0) \gamma_{\rho} G_Q^{\dagger}(\vec{x},t; \vec{0},0) \right\rangle, \\
	C_{2,\rho\mu}^-(\vec{p}, t)= &~a^3 \sum_{\vec{x}} \left\langle G_q(\vec{x},t; \vec{0},0) \gamma_{(\rho} \left[\overleftarrow{D}_{\mu)}-\overrightarrow{D}_{\mu)}\right] \right. \nonumber\\
	& \left. G_Q^{\dagger}(\vec{x},t; \vec{0},0) \right\rangle, \\
	C_{2,\rho\mu\nu}^{\pm}(\vec{p}, t)= &~ a^3 \sum_{\vec{x}} \left\langle G_q(\vec{x},t; \vec{0},0) \gamma_{(\rho}  \left[\overleftarrow{D}_{\mu} \overleftarrow{D}_{\nu)} \right.\right. \nonumber\\
	& \left.\left.  \pm 2 \overleftarrow{D}_{\mu} \overrightarrow{D}_{\nu)}+\overrightarrow{D}_{\mu} \overrightarrow{D}_{\nu)}\right] G_Q^{\dagger}(\vec{x},t; \vec{0},0) \right\rangle.
\end{align}
On the lattice, each covariant derivative $D_\mu$ is implemented with a symmetric discretization. As mentioned in Sec.~II.C, we choose the index $\rho=4$ and average over the combinations symmetric in $\mu\nu$ to increase statistics and improve the signal. On the ensembles listed in the ``OPE moments'' column of Table~\ref{tab:ensembles}, we evaluate the above two-point functions numerically and thus obtain the matrix elements required for the moment extractions. The simulation setup including gauge ensembles, source construction, smearing strategy, etc., follows Sec.~III and is closely aligned with the correlator calculations in Sec.~IV.A, and is therefore not repeated here.

In the large-$t$ limit, these correlators are dominated by the lowest-lying heavy pseudoscalar meson state, and their spectral decomposition reduces to the desired vacuum-to-meson matrix elements of the local operators defined in Eqs.~\eqref{eq:OPEmoments1}--\eqref{eq:OPEmoments2}, up to excited-state contaminations that are exponentially suppressed in $t$.
Taking suitable ratios with the axial-current correlator $C_{2,4}(\vec{p},t)$, the overlap factors and the Euclidean time dependence cancel in the ground-state limit, and the remaining plateaus yield the moments of heavy meson QCD LCDA.

Concretely, we determine the bare moments from suitable ratios of two-point functions, following the strategy of Ref.~\cite{RQCD:2019osh}. For the first moment, we use the ratios
\begin{align}
	\mathcal{R}_{1,a}^-(\vec{p}, t)
    &=\frac{i}{3}\sum^3_{\mu=1}\frac{1}{{p}_{\mu}}\frac{C_{2, 4\mu}^-(\vec{p},t)}{C_{2,4}(\vec{p},t)} \xrightarrow{t\to\infty} \langle\xi\rangle^{(0)}, \label{eq:R1a}\\
    \mathcal{R}_{1,b}^-(\vec{p}, t) &=\frac{4E}{3E^2+\vec{p}^2}\frac{C_{2, 44}^-(\vec{p},t)}{C_{2,4}(\vec{p},t)} \xrightarrow{t\to\infty} \langle\xi\rangle^{(0)}, \label{eq:R1b}
\end{align}
where $E\equiv E(\vec p)$ is the meson energy extracted from the corresponding two-point
spectroscopy, and the average over $\mu=1,2,3$ improves the statistical precision.
Similarly, the zeroth and second moments are obtained from
\begin{align}
	\mathcal{R}_{2,a1}^-(\vec{p}, t)=&-\frac{1}{3}\sum_{\mu\neq \nu}^3\frac{1}{{p}_{\mu}{p}_{\nu}}\frac{C_{2,4\mu\nu}^-(\vec{p},t)}{C_{2,4}(\vec{p},t)} \xrightarrow{t\to\infty} \langle\xi^2\rangle^{(0)}, \label{eq:R2a-}\\
	\mathcal{R}_{2,a2}^-(\vec{p}, t)=& -\frac{1}{3}\sum_{\mu=1}^3\frac{{p}_{\mu}}{{p}_1{p}_2{p}_3}\frac{C_{2,123}^-(\vec{p},t)}{C_{2,\mu}(\vec{p},t)} \xrightarrow{t\to\infty} \langle\xi^2\rangle^{(0)}, \label{eq:R2b-}\\
  \mathcal{R}_{2,a1}^+(\vec{p}, t)=&-\frac{1}{3}\sum_{\mu\neq \nu}^3\frac{1}{{p}_{\mu}{p}_{\nu}}\frac{C_{2,4\mu\nu}^+(\vec{p},t)}{C_{2,4}(\vec{p},t)} \xrightarrow{t\to\infty} \langle1\rangle^{(0)}, \label{eq:R2a+}\\
  \mathcal{R}_{2,a2}^+(\vec{p}, t)=& -\frac{1}{3}\sum_{\mu=1}^3\frac{{p}_{\mu}}{{p}_1{p}_2{p}_3}\frac{C_{2,123}^+(\vec{p},t)}{C_{2,\mu}(\vec{p},t)} \xrightarrow{t\to\infty} \langle1\rangle^{(0)}, \label{eq:R2b+}
\end{align}
where the summations in Eq.~\eqref{eq:R2a-} and \eqref{eq:R2a+} are taken over the three independent unordered pairs $(\mu,\nu)=(1,2),(1,3),(2,3)$, and the superscript $(0)$ indicates bare moments before the RI/SMOM renormalization and subsequent conversion to the $\overline{\rm MS}$ scheme.

In our lattice implementation, the choice of external three-momentum $\vec{p}$ is dictated by the kinematic prefactors appearing in the ratios $\mathcal{R}$. In particular, whenever a ratio contains explicit factors $1/p_{\mu}$ or $1/(p_{\mu}p_{\nu})$, the corresponding momentum components must be nonzero in order to avoid kinematic singularities and to ensure a clean ground-state plateau. For the first moment, we employ $\mathcal{R}_{1,a}^-$ and $\mathcal{R}_{1,b}^-$. The $\mathcal{R}_{1,a}^-$ needs a nonzero momentum input, we evaluate it at the minimal nonzero lattice momenta with a single non-zero component, $\vec{p}=(1,0,0),~(0,1,0)$ and $(0,0,1)$ in units of $2\pi/L$, and average over the three equivalent directions to improve statistics and reduce hypercubic artifacts. By contrast, $\mathcal{R}_{1,b}^-$ does not require nonzero momentum and is computed at $\vec{p}=\mathbf{0}$, providing a more accurate determination of $\langle\xi\rangle$. For the zeroth and second moment, the ratio $\mathcal{R}_{2,a1}^{\pm}(\vec{p}, t)$ is proportional to $1/(p_{\mu}p_{\nu})$ with $\mu\neq\nu$, and we therefore choose momenta with two nonvanishing components, $\vec{p}=(1,1,0),~(1,0,1)$ and $(0,1,1)$ in units of $2\pi/L$, again averaging over permutations related by cubic symmetry. The alternative determination $\mathcal{R}_{2,a2}^{\pm}$ involves $p_{\mu}/(p_1p_2p_3)$ and thus requires all three components to be nonzero. Accordingly, we compute $\mathcal{R}_{2,a2}^{\pm}$ at $\vec{p}=(1,1,1)\times2\pi/L$. 

After obtaining the bare moments from the ratios in Eqs.~\eqref{eq:R1a}--\eqref{eq:R2b+}, we apply a nonperturbative renormalization to the local operators. We adopt the RI/SMOM scheme, imposing renormalization conditions at a symmetric momentum point as \cite{Sturm:2009kb,Constantinou:2014fka,RQCD:2019osh,Li:2025zfw}
\begin{align}
p_1^2=p_2^2=(p_1- p_2)^2=\mu^2, 
\label{smom_kinematics}
\end{align}
where the renormalization and mixing factors are determined from amputated Green's functions with off-shell quark external states \cite{Sturm:2009kb,Constantinou:2014fka}.

Since the RI/SMOM scheme is defined through off-shell quark Green’s functions, it is gauge dependent and requires gauge fixing on each configuration. Hence, we calculate the renormalization factors in Landau gauge, which is the standard choice in lattice nonperturbative renormalization studies because it can be implemented straightforwardly on the lattice and matched consistently to continuum perturbation theory \cite{Sturm:2009kb,Constantinou:2014fka,Bali:2020isn}. The renormalization scale is scanned over a range of $\mu^2$, and we identify a Rome--Southampton window satisfying
\begin{align}
\Lambda_{\rm QCD}^2 \ll \mu^2 \ll (\pi/a)^2,
\end{align}
so that nonperturbative infrared effects and ultraviolet discretization artifacts are both parametrically suppressed \cite{Martinelli:1994ty,Sturm:2009kb,Constantinou:2014fka}.

The operator mixing effect is handled explicitly by choosing a finite operator basis that is closed under the lattice symmetries. In our case, the second-moment operator $\mathcal{O}_{\rho\mu\nu}^-$ can mix with $\mathcal{O}_{\rho\mu\nu}^+$ because they carry the same quantum numbers under the reduced hypercubic symmetry. The mixing coefficients are determined nonperturbatively by imposing RI/SMOM renormalization conditions on the full operator multiplet \cite{Martinelli:1994ty,Sturm:2009kb,Constantinou:2014fka,RQCD:2019osh,Bali:2020isn}. The projection operators $P^{(i)}$ are constructed to isolate the independent tree-level tensor structures, so that different Dirac components are disentangled and the resulting linear system determines the renormalization matrix $Z_{mm'}$ unambiguously \cite{Sturm:2009kb,Constantinou:2014fka}. Concretely, we impose the renormalization conditions in the form
\begin{align}
&\sum_{i=1}^{d}\mathrm{tr}\!\left[\Lambda^{(i)}_{B,m,{\rm tree}}(p_1,p_2)\,{P}^{(i)}\right]
\nonumber\\
&\qquad={Z}_q^{-1}\sum_{m^{\prime}=1}^{M}{Z}_{mm^{\prime}}
\sum_{i=1}^{d}\mathrm{tr}\!\left[\Lambda^{(i)}_{B,m^{\prime}}(p_1,p_2)\,{P}^{(i)}\right],
\label{NPR_condition}
\end{align}
evaluated at the symmetric kinematics in Eq.~\eqref{eq:smom_kinematics}. Here $Z_q$ denotes the quark-field renormalization constant in RI/SMOM, $\Lambda^{(i)}_{B,m}(p_1,p_2)$ is the amputated Green's function of the bare operator $\mathcal{O}_{B,m}$ with external off-shell quark momenta $p_1$ and $p_2$, and $P^{(i)}$ is the corresponding projector. The index $m$ labels the operators in the mixing basis with the size $M$, while the index $i$ runs over the independent components of the multiplet used to fully constrain the system \cite{Sturm:2009kb,Constantinou:2014fka,Bali:2020isn}. Solving Eq.~\eqref{NPR_condition} yields the full mixing-renormalization matrix $Z_{mm'}$, which is then used to renormalize the bare moments and to convert them to the $\overline{\rm MS}$ scheme at the scale $\mu$.

After determining the renormalization matrix $Z_{mm'}$ in the RI/SMOM scheme, we convert the renormalized operators to the $\overline{\rm MS}$ scheme using continuum perturbation theory. Concretely, we employ the available RI/SMOM$\to\overline{\rm MS}$ conversion factors at three-loop accuracy for the one-derivative operator $\mathcal{O}_{\rho\mu}^{-}$ and at two-loop accuracy for the two-derivative operator multiplet $\{\mathcal{O}_{\rho\mu\nu}^{-},\mathcal{O}_{\rho\mu\nu}^{+}\}$, including the mixing matrix, following the prescriptions in Refs.~\cite{Sturm:2009kb,Constantinou:2014fka,RQCD:2019osh,Li:2025zfw}. The converted results are quoted at the hard scale $\mu=m_D$, which is also the matching scale adopted in the LaMET analysis, so that the OPE and LaMET determinations can be compared at a common short-distance scale.

In practice, we present the renormalization factors normalized by the axial-current renormalization constant $Z_A$, which largely cancels the quark-field renormalization and reduces the sensitivity to residual gauge-fixing artifacts and discretization effects in the RI/SMOM setup \cite{Martinelli:1994ty,Sturm:2009kb,Constantinou:2014fka}. The resulting ratios $Z_{r2a}/Z_A$ and $Z_{r2b}/Z_A$ for the first-moment extractions exhibit a mild $a^2p^2$ dependence and are mutually consistent across all ensembles, as shown in Fig.~\ref{fig:RCs_first_moments_mD}. The corresponding second-moment renormalization matrix elements $Z_{11},Z_{12},Z_{21},Z_{22}$ are displayed in Fig.~\ref{fig:RCs_second_moments_mD}. We observe that the off-diagonal mixing is numerically small (in particular $Z_{21}\ll Z_{11},Z_{22}$), while the $Z_{12}$ mixing is moderate but well resolved, indicating that the operator mixing is under control within our chosen Rome--Southampton window. The final renormalization constants at $\mu=m_D$ are collected in Table~\ref{tab:renormalization_factor_RISMOM}.

\begin{table*}[http]
  \centering
  \renewcommand{\arraystretch}{1.8}
  \setlength{\tabcolsep}{2.5mm}
  \begin{tabular}{lccccc}
    \hline     
    Ensemble &C24P29 & C48P14 & F32P30 & G36P29 & H48P32 \\
    \hline     
     $Z_{r2a}/Z_A$ &1.29176(17) &1.28414(3) &1.35223(6) &1.37305(4) &1.40291(2)\\
     $Z_{r2b}/Z_A$ &1.29171(20) &1.28380(2) &1.34964(5) &1.37260(4) &1.40131(2)\\
     \hline
     $Z_{11}/Z_A$  &1.68625(47) &1.67085(5) &1.78173(15) &1.81155(9) &1.86087(4)\\
     $Z_{12}/Z_A$  &-0.06492(54) &-0.06237(9) &-0.09071(14) &-0.09757(11) &-0.11709(5)\\
     $Z_{21}/Z_A$  &0.00906(3) &0.00936(1) &0.00652(1) &0.00625(1) &0.00475(1)\\
     $Z_{22}/Z_A$  &1.44320(67) &1.43927(7) &1.43185(11) &1.41577(9) &1.39821(2)\\
    \hline  
  \end{tabular}
  \caption{Renormalization constants for the local OPE-moment operators, converted from the RI/SMOM
  scheme to the $\overline{\rm MS}$ scheme and quoted at the scale $\mu=m_D$. The factors are
  normalized by $Z_A$. Here $Z_{r2a}$ and $Z_{r2b}$ correspond to the renormalization of the
  one-derivative operator entering the first-moment extractions in
  Eqs.~\eqref{eq:R1a}--\eqref{eq:R1b}, while $Z_{ij}$ $(i,j=1,2)$ denote the $2\times2$ mixing matrix
  for the two-derivative operator multiplet $\{\mathcal{O}_{\rho\mu\nu}^{-},\mathcal{O}_{\rho\mu\nu}^{+}\}$
  relevant for the second moment.}
  \label{tab:renormalization_factor_RISMOM}
\end{table*}

With the renormalization factors in Table~\ref{tab:renormalization_factor_RISMOM}, we obtain the renormalized moments by applying the corresponding renormalization matrices to the bare ratios defined in Eqs.~\eqref{eq:R1a}--\eqref{eq:R2b+}.  For the first moment, the one-derivative operator renormalizes multiplicatively, and we obtain
\begin{align}
\langle \xi \rangle^{\overline{\rm MS}}(\mu)=\frac{Z_{r2}}{Z_A}\, \langle \xi \rangle^{(0)} \quad \text{with } \langle \xi \rangle^{(0)}  \text{ from }  \left\{\mathcal R_{1,a}^-,\,\mathcal R_{1,b}^-\right\},
\label{eq:xi_renorm}
\end{align}
where $Z_{r2}=Z_{r2a}$ or $Z_{2b}$ denotes the appropriate renormalization factor for the bare moment operator and we normalize by $Z_A$. For the second moment, the two-derivative operator multiplet $\{\mathcal O^-_{\rho\mu\nu},\mathcal O^+_{\rho\mu\nu}\}$ mixes under renormalization due to the reduced lattice symmetry, so that
\begin{align}
\begin{pmatrix}
\langle \xi^2 \rangle^{\overline{\rm MS}}(\mu) \\
\langle \mathbf 1 \rangle^{\overline{\rm MS}}(\mu)
\end{pmatrix}
=
\frac{1}{Z_A}
\begin{pmatrix}
Z_{11} & Z_{12}\\
Z_{21} & Z_{22}
\end{pmatrix}
\begin{pmatrix}
\langle \xi^2 \rangle^{(0)}\\
\langle \mathbf 1 \rangle^{(0)}
\end{pmatrix}, \nonumber\\
\qquad \text{with }\quad 
\left\{
\begin{array}{l}
\langle \xi^2 \rangle^{(0)} \text{ from } \left\{\mathcal R_{2,a1}^-,\,\mathcal R_{2,a2}^-\right\}\\
\langle \mathbf 1 \rangle^{(0)} \text{ from } \left\{\mathcal R_{2,a1}^+,\,\mathcal R_{2,a2}^+\right\}
\end{array}
\right..
\label{eq:xi2_renorm}
\end{align}
The above relations implement the full nonperturbative mixing subtraction in the RI/SMOM scheme and their subsequent conversion to $\overline{\rm MS}$ at the scale $\mu$ (here taken as $\mu=m_D$).

\begin{figure*}[http]
  \centering
  \includegraphics[width=0.49\linewidth]{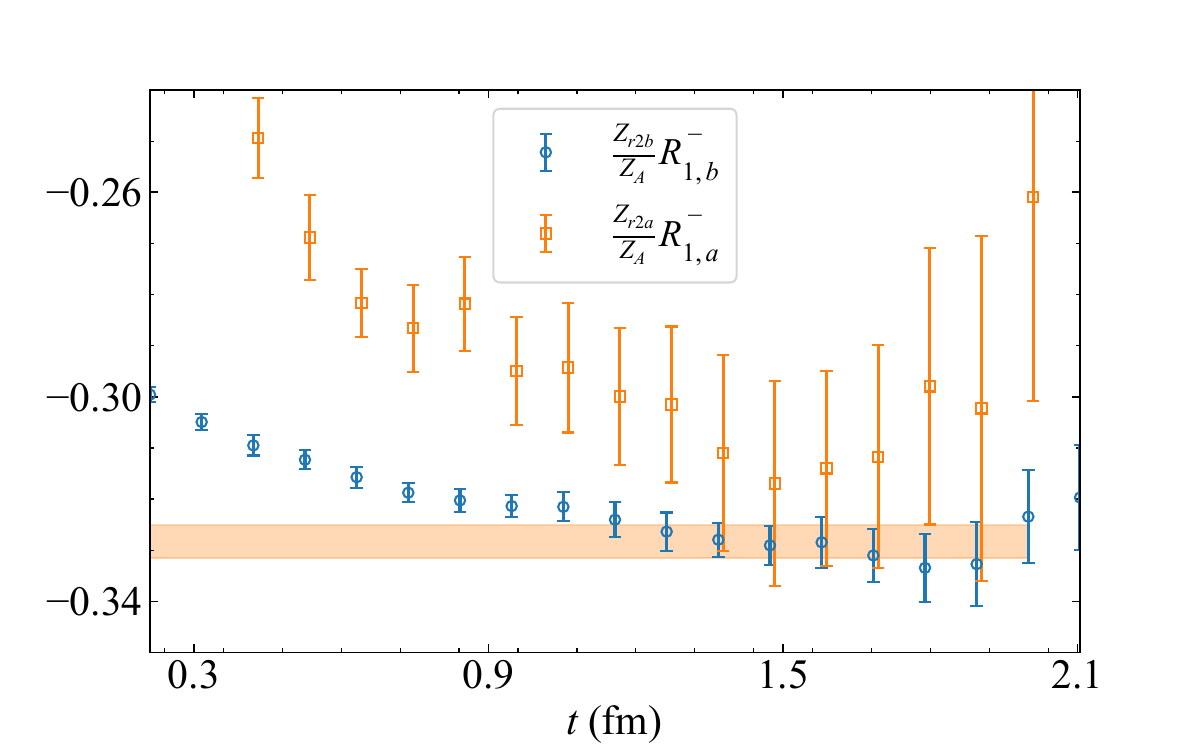}\hfill
  \includegraphics[width=0.49\linewidth]{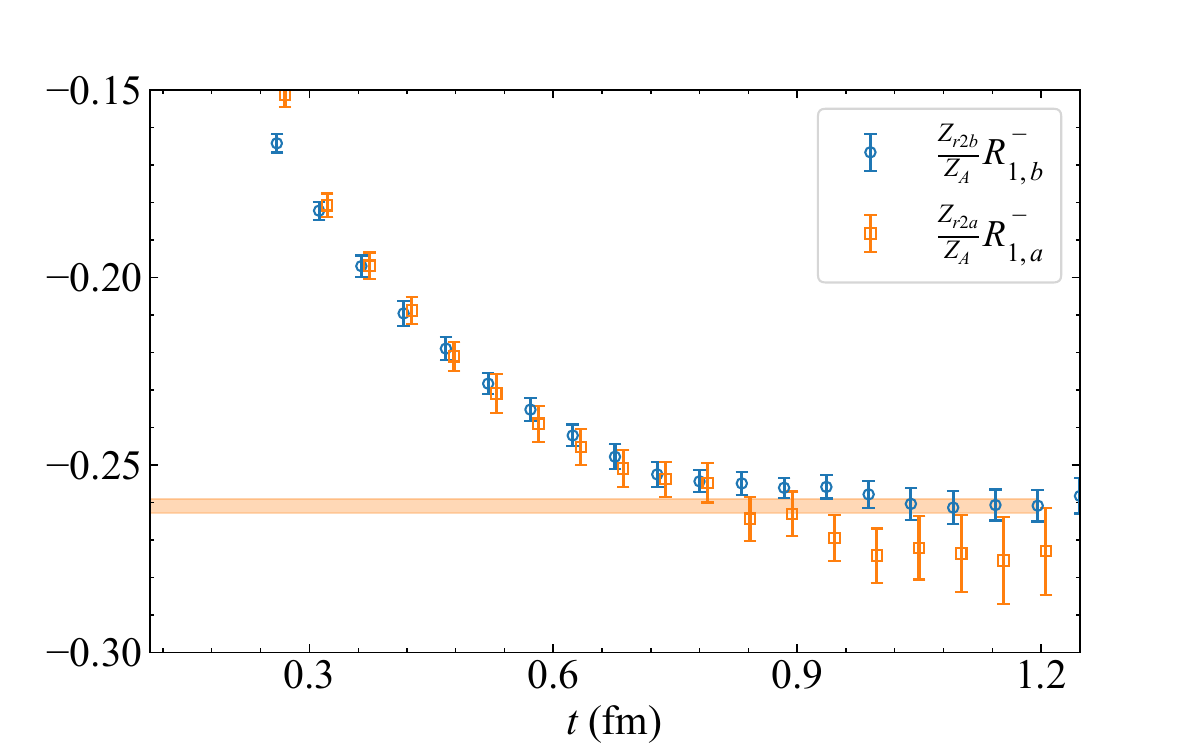}
  \caption{Renormalized ratios $\mathcal{R}^{-}_{1,b}$ (blue, $\vec p=\vec0$) and $\mathcal{R}^{-}_{1,a}$ (orange, smallest nonzero $\vec p$) versus source--sink separation $t$ for extracting $\langle\xi\rangle^{\overline{\rm MS}}$ on C48P14 (left) and H48P32 (right). The band denotes the model-averaged plateau fit to $\mathcal{R}^{-}_{1,b}$, which is consistent with that from with $\mathcal{R}^{-}_{1,a}$ within errors.}
  \label{fig:re_OPE_moments}
\end{figure*}

Applying Eqs.~\eqref{eq:xi_renorm}--\eqref{eq:xi2_renorm} to the bare lattice data gives the renormalized ratios shown in Fig.~\ref{fig:re_OPE_moments}.  Of the two ways to obtain the first moment, $\mathcal R_{1,a}^-$ requires nonzero momentum whereas $\mathcal R_{1,b}^-$ admits a kinematics choice with zero momentum. In practice, we find that the $\mathcal R_{1,b}^-$ channel exhibits a substantially better signal quality, consistent with the expectation that the zero-momentum setup minimizes both statistical fluctuations and discretization effects in the local-operator matrix elements. We therefore use primarily ${\cal R}^-_{1,b}$ to determine $\langle\xi\rangle^{\overline{\rm MS}}$, and perform the ground-state fit using a model-averaging analysis to eliminate the residual excited-state contamination and fit-range systematics \cite{Jay:2020jkz}. As also shown in Fig.~\ref{fig:re_OPE_moments}, the resulting fitted value is consistent with the $\mathcal R_{1,a}^-$ determination within uncertainties, providing a nontrivial cross-check of the moment extraction.

\begin{figure*}[http]
  \centering
  \includegraphics[width=0.49\linewidth]{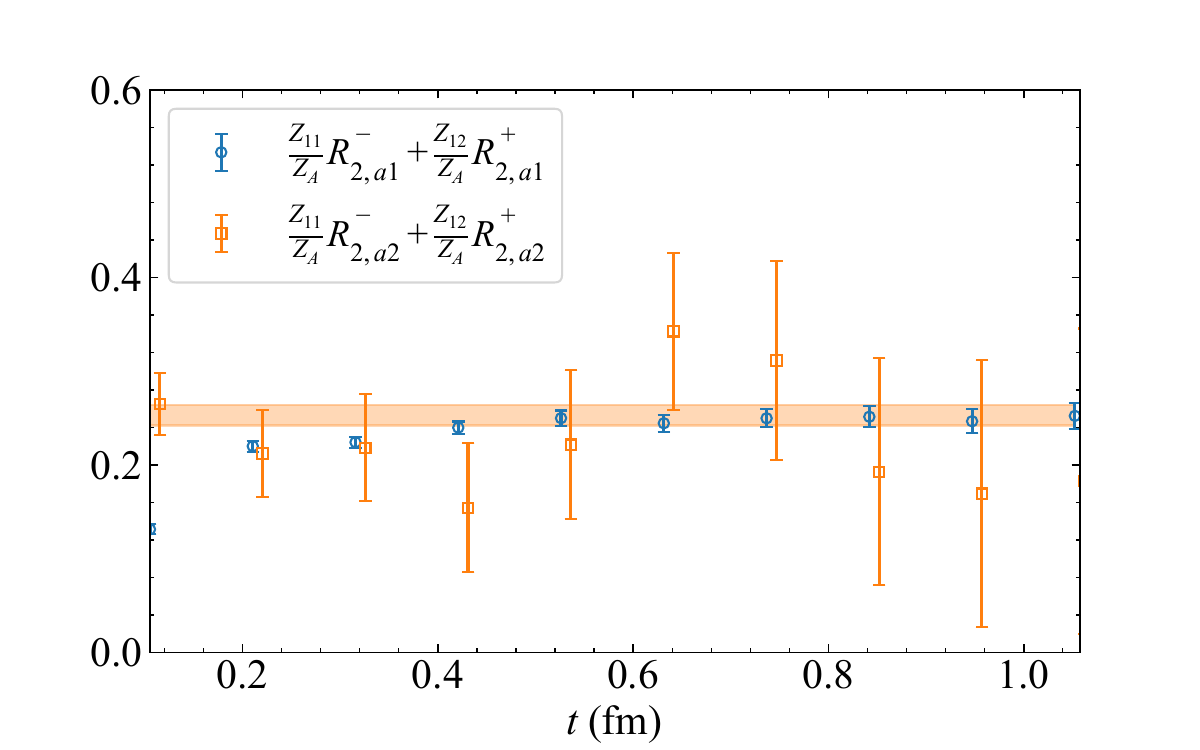}
  \includegraphics[width=0.49\linewidth]{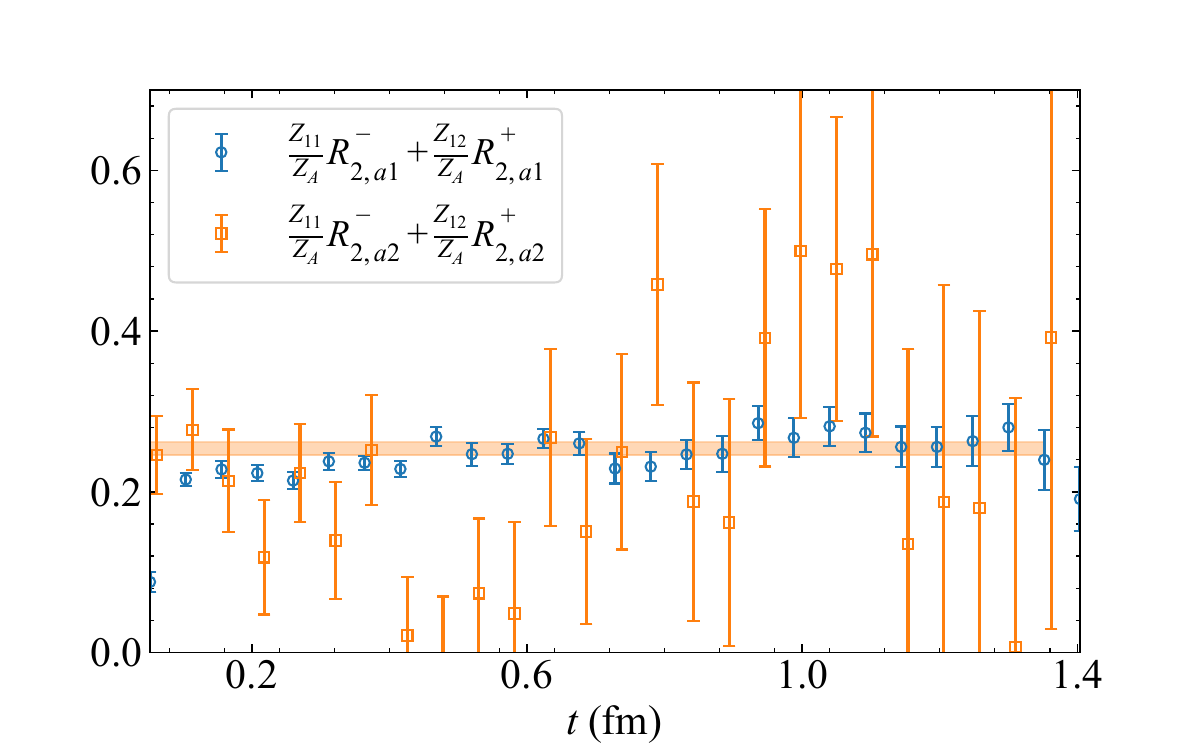}
  \caption{Renormalized ratio estimators for the second moment $\langle\xi^2\rangle$ (in the $\overline{\rm MS}$ scheme at $\mu=m_D$) on two representative ensembles. Blue points show $\mathcal{R}^{-}_{2,a1}$ and orange points show $\mathcal{R}^{+}_{2,a2}$ as functions of the source--sink separation $t$. The shaded band indicates the model-averaged plateau fit used to extract $\langle\xi^2\rangle^{\overline{\rm MS}}$.}
  \label{fig:re_OPE_moments_second}
\end{figure*}

For the second moment, we analyze two independent ratios defined in Eqs.~\eqref{eq:R2a-}--\eqref{eq:R2b+}. Figure~\ref{fig:re_OPE_moments_second} illustrates representative results for $\langle\xi^2\rangle$ on the physical-mass ensemble C48P14 and the fine ensemble H48P32. The ratio $\mathcal{R}^{-}_{2,a1}$ shows a clear and stable plateau at large-$t$, enabling a stable extraction with correlated fits. By contrast, $\mathcal{R}^{+}_{2,a2}$ exhibits substantially larger fluctuations and uncertainties at large Euclidean time, which is expected since it involves a noisier tensor structure and more severe kinematic prefactors. Nevertheless, within uncertainties it remains compatible with the plateau value inferred from $\mathcal{R}^{-}_{2,a1}$. In our final determination of $\langle\xi^2\rangle^{\overline{\rm MS}}$, we therefore fit the ground-state of $\mathcal{R}^{-}_{2,a1}$, while $\mathcal{R}^{+}_{2,a2}$ is kept as a consistency check and is incorporated through the nonperturbative mixing matrix.

\begin{figure*}[http]
  \centering
  \includegraphics[width=0.45\linewidth]{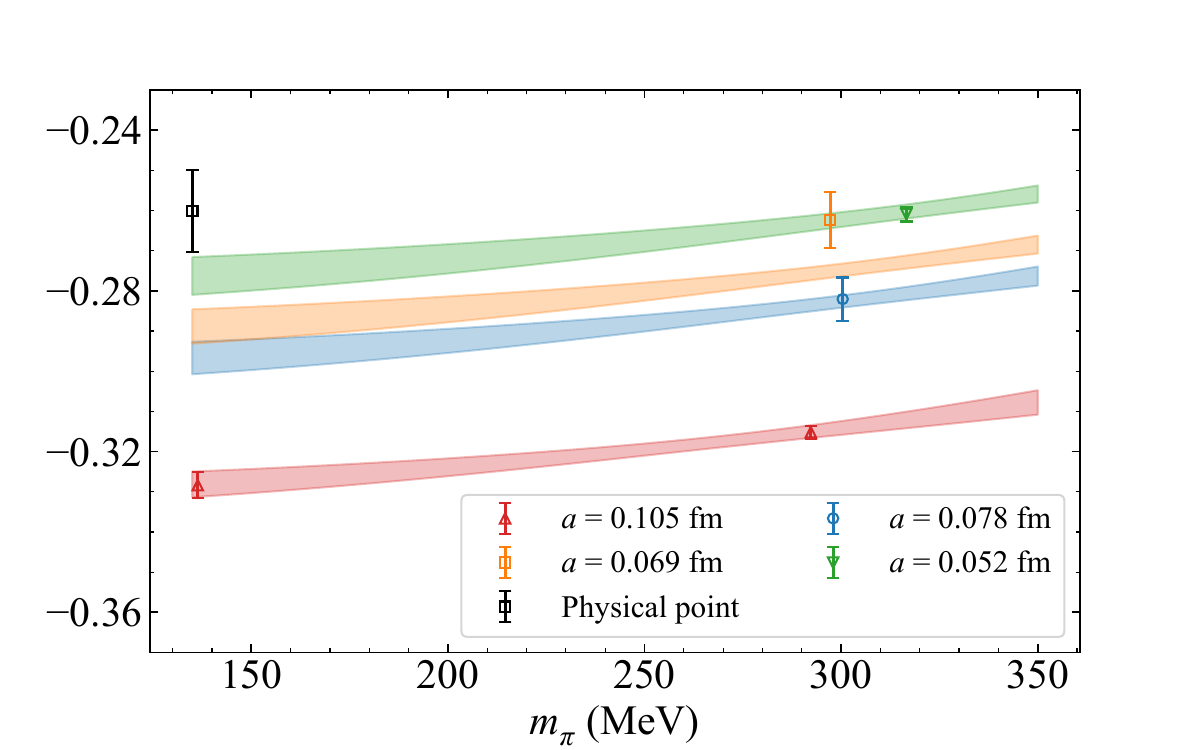}
  \includegraphics[width=0.45\linewidth]{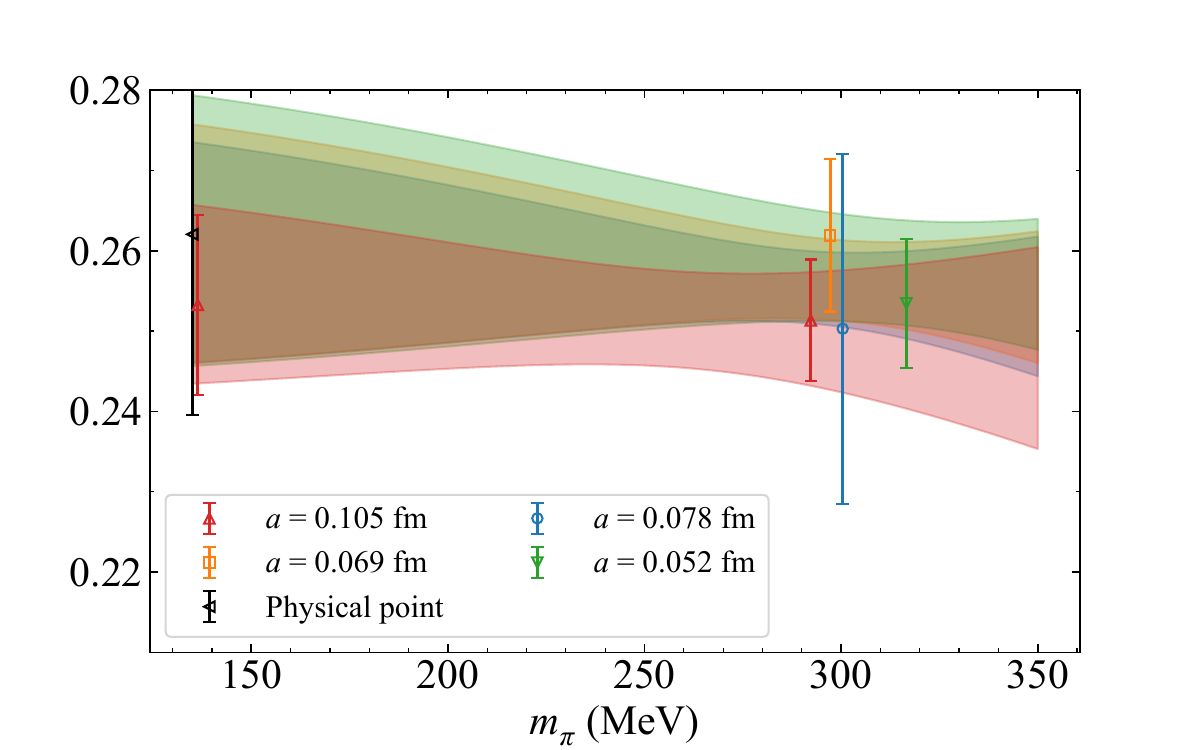}\\
  \includegraphics[width=0.45\linewidth]{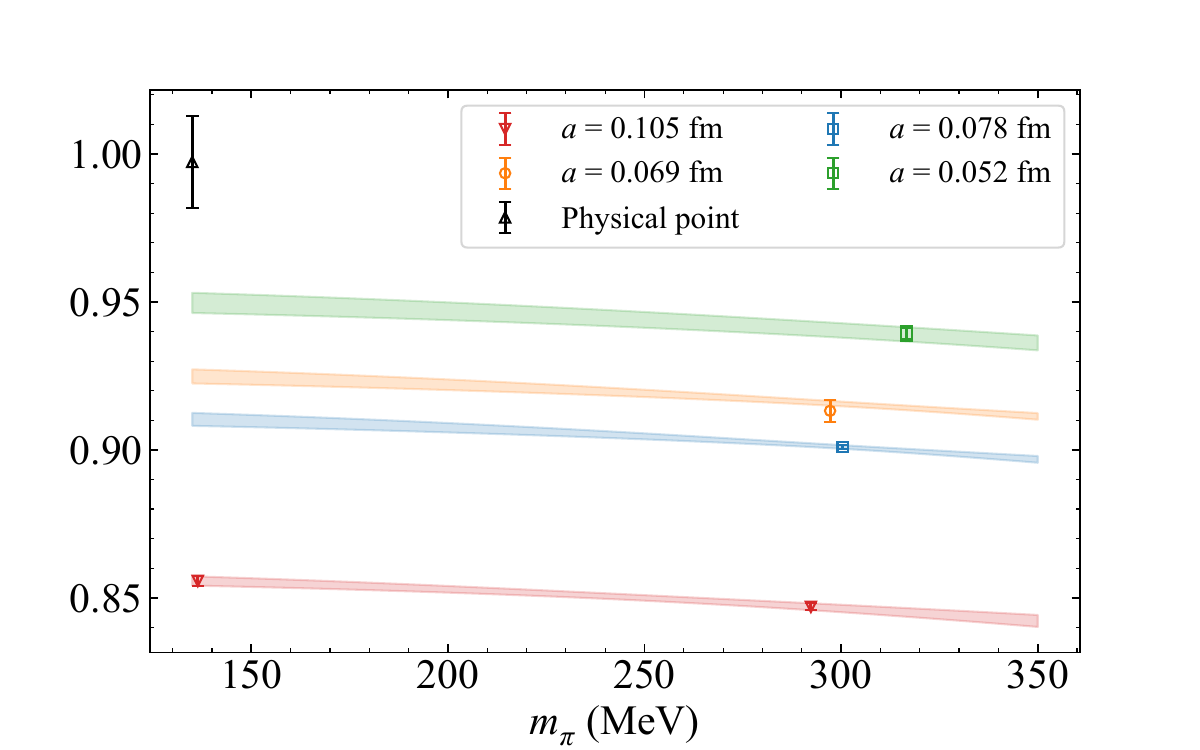}
  \caption{Combined continuum and physical-mass extrapolations of the renormalized lattice-OPE moments at $\mu=m_D$. 
  Data points correspond to the ensembles listed in the ``OPE moments'' column of Table~\ref{tab:ensembles} (including two ensembles at the same lattice spacing with different $m_\pi$), and the band shows the fit using Eq.~\eqref{eq:OPE_moment_extrap}. 
  The marker at the physical point indicates the extrapolated result. The upper-left, upper-right, and lower panels show the results for $\langle\xi\rangle$, $\langle\xi^2\rangle$, and $\langle1\rangle$, respectively.}
  \label{fig:OPE_moments_extrap}
\end{figure*}

After obtaining the renormalized moments on each ensemble, we perform a combined extrapolation to the continuum limit and the physical light-quark mass.  For each moment we adopt the ansatz
\begin{align}
\langle\xi^n\rangle^{\overline{\rm MS}}(a,m_\pi;\mu)=&\langle\xi^n\rangle^{\overline{\rm MS}}_{\rm phys}(\mu)
\left[
1 + c_1^{(n)} a + c_2^{(n)} a^2 \right. \nonumber\\
&\quad \left. + c_3^{(n)}\big(m_\pi^2-(m_\pi^{\rm phys})^2\big)
\right],
\label{eq:OPE_moment_extrap}
\end{align}
where the linear-$a$ term accounts for residual $\mathcal{O}(a)$ artifacts induced by the finite-difference realization of covariant derivatives in the local moment operators, while the $a^2$ term parametrizes the remaining leading discretization effects for our $\mathcal{O}(a)$-improved lattice action.  The light-quark mass dependence is modeled as being linear in $m_\pi^2$ in the range of pion masses used here.  The extrapolation fits for the zeroth, first, and second moments are shown in Fig.~\ref{fig:OPE_moments_extrap}, and the corresponding renormalized values on each ensemble and the extrapolated results are summarized in Table~\ref{tab:OPEmoments_mD}.

\begin{table}[http]
  \renewcommand{\arraystretch}{1.8}
  \setlength{\tabcolsep}{2.5mm}
  \centering
  \begin{tabular}{lccc}
    \hline 
     & $\mathbf{\langle1\rangle}$ & $\mathbf{\langle\xi\rangle}$ & $\mathbf{\langle\xi^2\rangle}$ \\
    \hline
    C24P29  & 0.8469(11) & -0.3152(16) & 0.2513(76)\\
    C48P14  & 0.8558(14) & -0.3283(32) & 0.253(11)  \\
    F32P30  & 0.9014(7)  & -0.2820(54) & 0.250(22) \\
    G36P29  & 0.9145(46) & -0.2624(70) & 0.2619(95) \\
    H48P32  & 0.9392(30) & -0.2609(18) & 0.2535(80) \\
    \hline
    Extrapolation & 0.997(61) & -0.260(10) & 0.262(23) \\
    \hline
  \end{tabular}
  \caption{Renormalized lattice-OPE moments in the $\overline{\rm MS}$ scheme at $\mu=m_D$ for each ensemble, together with the combined extrapolation to $a\to0$ and $m_\pi\to m_\pi^{\rm phys}$ using Eq.~\eqref{eq:OPE_moment_extrap}.}
  \label{tab:OPEmoments_mD}
\end{table}

\subsection{Comparison of Moments from LaMET and OPE}

As discussed in Sec.~II.C, an efficient way to benchmark the leading finite-$P^z$ effect, $\mathcal{O}\left(m_H^2/(P^z)^2\right)$, is to confront two complementary lattice determinations of the same QCD LCDA moments: the moments inferred from the LaMET-reconstructed distribution $\phi(y,\mu)$ obtained at large but finite $P^z$, and the moments computed directly from matrix elements in the lattice OPE approach, which does not rely on highly boosted external states.

Based on the results of OPE moments summarized in Table~\ref{tab:OPEmoments_mD}, we obtain the first two Gegenbauer moments by converting the OPE moments $\langle1\rangle$, $\langle\xi\rangle$ and $\langle\xi^2\rangle$ through the linear relations in Eq.~\eqref{eq:GandLmoments}, 
\begin{align}
a_1^{\rm OPE}(\mu)=-0.434(17),\qquad a_2^{\rm OPE}(\mu)=0.183(73),
\end{align}
here we choose the same scale $\mu=m_D$ at the $\overline{\rm MS}$ scheme as the LaMET-extracted QCD LCDA.

\begin{table*}[http]
  \centering
  \renewcommand{\arraystretch}{1.8}
  \setlength{\tabcolsep}{2.5mm}
  \caption{Gegenbauer-moment fits to the LaMET-reconstructed $\phi(y,\mu)$ in the window
  $y\in[0.10,0.90]$, using a truncation order $N$.}
  \begin{tabular}{lccccccc}
    \hline\hline
    $N$ & $a_0$ & $a_1$ & $a_2$ & $a_3$ & $a_4$ & $a_6$ & $a_8$ \\
    \hline
    2 & 0.957(39) & -0.451(39) & 0.144(34) & -- & -- & -- & -- \\
    3 & 0.955(40) & -0.446(42) & 0.139(38) & 0.009(30) & -- & -- & -- \\
    4 & 0.956(41) & -0.447(43) & 0.142(44) & 0.006(38) & 0.003(23) & -- & -- \\
    5 & 0.954(41) & -0.435(45) & 0.130(46) & 0.030(46) & -0.018(32) & -- & -- \\
    6 & 0.952(41) & -0.438(45) & 0.125(48) & 0.028(46) & -0.025(37) & -0.004(11) & -- \\
    7 & 0.953(41) & -0.434(50) & 0.126(48) & 0.033(53) & -0.024(37) & -0.004(11) & -- \\
    8 & 0.954(49) & -0.433(52) & 0.128(66) & 0.033(53) & -0.021(64) & -0.003(36) & 0.0006(132) \\
    \hline\hline
  \end{tabular}
  \label{tab:gegenbauer_fits}
\end{table*}

Independently, we can also determine the Gegenbauer moments $a_n(\mu)$ by fitting the LaMET-reconstructed QCD LCDA $\phi(y,\mu)$ based on Eq.~\eqref{eq:gegenbauer_expansion}. To reduce the sensitivity to endpoint regions where finite-$P^z$ and other power corrections are enhanced, we perform the fits in the window $y\in[0.10,\,0.90]$.  In the fitting we keep the parameter $a_0$, which is expected to be 1 at the conformal expansion, to validate the normalization of the QCD LCDA. Truncating the expansion at different maximal orders $N$, we obtain the coefficients shown in Table~\ref{tab:gegenbauer_fits}. One can see that the extraction of $a_1$ and $a_2$ is stable as $N$ is increased, while higher moments are consistent with zero and poorly constrained by the available precision. These results indicate that, in the moderate-$y$ region relevant for our fits, the shape is already well captured by the lowest moments, and the truncation uncertainty for $a_{1,2}$ is small.

\begin{figure}[http]
  \centering
  \includegraphics[width=1.00\linewidth]{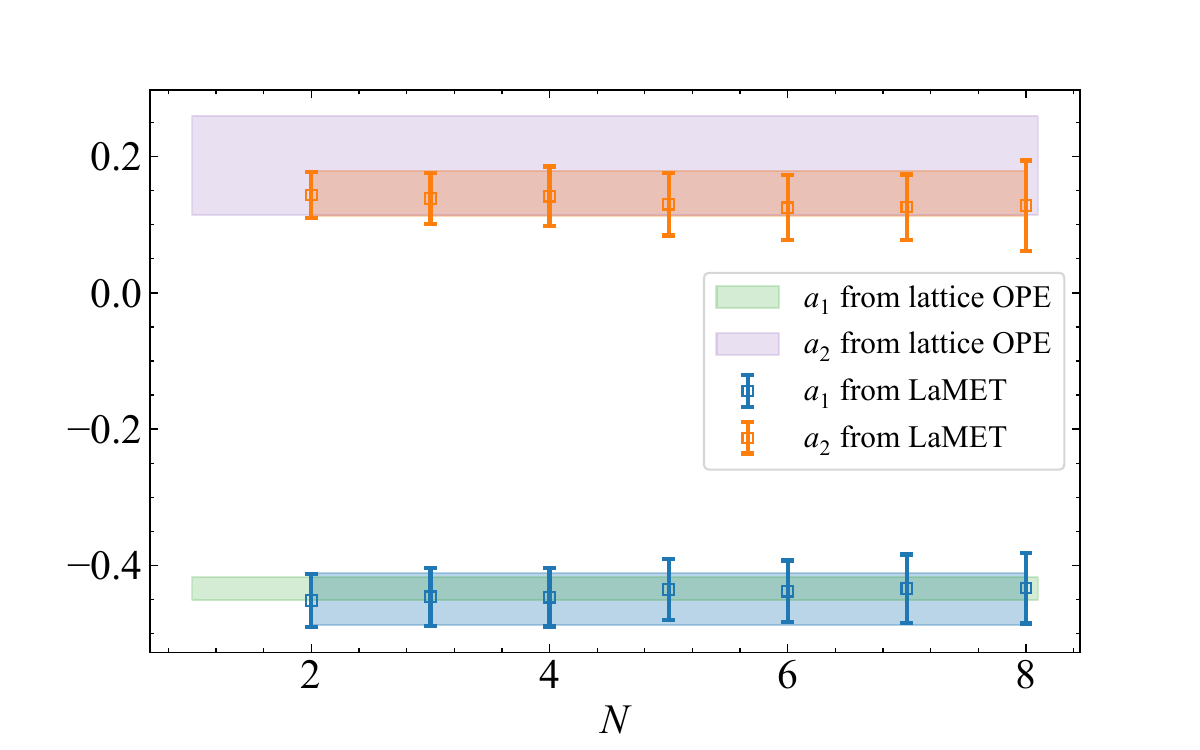}
    \caption{Benchmark of the lowest Gegenbauer moments. The blue (orange) points show $a_1$ ($a_2$) extracted
  from fits to the LaMET-reconstructed $\phi(y,\mu)$ truncated at different maximal orders $N$; the orange and blue bands
  indicate constant fits used to obtain the final LaMET values. The green and purple bands denote the corresponding moments
  obtained from lattice-OPE after conversion via Eq.~\eqref{eq:GandLmoments}.}
  \label{fig:moment_compare_LaMET_OPE}
\end{figure}

The stability of the extractions for $a_{1,2}$ with respect to $N$ is illustrated in Fig.~\ref{fig:moment_compare_LaMET_OPE}, where the blue and orange points show the extracted $a_1$ and $a_2$ at each truncation order. A constant fit over $N$ yields our final LaMET determinations,
\begin{align}
a_1^{\rm LaMET}(\mu)=-0.449(38),\quad a_2^{\rm LaMET}(\mu)=0.146(33),
\end{align}
at $\mu=m_D$. The green and red markers in the figure show the corresponding OPE-based results, and one can see that the two determinations agree within uncertainties.

Given that the OPE strategy does not rely on large boosts, this agreement provides a nontrivial validation that the finite-$P^z$ systematics in the LaMET reconstruction are under quantitative control for the momenta used in this work. This benchmark therefore supports the reliability of our LaMET determination of the $D$ meson QCD LCDA in the moderate-$y$ region, which forms the essential input for the subsequent HQLaMET matching to the HQET LCDA.

\section{ HQET LCDA and Phenomenological Discussions}
\label{sec:HQET_and_Pheno}
\subsection{Determination of HQET LCDA}

\begin{figure}[http]
\centering
\includegraphics[width=\linewidth]{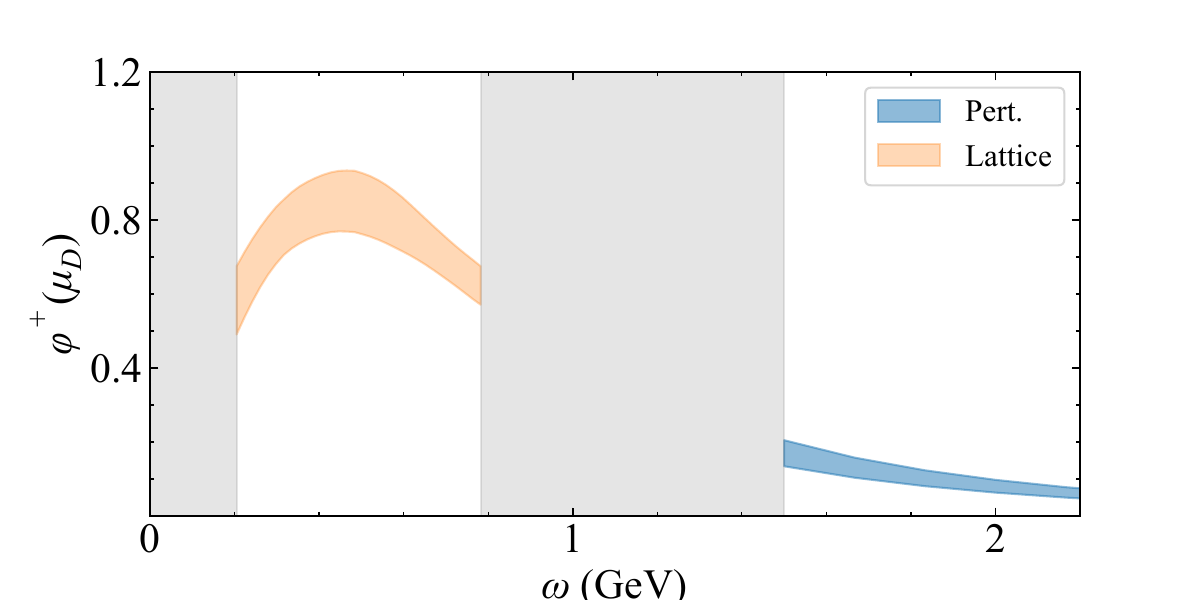}
\caption{HQET LCDA $\varphi^+(\omega,\mu)$ reconstructed from the QCD LCDA in the peak region (orange band) and from the perturbative tail formula at large $\omega$ (blue band). The gray-shaded interval indicates the intermediate and endpoint regions where neither approach is fully predictive.}
\label{fig:HQET_LCDA_peak_tail}
\end{figure}

With the physical $D$ meson QCD LCDA $\phi(y,\mu=m_D)$ determined in Sec.~IV, we can now construct the HQET LCDA $\varphi^+(\omega,\mu)$ by exploiting the ``peak-and-tail'' factorization discussed in Sec.~II.B.  In the peak region, the relevant momentum fraction in the heavy meson QCD LCDA is parametrically small,
\begin{align}
\frac{\Lambda_{\rm QCD}}{P^z}\ll y \sim \frac{\Lambda_{\rm QCD}}{m_H}\ll 1,
\end{align}
so that the light-quark momentum $\omega$ in HQET LCDA satisfies
\begin{align}
\frac{\Lambda_{\rm QCD}}{P^z} m_H\ll \omega \ll m_H.
\end{align}
In this region, the heavy-quark-mass dependence factorizes and the HQET LCDA can be obtained from the QCD LCDA through the peak-region matching formula in Eq.~\eqref{eq:HQET_matching_peak}.  With our numerical setups, this matching yields a reliable window of roughly $\omega\simeq 0.2-1~{\rm GeV}$, shown as the orange band in Fig.~\ref{fig:HQET_LCDA_peak_tail}. A pronounced peak is observed in this region around $\omega\simeq 0.2-0.8~{\rm GeV}$, consistent with the qualitative expectations of standard HQET-LCDA model shapes \cite{Grozin:1996pq,Braun:2003wx,Lange:2003ff,Beneke:2018wjp}.

For the tail region at $\omega\gtrsim m_H$, the HQET LCDA is generated by short-distance radiation and is perturbatively calculable. We use the one-loop expression in Eq.~\eqref{eq:HQET_matching_tail}, displayed as the blue band in Fig.~\ref{fig:HQET_LCDA_peak_tail}. To provide a conservative estimate of higher-power effects entering the tail formula, we vary $\bar{\Lambda}$ in the range $\bar{\Lambda}\in[0.4,0.8]~{\rm GeV}$, and take the resulting spread as the perturbative tail uncertainty band.

\begin{figure}[http]
  \centering
    \includegraphics[width=\linewidth]{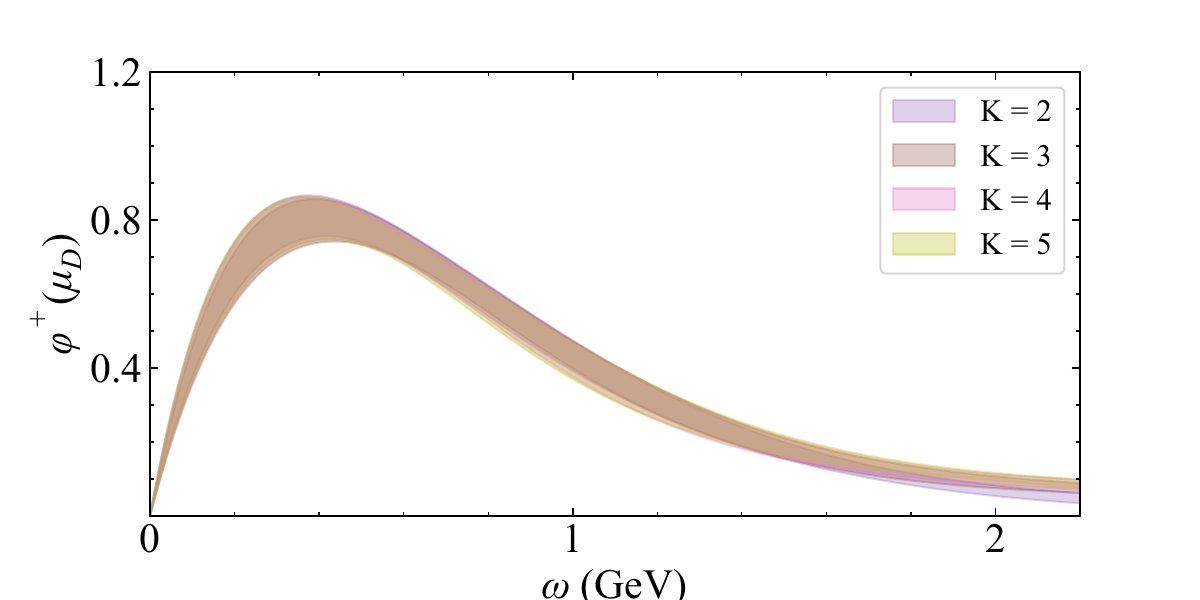}
    \caption{Reconstruction of the HQET LCDA $\varphi^+(\omega,\mu)$ using the Laguerre-polynomial parametrization in Eq.~\eqref{eq:HQETLCDAsmallomegaexpansion2}. The bands show fits with truncation orders $K=2,3,4,5$ constrained by the first-principles information in the peak region and in the perturbative tail.}
    \label{fig:HQET_LCDA_param}
\end{figure}

Between the peak region and the perturbative tail, there exists an intermediate window in which neither a direct lattice determination nor a fixed-order perturbative description is quantitatively reliable. In addition, in the very small-$\omega$ endpoint region, $\omega \lesssim {\Lambda_{\rm QCD}}\times m_H/P^z$, the LaMET power corrections become uncontrolled and the extraction ceases to be predictive. We therefore treat these domains as non-predictive from first principles, and indicate them by the gray shading in Fig.~\ref{fig:HQET_LCDA_peak_tail}.

To obtain a continuous distribution of the HQET LCDA over the full $\omega$ range, we adopt a model-independent parametrization proposed in Ref.~\cite{Feldmann:2022uok}, based on an expansion in generalized Laguerre polynomials,
\begin{align}
	\varphi^+(\omega,\mu)= \frac{\omega\, e^{-\omega / \omega_0}}{\omega_0^2}
	\sum_{k=0}^K \frac{b_k(\mu)}{1+k}\,L_k^{(1)}\!\left(\frac{2\omega}{\omega_0}\right),
	\label{eq:HQETLCDAsmallomegaexpansion2}
\end{align}
where $L_k^{(1)}$ are associated Laguerre polynomials and $\omega_0$ sets the characteristic falloff scale of the distribution. This basis is orthogonal on $\omega\in[0,\infty)$ with the weight $\omega e^{-\omega/\omega_0}$, so that this parametrization satisfies small-$\omega$ behavior $\varphi^+(\omega)\propto \omega$ and provides a controlled interpolation between the nonperturbative peak window and the perturbative large-$\omega$ tail. 
In practice, we determine the parameters $\{\omega_0,b_k\}$ from a combined fit to the first-principles information available in the peak and tail regions, which is collected in Table. \ref{tab:para_fit_result}

\begin{table*}[http]
  \centering
  \renewcommand{\arraystretch}{1.8}
  \setlength{\tabcolsep}{2.5mm}
  \begin{tabular}{lccccccc}
    \hline     
    K &$\omega_0$ & $b_0$ & $b_1$ & $b_2$ & $b_3$ & $b_4$ & $b_5$\\
    \hline     
     $2$ &0.544(98) &1.16(16) &0.38(19) &0.093(57) &-- &-- &-- \\
     $3$ &0.88(14) &1.62(16) &1.20(34) &0.58(13) &0.52(35) &-- &-- \\
     $4$ &1.00(14) &1.77(14) &1.46(29) &1.03(30) &0.49(36) &0.53(32) &-- \\
     $5$ &1.09(13) &1.86(12) &1.66(26) &1.24(36) &0.75(36) &0.49(34) &0.42(35) \\

    \hline  
  \end{tabular}
  \caption{The results for parameters based on the parametrizatio in Eq.~\eqref{eq:HQETLCDAsmallomegaexpansion2}}
  \label{tab:para_fit_result}
\end{table*}

The stability of the reconstruction with respect to the truncation order provides an internal convergence check. As shown in Fig.~\ref{fig:HQET_LCDA_param}, truncating the expansion at $K=2,3,4,5$ yields mutually consistent results within uncertainties over the full $\omega$, and the near overlap of the $K=2-5$ bands indicates that the present peak and tail inputs already constrain the dominant shape degrees of freedom, and residual truncation effects are subleading at our current precision. In addition, for each $K$, we perform fits over multiple fitting windows—$\omega\in[0.2,0.8], [0.3,0.8], [0.2,0.7],$ \text{and} $[0.3,0.7]$—to obtain stable estimates of the fit parameters.

\subsection{Inverse moments}

A central set of nonperturbative inputs encoded in the leading HQET LCDA $\varphi^+(\omega,\mu)$ are its inverse moments defined by Eq.~\eqref{eq:HQETmoments}, which play a pivotal role in QCD factorization theorems for $B$ decays and light-cone sum rule studies in heavy flavor physics.
While the first inverse moment $\lambda_B^{-1}$ and the first inverse-logarithmic moments $\sigma_B^{(1)}$ have been estimated in various models~\cite{Khodjamirian:2020hob,Lee:2005gza,Braun:2003wx,Grozin:1996pq}, there is considerable scope for enhancing their reliability and precision. In the case of $\sigma^{(2)}_B$ and subsequent orders, no existing results are currently available.

Results for the first inverse moment $\lambda_B$ and inverse-logarithmic moments $\sigma_B^{(1,2)}$ of the HQET LCDA,  including our lattice-derived values at $\mu=m_D$ and $\mu=1$ GeV (after evolution with the renormalization group equation), are collected in  Tab.~\ref{tab:lambdaB_sigmaB_summary}, alongside corresponding results from experimental constraints (e.g., $B\to \gamma\ell\nu$  measurements), QCD sum rules, and phenomenological models.  This comparison highlights that our $\lambda_B =0.340(20)$ GeV (at $\mu=1$ GeV) is consistent with the experimental lower bound $\lambda_B >0.24$ GeV  and aligns closely with recent theoretical determinations (0.338(68)-0.48(11)GeV), while our  $\sigma_B^{(1)} = 1.685(63) $ (at $\mu=1$ GeV)  agrees well with model-based predictions (1.4(4)-1.6(2)). The consistency underscores the reliability of our first-principles approach in reducing LCDA-related uncertainties that dominate many heavy flavor physics predictions.

\begin{table*}[http]
  \centering
  \renewcommand{\arraystretch}{1.8}
  \setlength{\tabcolsep}{2.5mm}
  \begin{tabular}{l l c c c}
    \hline
    $\mu$ & Reference (Method) & $\lambda_B~(\mathrm{GeV})$ & $\sigma_B^{(1)}$ & $\sigma_B^{(2)}$ \\
    \hline
    $m_D$  & This work & 0.423(28) & 2.041(62) & 5.50(26) \\
           & Ref.~\cite{LatticeParton:2024zko} (LQCD) & 0.420(71) & 2.17(16) & 6.33(80) \\
    \hline
    $1~\mathrm{GeV}$    & This work & 0.340(20) & 1.685(63) & -- \\
           & Ref.~\cite{LatticeParton:2024zko} (LQCD) & 0.376(63) & 1.66(13) & -- \\
           & Ref.~\cite{Belle:2018jqd} (Experiment) & $>0.24$ & -- & -- \\
           & Ref.~\cite{Gao:2019lta} (QCD sum rule)  & $0.343^{+0.064}_{-0.079}$ & 1.4(4) & -- \\
           & Ref.~\cite{Braun:2003wx} (QCD sum rule)  & 0.46(11) & 1.4(4) & -- \\
           & Ref.~\cite{Khodjamirian:2020hob} (QCD sum rule) & 0.383(153) & -- & -- \\
           & Ref.~\cite{Lee:2005gza} (OPE)  & 0.48(11) & 1.6(2) & -- \\
           & Ref.~\cite{Grozin:1996pq} (Asymptotic behavior)  & 0.35(15) & -- & -- \\
           & Ref.~\cite{Mandal:2023lhp} (Global Fit) & 0.338(68) & -- & -- \\
    \hline
  \end{tabular}
  \caption{Summary of $\lambda_B$ and logarithmic moments $\sigma_B^{(n)}$.}
  \label{tab:lambdaB_sigmaB_summary}
\end{table*}

As shown in Eq.~\eqref{eq:ErrorsOfBtoKFF}, the inverse moments of the heavy meson dominate the systematic uncertainties of the weak decay form factors of $B$ mesons at large recoil. 
At leading power, spectator-scattering terms typically scale as $\sim 1/\lambda_B$,
so that $\lambda_B$ constitutes one of the dominant hadronic uncertainties in precision predictions,
while $\sigma_B^{(n)}$ parameterize subleading sensitivity to the logarithmic $\omega$-dependence of the kernels and thus affect the shape and normalization in a correlated way.
Therefore, a first-principles determination of
$\lambda_B$ and $\sigma_B^{(n)}$ provides a direct bridge between the nonperturbative structure of the heavy meson and precision phenomenology for $B$ decays.

As can be seen from the error budget in Eq.~\eqref{eq:ErrorsOfBtoKFF}, the dominant uncertainties in the $B\to K^*$ form-factor determination arise from two sources: (i) the uncertainty of the first inverse moment $\lambda_B$ itself (the terms labeled by the subscript $\lambda_B$), and (ii) the uncertainty propagated from the model dependence of the HQET LCDA $\varphi^+(\omega)$ (the terms labeled by the subscript $\varphi_B$). With the first-principles reconstruction of $\varphi^+(\omega)$ obtained in this work, the model-dependent contribution can be removed altogether. In addition, the comparison in Table.~\ref{tab:lambdaB_sigmaB_summary} shows that our determination of $\lambda_B$ is significantly more precise, which directly translates into a more accurate LCSR prediction for the $B\to K^*$ form factors; a detailed discussion can be found in Ref.~\cite{HeavymesonDA_short_paper}.

\section{Summary and Prospect}

In this work, we have presented a comprehensive and refined framework for the first-principles determination of heavy meson light-cone distribution amplitudes, advancing our earlier pioneering study by addressing key limitations and delivering robust results for both QCD LCDAs and HQET LCDAs, two critical nonperturbative inputs for heavy flavor physics.
Building on the heavy-quark large-momentum effective theory framework, we extend the analysis to six lattice QCD ensembles with varying lattice spacings (0.0519fm, 0.0775fm, 0.0683fm and 0.1053 fm) and pion masses (from 135.5 to 317.2 MeV), enabling controlled continuum, chiral, and infinite-momentum extrapolations to the physical point. We refine the lattice simulation pipeline with momentum-smeared sources, Hypercubic  smearing for Wilson lines, and optimized interpolating operators, significantly enhancing the signal-to-noise ratio of nonlocal correlators. 

For QCD LCDAs, we have extracted bare quasi-distribution amplitudes, applied nonperturbative hybrid renormalization, performed $\lambda$-extrapolation to resolve long-distance tails, and matched to QCD LCDAs via LaMET with renormalon resummation—yielding asymmetric distributions peaked at  (light-quark momentum fraction) with total uncertainties $\le 30\%$  in the physically relevant range $0.1<y<0.9$. These QCD LCDA results are validated by cross-comparison with two lowest moments from operator product expansion of local twist-two operators, confirming that power corrections in LaMET are likely  well-controlled.

For HQET LCDAs, we leverage the ``peak-and-tail" factorization: the nonperturbative peak region ($\omega \in [0.2,0.8]$ GeV, $\omega=ym_H$) is derived from lattice QCD-based QCD LCDAs, while the perturbative tail region ($\omega\ge1.0$ GeV) is computed via one-loop HQET. A model-independent Laguerre polynomial parametrization merges these regions smoothly, and we extract the key inverse moments $\lambda_B=0.340(20)$ GeV at $\mu=1$ GeV and inverse-logarithmic moment $\sigma_B^{(1)}=1.685(63)$, consistent with experimental constraints and phenomenological expectations. In a companion paper~\cite{HeavymesonDA_short_paper}, we have demonstrated  the phenomenological utility of our results by predicting branching ratios for $W\to D\gamma$ and $W\to B\gamma$, highlighting the impact of precise QCD and HQET LCDA inputs on high-precision tests of the Standard Model.
Collectively, this work and Ref.~\cite{HeavymesonDA_short_paper} overcome the single-lattice-spacing limitation of our earlier study, strengthen theoretical consistency via cross-validation, and provides a unified, precision framework for accessing both QCD and HQET LCDAs—laying the groundwork for next-generation heavy flavor physics phenomenology.

Several promising directions remain to further refine and expand our already rather advanced framework:
\begin{itemize}
\item  Higher-Order Perturbative Corrections: Current results rely on next-to-leading order (NLO) perturbative matching kernels and one-loop jet functions for HQET LCDAs. Extending these to two-loop accuracy will reduce perturbative uncertainties, particularly in the tail region of HQET LCDAs and the LaMET matching for QCD LCDAs, further stabilizing endpoint behavior.

\item Extended Lattice Ensembles: Future simulations with finer lattice spacings ($a\le 0.04$ fm) and higher boost momenta ($P_z\sim 4$ GeV) will further suppress discretization artifacts and power corrections in $1/P^z$, enabling more precise continuum extrapolations. Incorporating ensembles with physical pion masses directly will also reduce chiral extrapolation uncertainties.

\item 
Heavy vector mesons and Higher-Twist LCDAs: Leveraging heavy quark spin symmetry on LCDAs~\cite{Deng:2024dkd,Wang:2024wwa}, we plan to extend the framework to heavy vector mesons $D^*$, whose HQET LCDAs are expected to share universal features with pseudoscalar heavy mesons. Additionally, exploring higher-twist LCDAs will enable a more complete description of heavy meson structure, critical for subleading-power corrections in factorization theorems.

\item 
Model-Independent Moment Extraction: Developing improved model-independent parametrizations for QCD and HQET LCDAs, for example, extending the Laguerre polynomial basis or adopting conformal field theory-inspired expansions, will enhance the robustness of moment extractions and reduce reliance on functional form assumptions.

\item Another future milestone would be the development of robust techniques for the direct simulation of HQET heavy-quark fields $h_v$ on the lattice. Unlike the sequential matching from QCD  to HQET adopted in this work, direct lattice implementation of HQET fields would bypass the need for intermediate QCD LCDAs, directly accessing HQET quasi-DAs and reducing theoretical uncertainties associated with two-step matching. This requires overcoming technical challenges such as stabilizing large-velocity HQET propagators and mitigating signal-to-noise degradation, but would enable a more direct, first-principles route to HQET LCDAs and validate the consistency of the sequential HQLaMET approach.

\end{itemize}

By pursuing these directions, we aim to further solidify the framework as a gold-standard first-principles tool for accessing heavy meson LCDAs, enabling more stringent tests of the standard model and deeper insights into strong interaction dynamics in the heavy-quark sector.

\section*{Acknowledgement}
We thank the CLQCD collaborations for providing us the gauge configurations with dynamical fermions~\cite{CLQCD:2023sdb,CLQCD:2024yyn}, which are generated on the HPC Cluster of ITP-CAS, the Southern Nuclear Science Computing Center(SNSC), the Siyuan-1 cluster supported by the Center for High Performance Computing at Shanghai Jiao Tong University, and the Dongjiang Yuan Intelligent Computing Center.

This work is supported in part by National Natural Science Foundation of China under grants No.12125503, 12305103, 12375069, 12375080, 12525504, 12435002, 12293060, 12293062, 12275277, 12435004 and 12447101. CDL is also is partly supported by the National Key Research and Development Program of China (2023YFA1606000). YBY is also supported in part by National Key R\&D Program of China No.2024YFE0109800, and the Strategic Priority Research Program of Chinese Academy of Sciences, Grant No. YSBR-101. QAZ is also supported by the Fundamental Research Funds for the Central Universities. JHZ is also supported by Shenzhen Fundamental Research Grant No. JCYJ20250604141224032, the Ministry of Science and Technology of China under Grant No. 2024YFA1611004, and by CUHK-Shenzhen under grant No. UDF01002851.

\clearpage
\begin{appendix}
\begin{widetext}
\section{Nonperturbative Renormalization of OPE Moments in RI/SMOM Scheme}

The local operators in Eqs.~\eqref{eq:localoperformoments1}--\eqref{eq:localoperformoments2} require nonperturbative renormalization. We renormalize them in the regularization-independent symmetric momentum-subtraction (RI/SMOM) scheme, which is designed to suppress exceptional-momentum infrared contaminations and has become a standard choice for lattice determinations of moments \cite{Sturm:2009kb,Lehner:2011fz,Constantinou:2014fka}. The renormalized quark field is defined by $\psi_R = Z_q^{1/2}\psi_B$, and the renormalized operators are related to the bare ones through 
\begin{align}
\mathcal{O}_{m}^{R}(\mu)=\sum_{m'} Z_{mm'}(\mu,a)\,\mathcal{O}_{m'}^{B}(a),
\end{align}
where $m,m'$ label operators within the same mixing multiplet. In practice, we compute the Landau-gauge momentum-space Green's function with an insertion of the bare operator $\mathcal{O}_m$,
\begin{align}
G_{\mathcal{O}_m}(p_1,p_2) =\sum_{x,y} e^{-i(p_1\cdot x-p_2\cdot y)} \left\langle \psi(x)\,\mathcal{O}_m(0)\,\bar\psi(y)\right\rangle,
\end{align}
and form the amputated vertex function
\begin{align}
\Lambda_{\mathcal{O}_m}(p_1,p_2)=S^{-1}(p_1)\,G_{\mathcal{O}_m}(p_1,p_2)\,S^{-1}(p_2),
\end{align}
with $S(p)$ the quark propagator in momentum space. The RI/SMOM renormalization condition is imposed at a symmetric kinematic point,
\begin{align}
p_1^2=p_2^2=(p_1-p_2)^2=\mu^2,
\label{eq:smom_kinematics}
\end{align}
so that no momentum channel is exceptional. We then project $\Lambda_{\mathcal{O}_m}$ onto the desired Dirac/Lorentz structure using a set of projectors $\{P^{(i)}\}$ and require the projected amputated Green's function to reproduce its tree-level value at the subtraction scale $\mu$,
\begin{align}
\sum_i {\rm Tr}&\!\left[P^{(i)}\,\Lambda_{\mathcal{O}_m}^{R}(p_1,p_2)\right]=\sum_i {\rm Tr}\!\left[P^{(i)}\,\Lambda_{\mathcal{O}_m}^{\rm tree}(p_1,p_2)\right], \nonumber\\
&\qquad \Lambda_{\mathcal{O}_m}^{R}\equiv Z_q^{-1}\sum_{m'} Z_{mm'}\,\Lambda_{\mathcal{O}_{m'}}^{B}.
\label{eq:smom_condition}
\end{align}
For the operators relevant to the first and second moments, the projectors are chosen as the Hermitian conjugates of the corresponding tree-level vertices, which leads to a compact matrix equation for the mixing-renormalization matrix $Z_{mm'}$ \cite{Sturm:2009kb,Constantinou:2014fka}. 
And then, the RI/SMOM-renormalized operators are converted to the $\overline{\rm MS}$ scheme at the
same scale $\mu$ using continuum perturbation theory; the results for the first and second moments are shown in Figs.\ref{fig:RCs_first_moments_mD} and \ref{fig:RCs_second_moments_mD}, respectively, and the resulting renormalized matrix elements are used to extract the moments. \cite{Sturm:2009kb,Constantinou:2014fka}.

\begin{figure*}[http]
  \centering
  \includegraphics[width=1.0\linewidth]{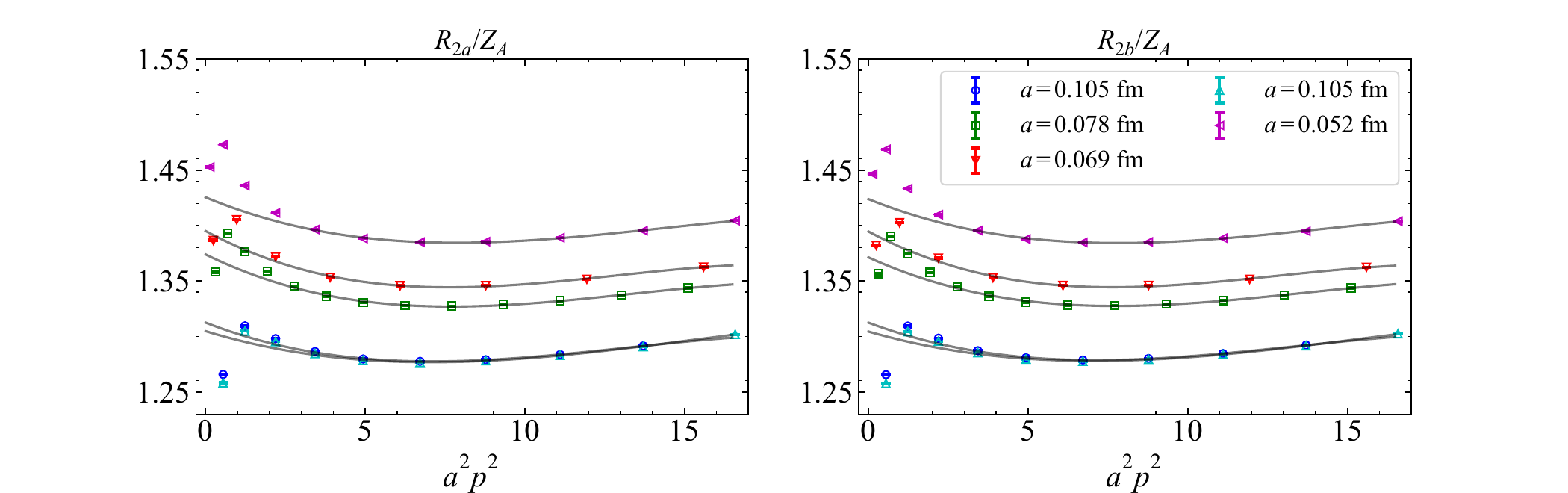}
  \caption{RI/SMOM$\to\overline{\rm MS}$ renormalization factors for the first-moment operators,
  shown as $Z_{r2a}/Z_A$ (left) and $Z_{r2b}/Z_A$ (right) versus $a^2p^2$ for all ensembles.}
  \label{fig:RCs_first_moments_mD}
\end{figure*}

\begin{figure*}[http]
  \centering
  \includegraphics[width=1.0\linewidth]{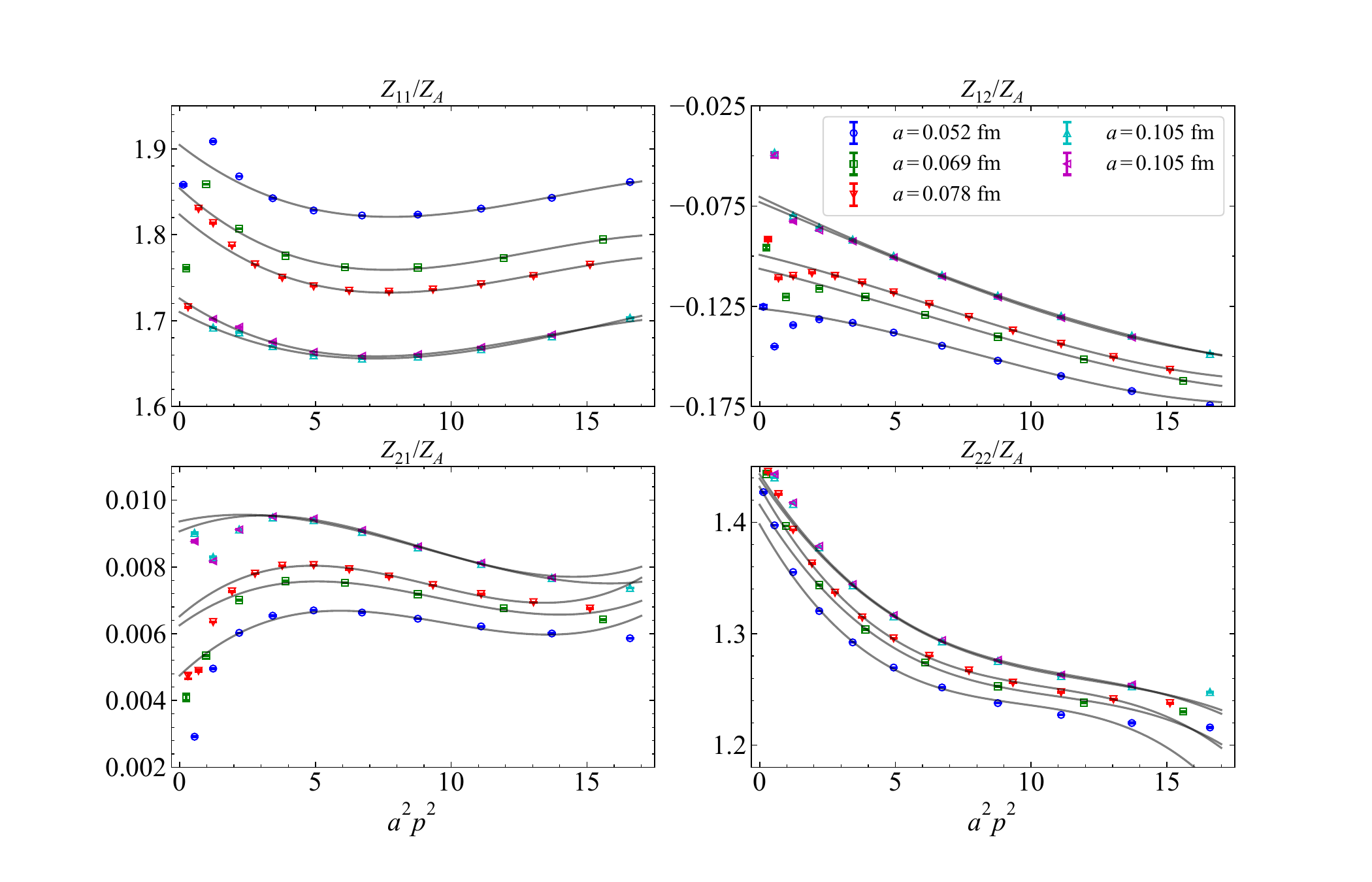}
  \caption{RI/SMOM$\to\overline{\rm MS}$ renormalization matrix elements for the second-moment
  operator multiplet, shown as $Z_{11}/Z_A$, $Z_{12}/Z_A$, $Z_{21}/Z_A$, and $Z_{22}/Z_A$ versus
  $a^2p^2$ for all ensembles.}
  \label{fig:RCs_second_moments_mD}
\end{figure*}

\section{Results for fits to the bare matrix element}
\label{ax:MEs_fits}
In this Appendix, we present additional examples of the correlated fits used to extract the bare matrix elements for several ensembles and kinematic settings, as shown in Figs.~\ref{fig:MEs_F32P30} and \ref{fig:MEs_H48P32}. These plots illustrate the fit quality for both the real and imaginary parts, as well as the practical choice between one-state and two-state ans\"atze adopted in the analysis. They further demonstrate the stability of the extracted ground-state matrix elements across different source-sink separations and momenta within the fitting strategy described in the main text.

\begin{figure*}[http]
  \centering
  \includegraphics[width=0.4\linewidth]{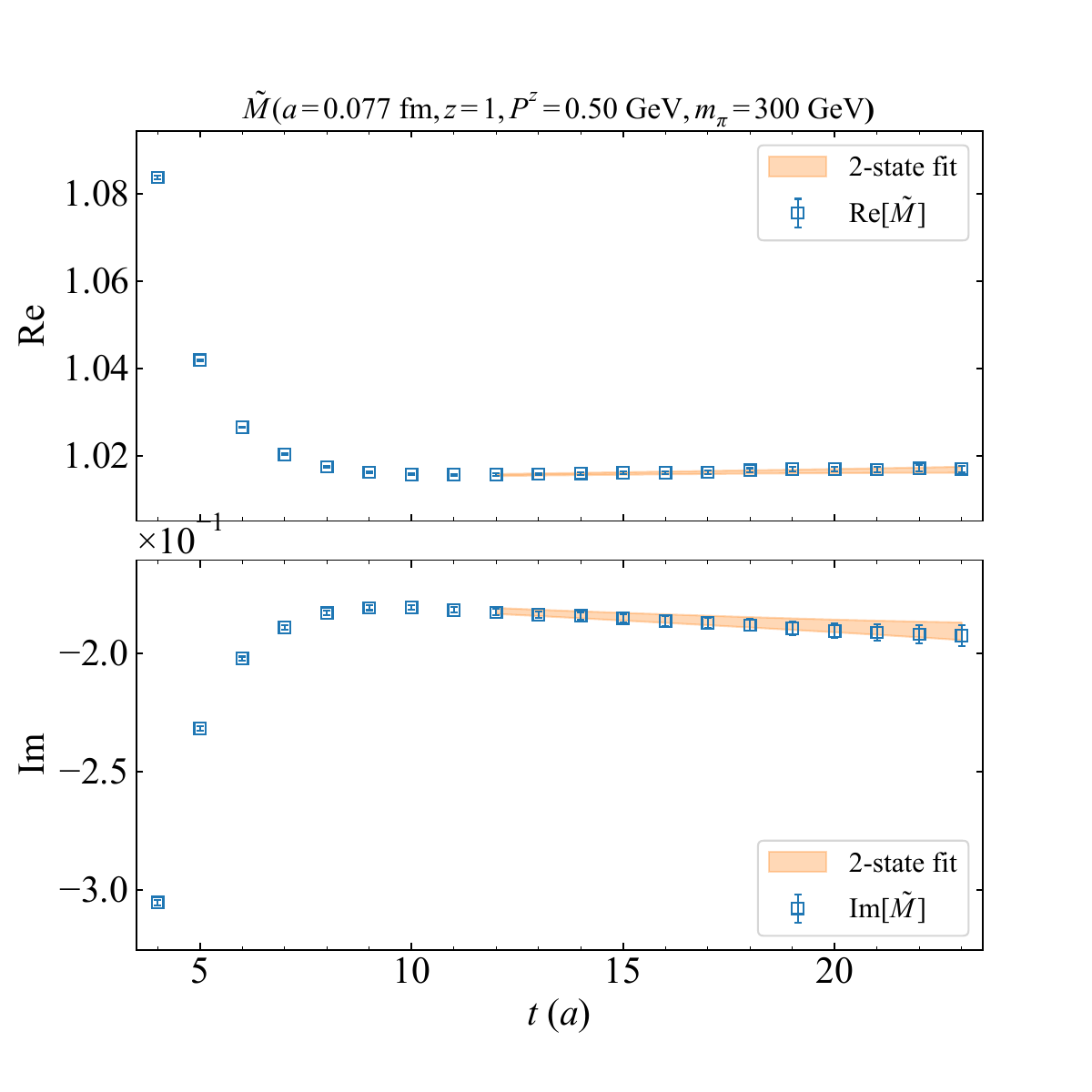}
    \includegraphics[width=0.4\linewidth]{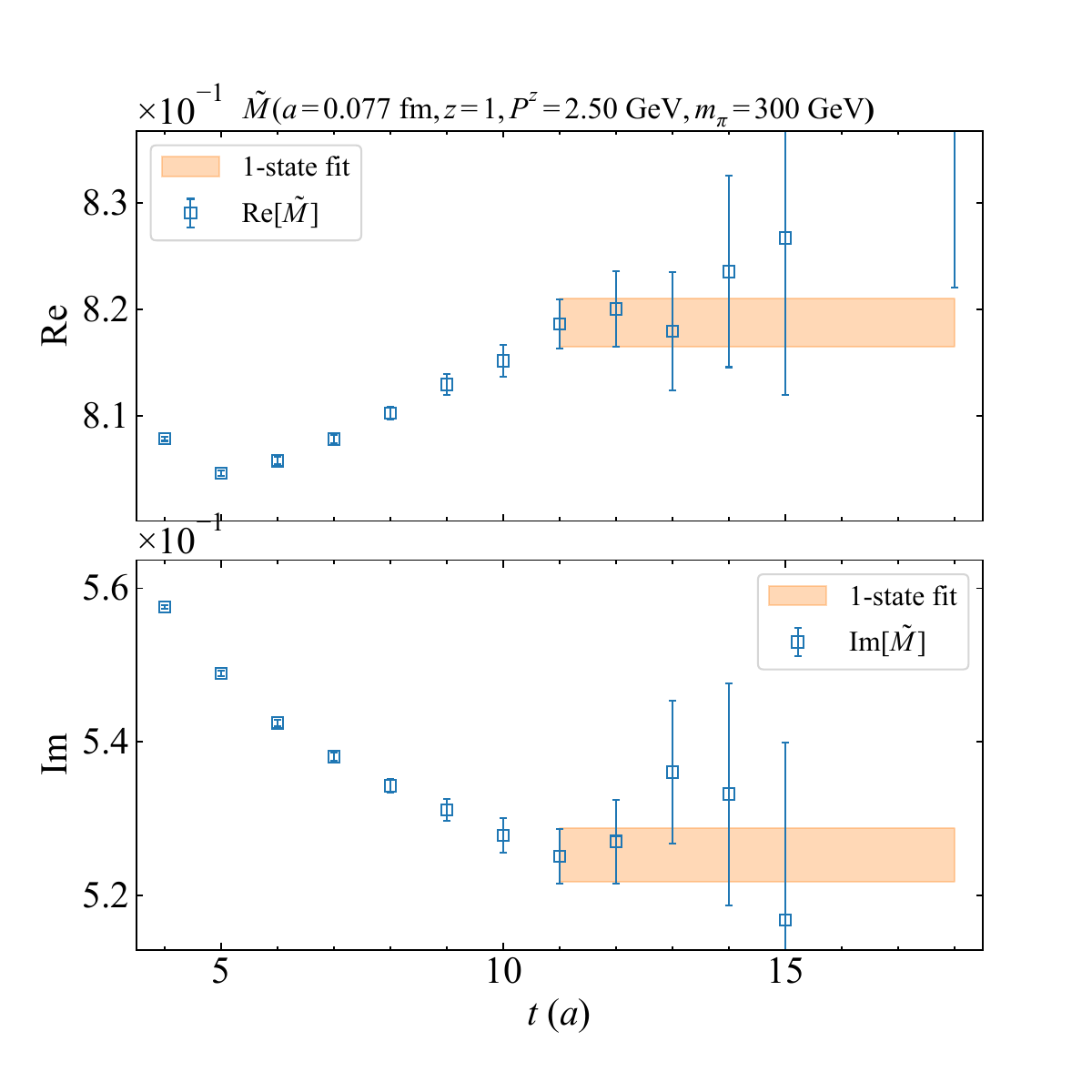}
  \includegraphics[width=0.4\linewidth]{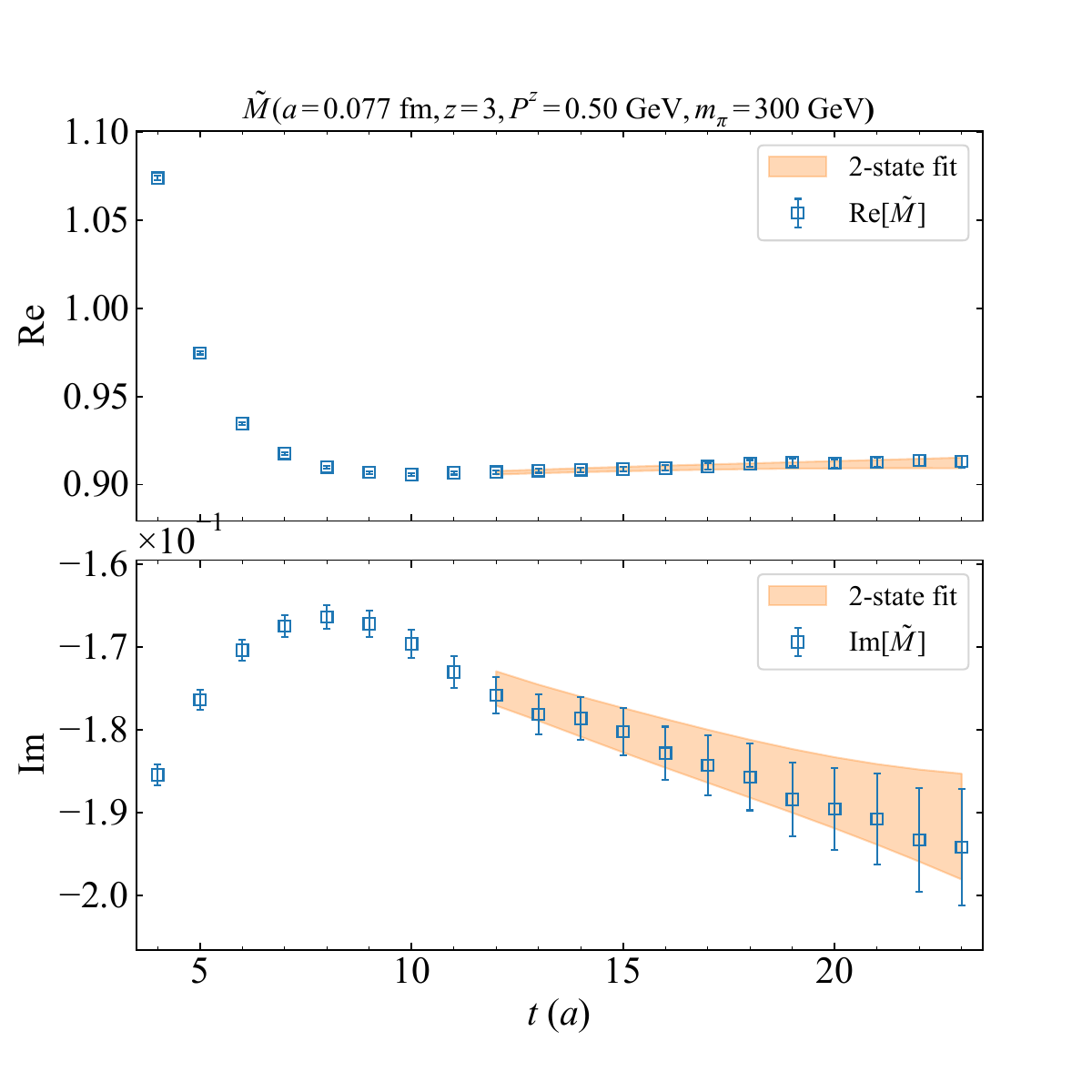}
  \includegraphics[width=0.4\linewidth]{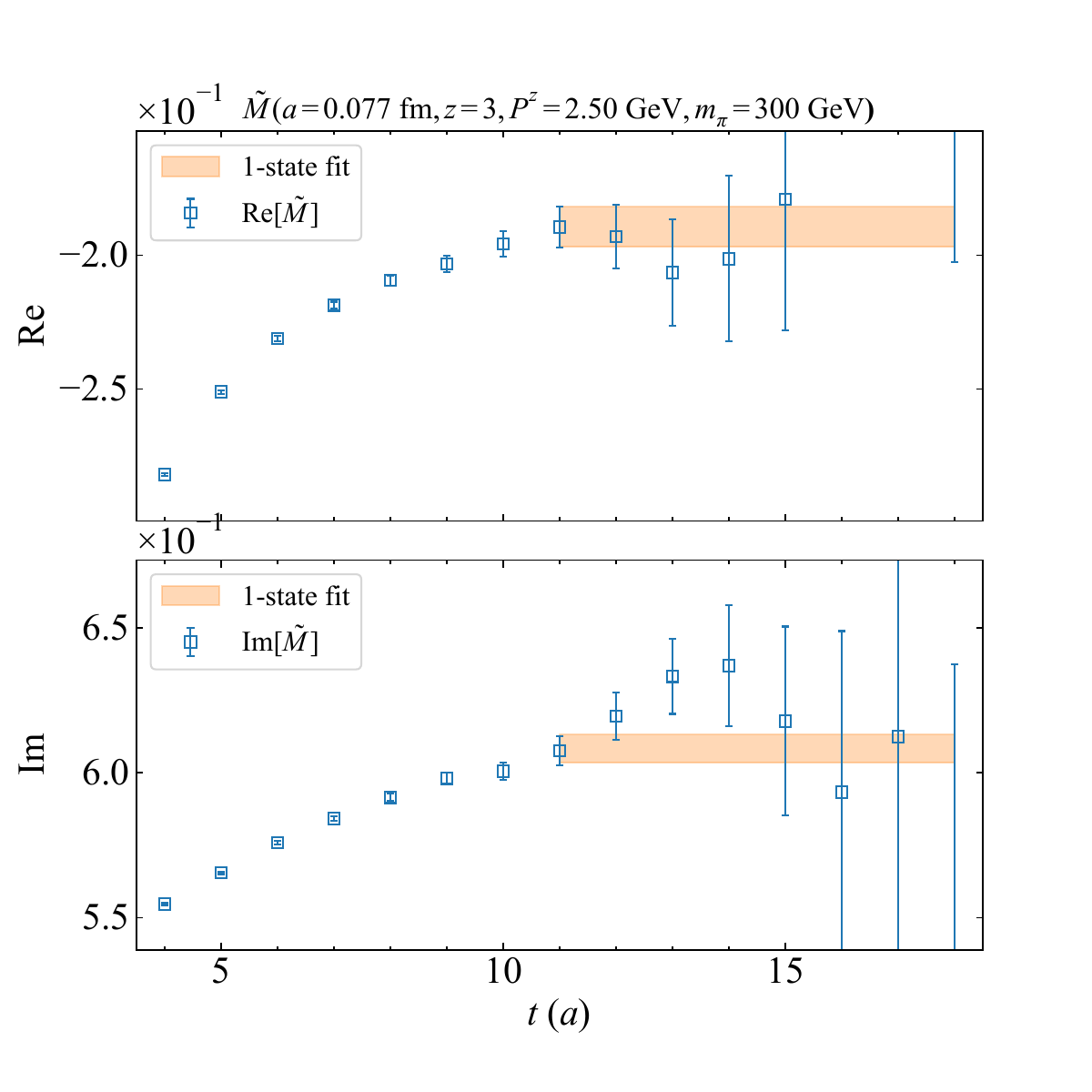}
  \includegraphics[width=0.4\linewidth]{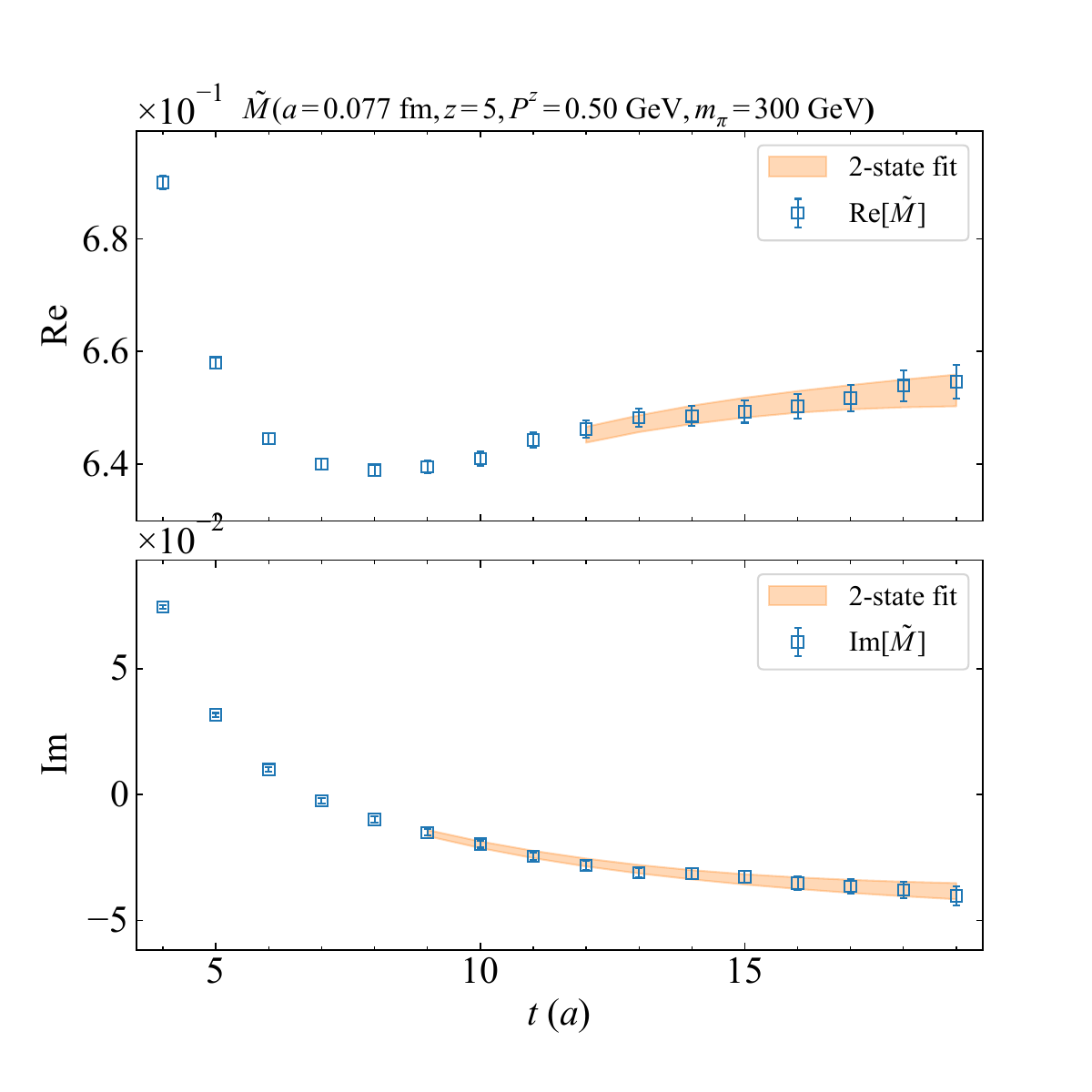}
  \includegraphics[width=0.4\linewidth]{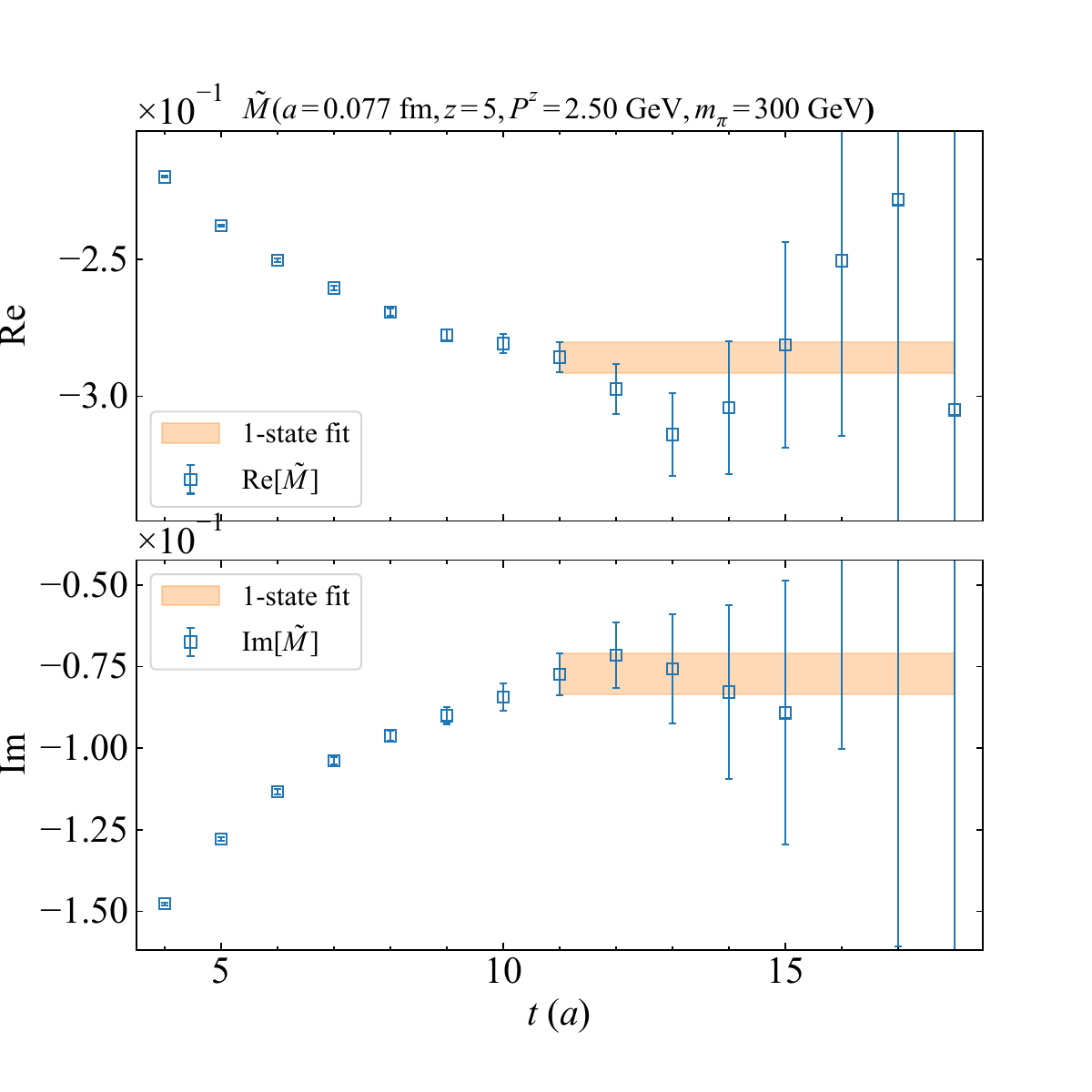}
  \caption{The figures present results from the F32P30 ensemble and compare the combinations for \{$\it z,P^z$\}=\{1,0.50 GeV\},\{3,0.50 GeV\},\{5,0.50 GeV\},\{1,2.50 GeV\},\{1,2.50 GeV\}, and \{1,2.50 GeV\}, respectively. The upper and lower panels correspond to the real and imaginary parts of the matrix elements, respectively. For z=1 we employ a two-state fit, while for all other z values we use a one-state fit.}
  \label{fig:MEs_F32P30}
\end{figure*}

\begin{figure*}[http]
  \centering
  \includegraphics[width=0.4\linewidth]{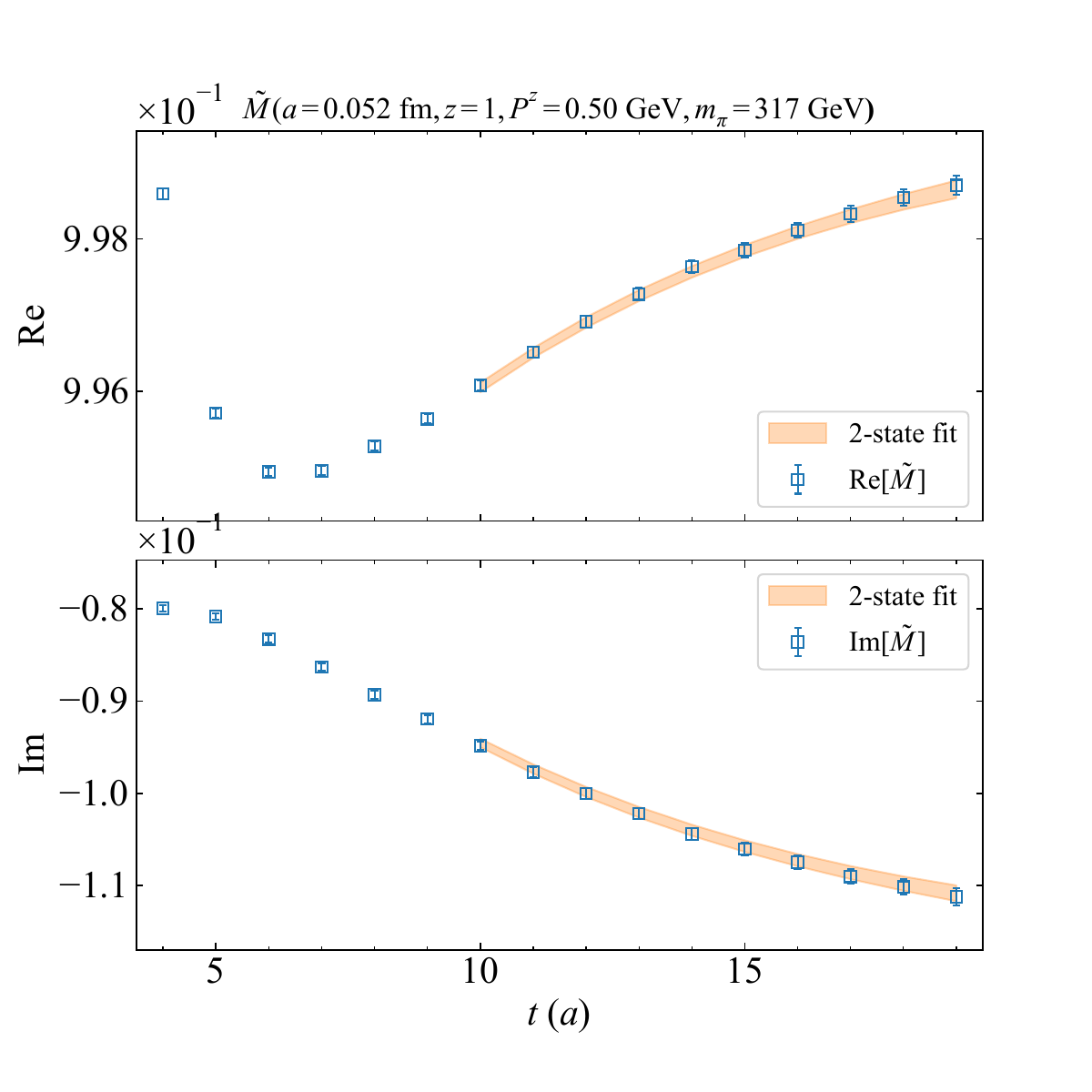}
  \includegraphics[width=0.4\linewidth]{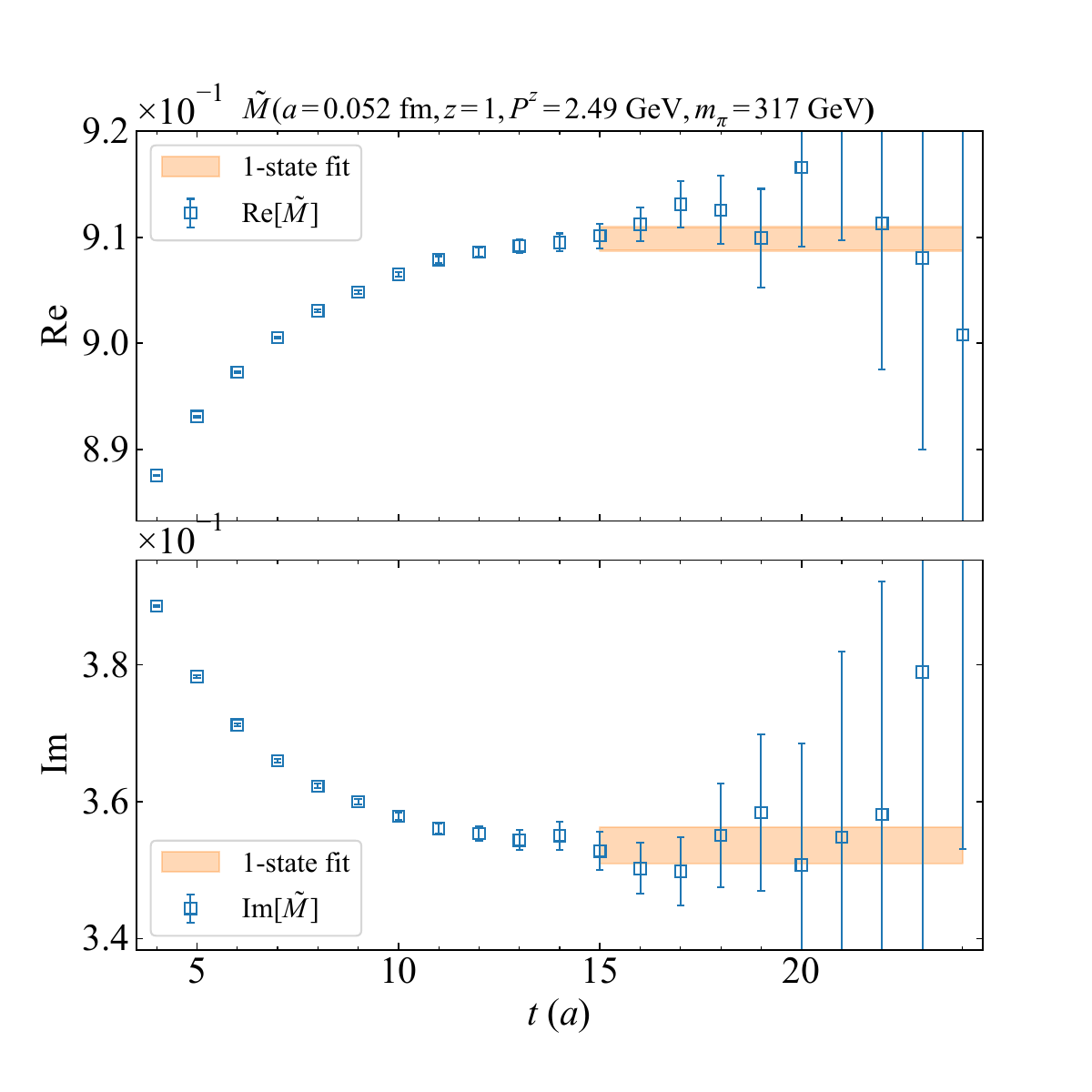}
  \includegraphics[width=0.4\linewidth]{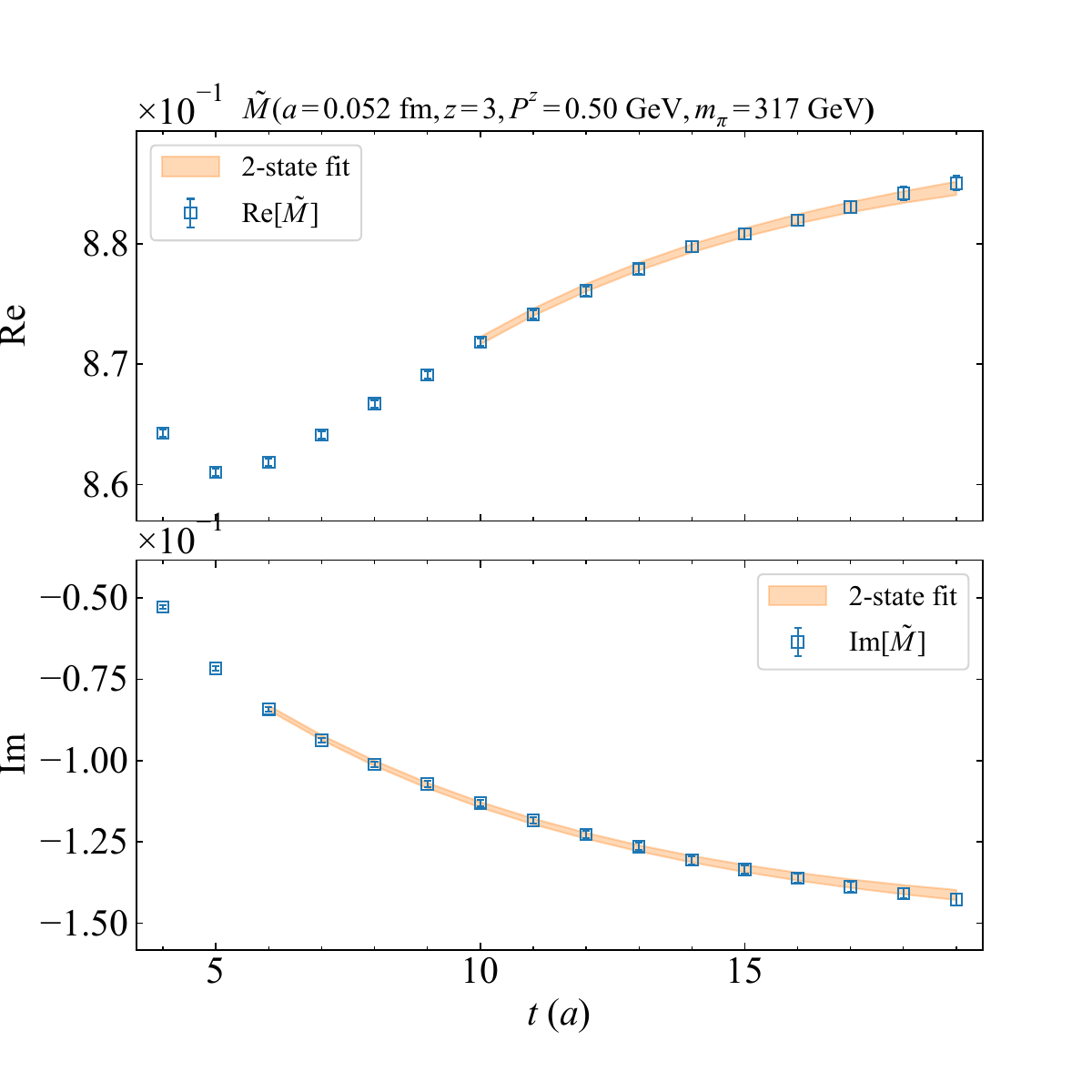}
  \includegraphics[width=0.4\linewidth]{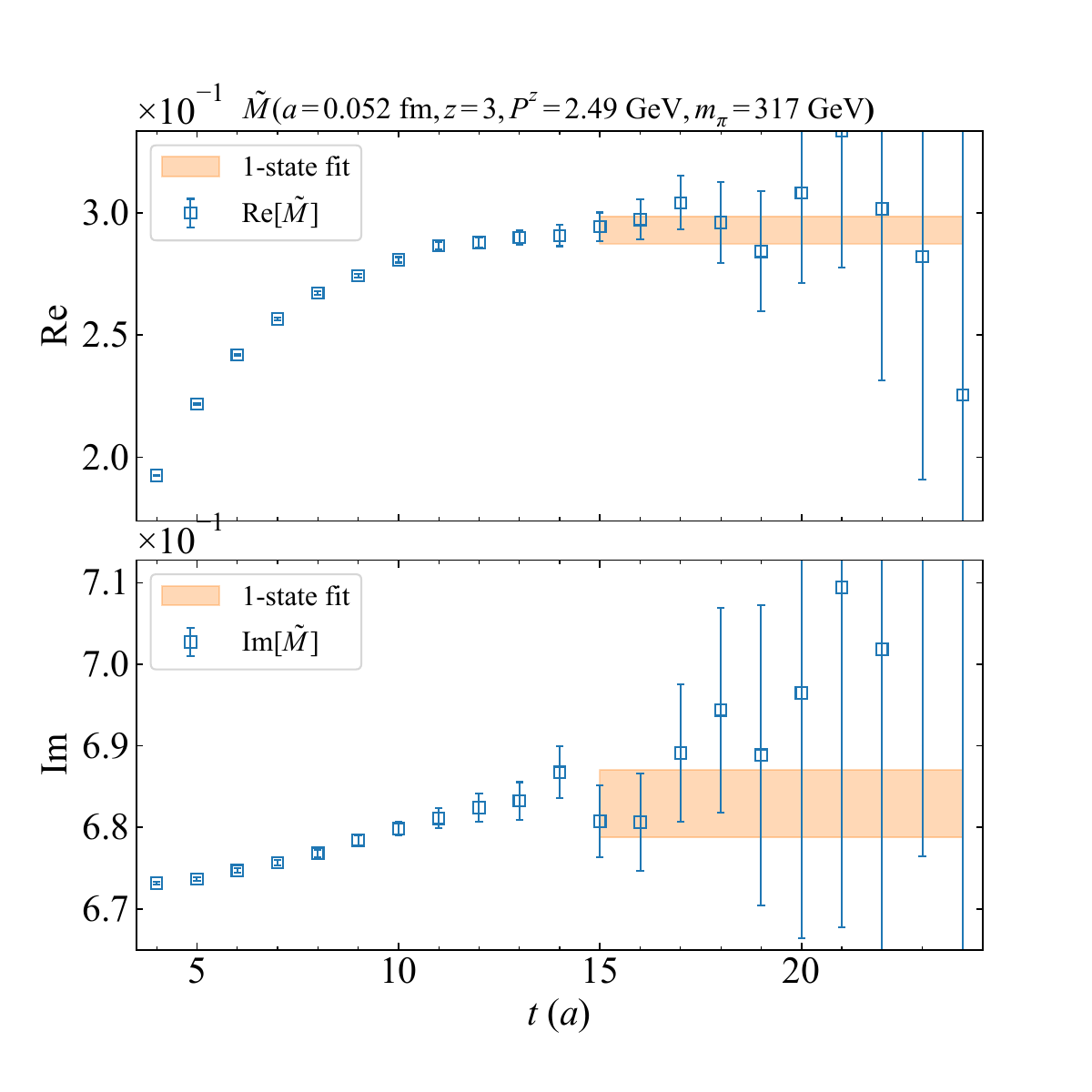}
  \includegraphics[width=0.4\linewidth]{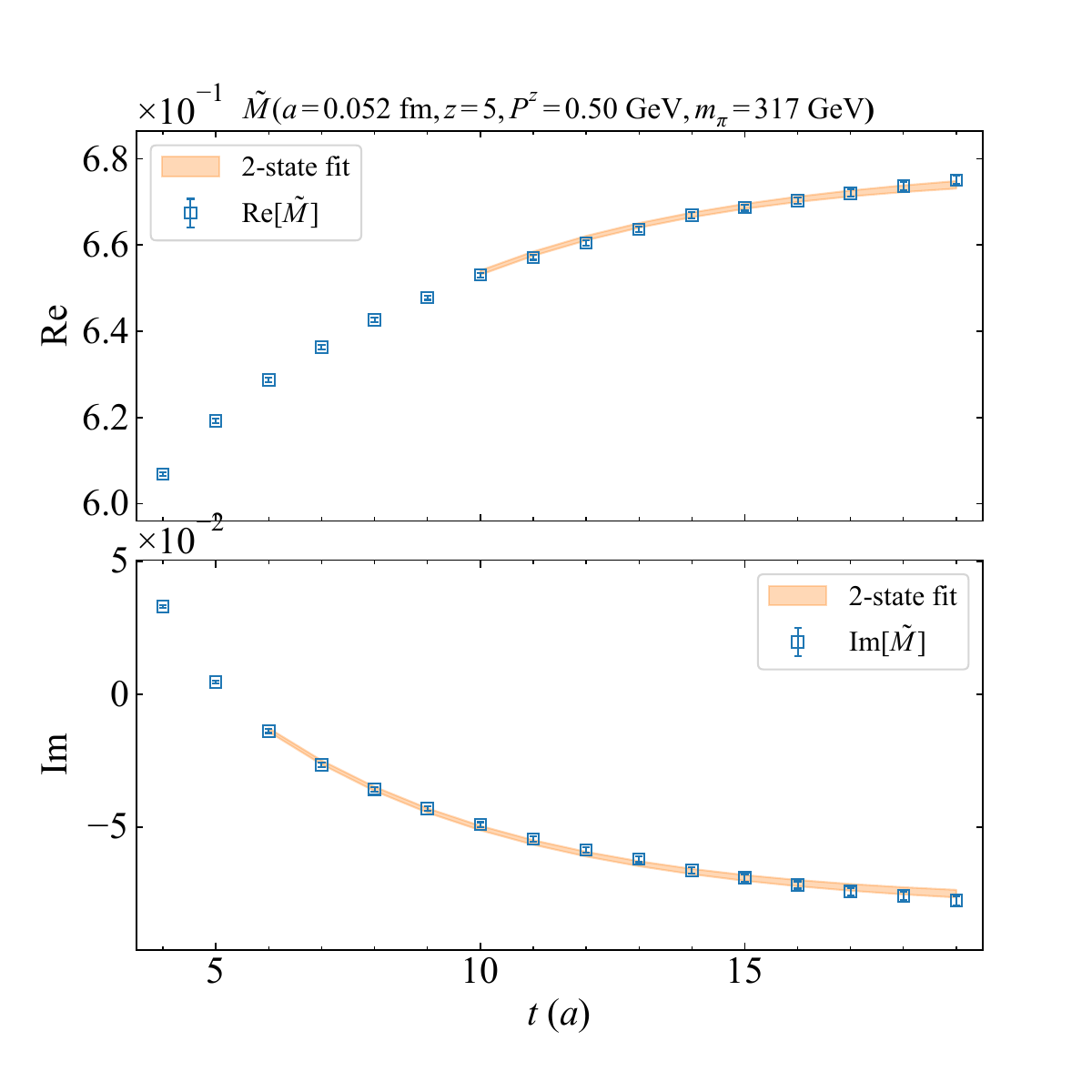}
  \includegraphics[width=0.4\linewidth]{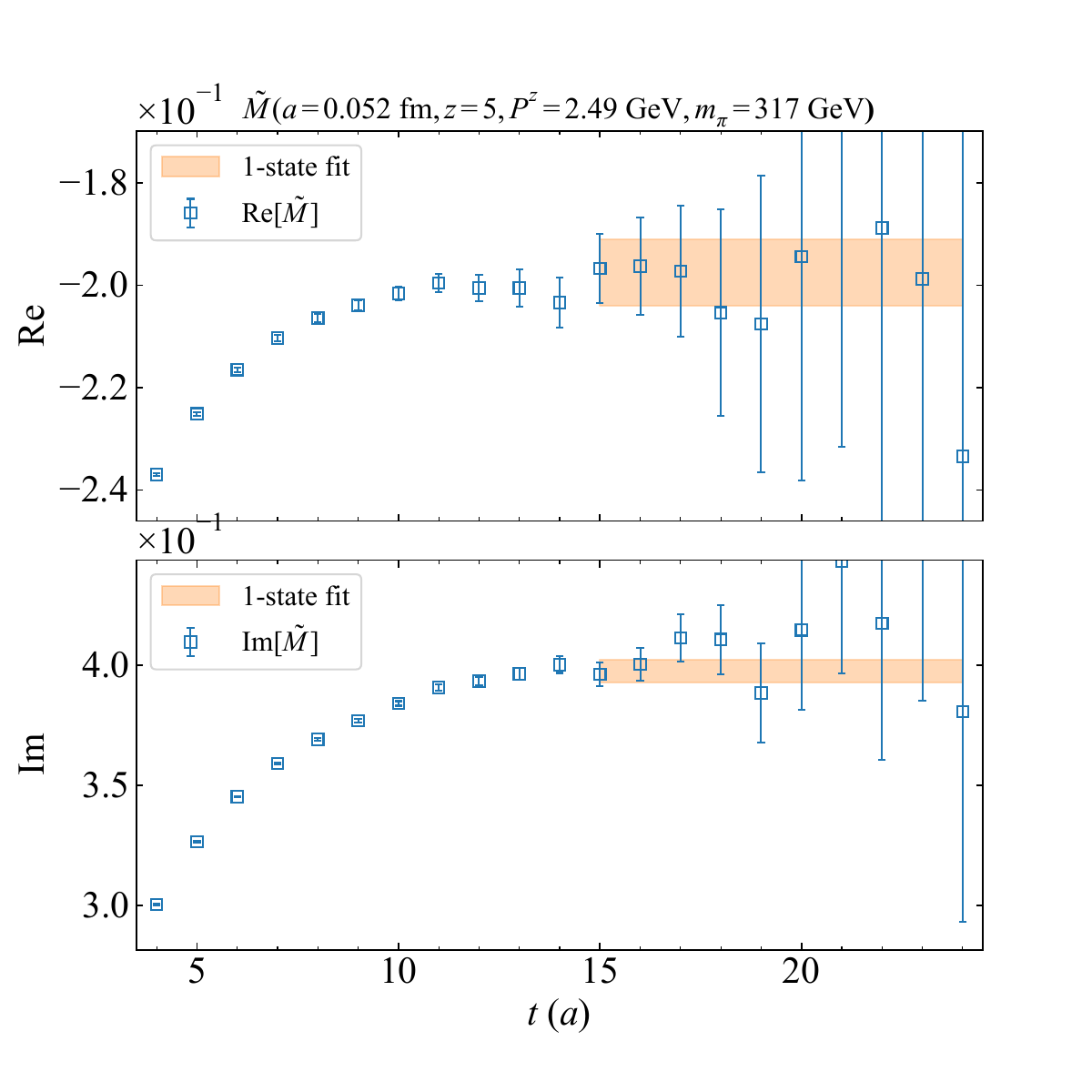}
  \caption{The figures present results from the H48P32 ensemble and compare the combinations for \{$\it z,P^z$\}=\{1,0.50 GeV\},\{3,0.50 GeV\},\{5,0.50 GeV\},\{1,2.49 GeV\},\{1,2.49 GeV\}, and \{1,2.49 GeV\}, respectively. The upper and lower panels correspond to the real and imaginary parts of the matrix elements, respectively. For z=1 we employ a two-state fit, while for all other z values we use a one-state fit.}
  \label{fig:MEs_H48P32}
\end{figure*}

\section{More Results and  for $\lambda$-extrapolation of the renormalized quasi-DAs}
\label{ax:more_lambda_extrapolation}
The $\lambda$-extrapolation is used to reconstruct the long-distance behavior of the renormalized coordinate-space matrix elements and thereby extend the accessible range of $\lambda$ beyond that directly reached on the lattice. To test the robustness of this procedure, we present additional examples obtained by varying the lattice spacing, the separation scale $z_s$ in the hybrid renormalization scheme, and the lower bound of the extrapolation window. Additional results for the $\lambda$-extrapolation of the renormalized quasi-DAs are shown in Figs.~\ref{fig:more_lambda_extrapolation_1}--\ref{fig:more_lambda_extrapolation_4}. These comparisons allow for a visual assessment of the stability under reasonable analysis variations and support the systematic uncertainty assigned to the extrapolation in the main analysis.

\begin{figure*}[http]
  \centering
  \includegraphics[width=0.45\linewidth]{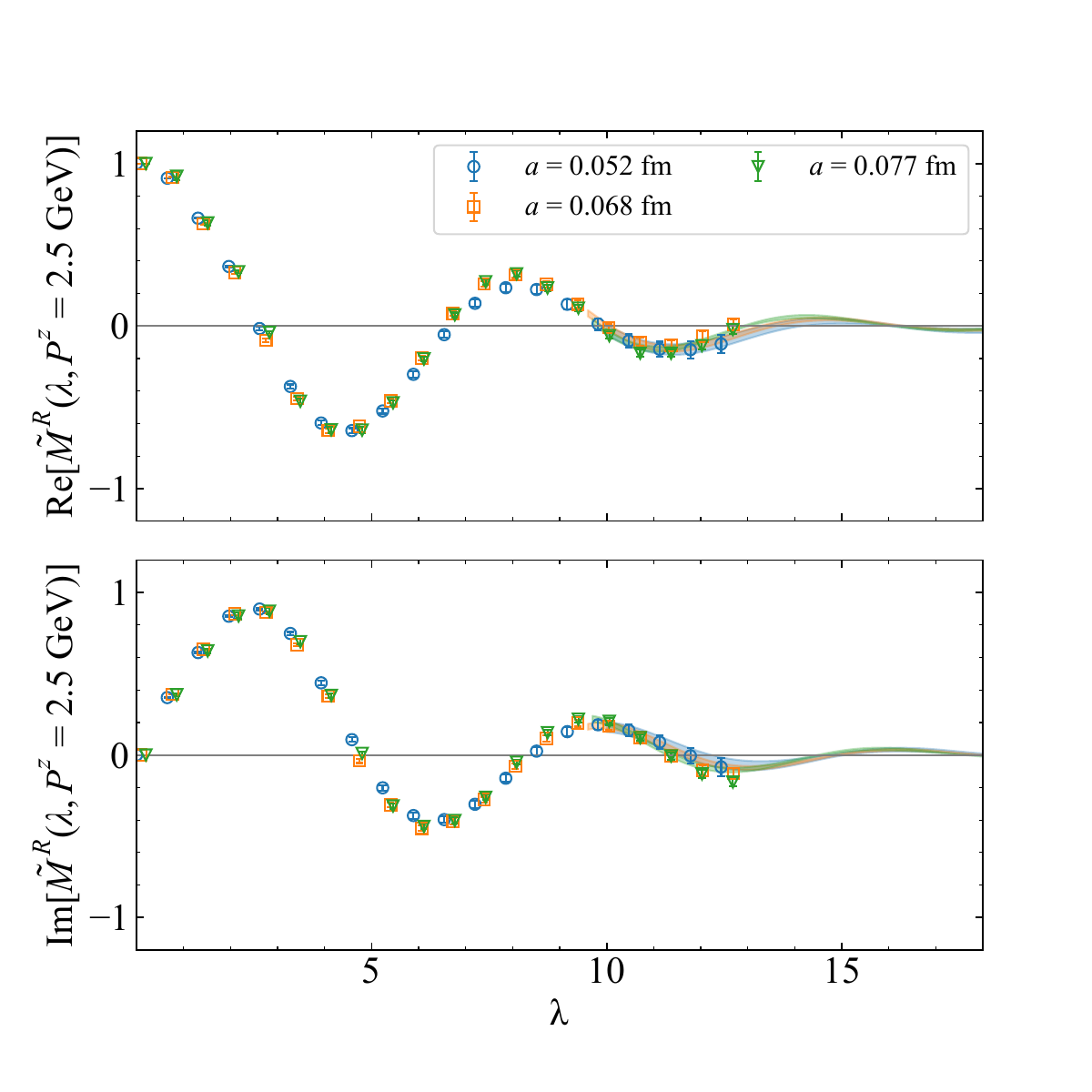}
  \includegraphics[width=0.45\linewidth]{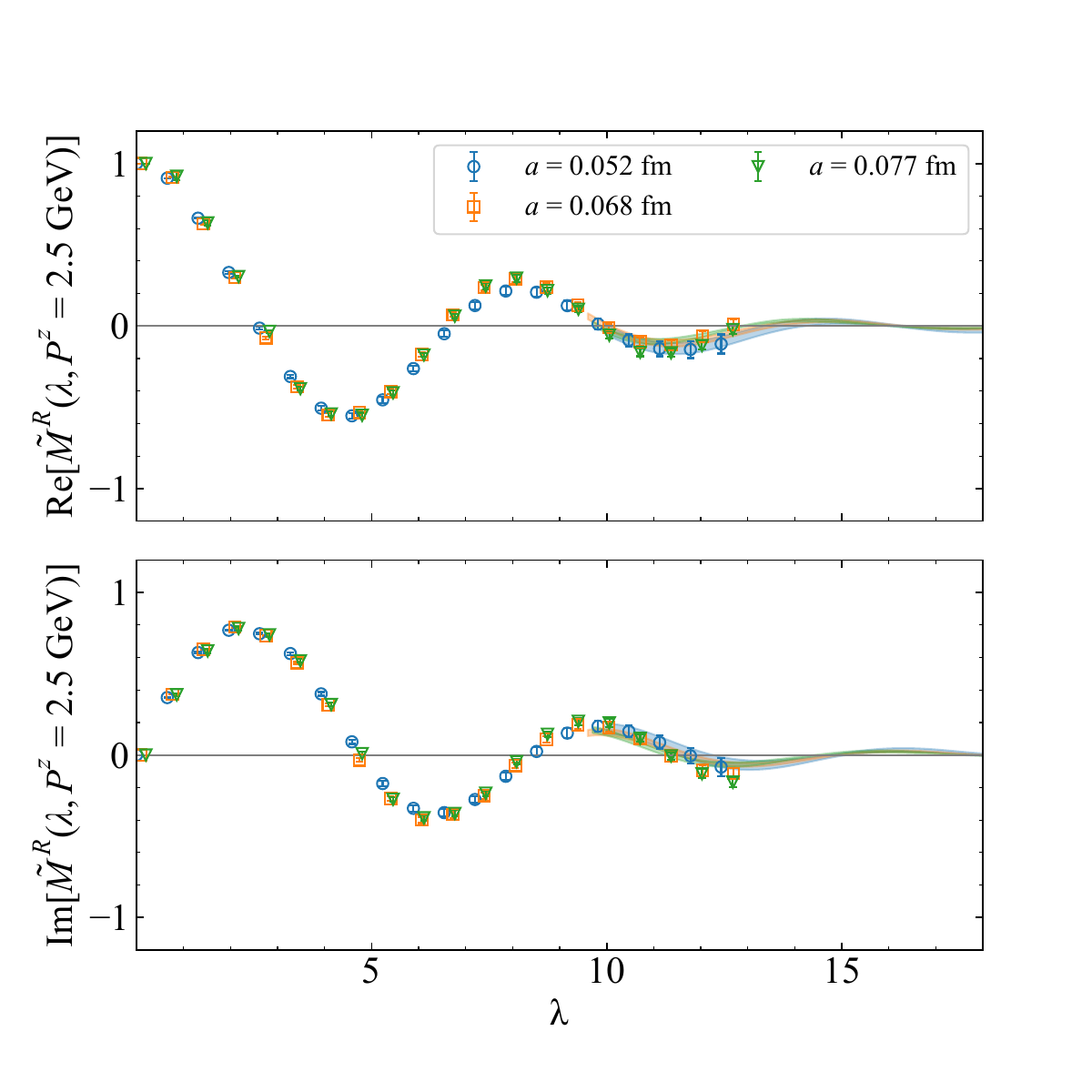}
  \caption{$\lambda$-extrapolation results (colored bands) at $P^z\simeq2.5\,{\rm GeV}$ with different lattice spacings(upper: real part; lower: imaginary part). The left and right panels correspond to $z_s=0.10$ fm and $z_s=0.21$ fm, respectively.}
  \label{fig:more_lambda_extrapolation_1}
\end{figure*}

\begin{figure*}[http]
  \centering
  \includegraphics[width=0.45\linewidth]{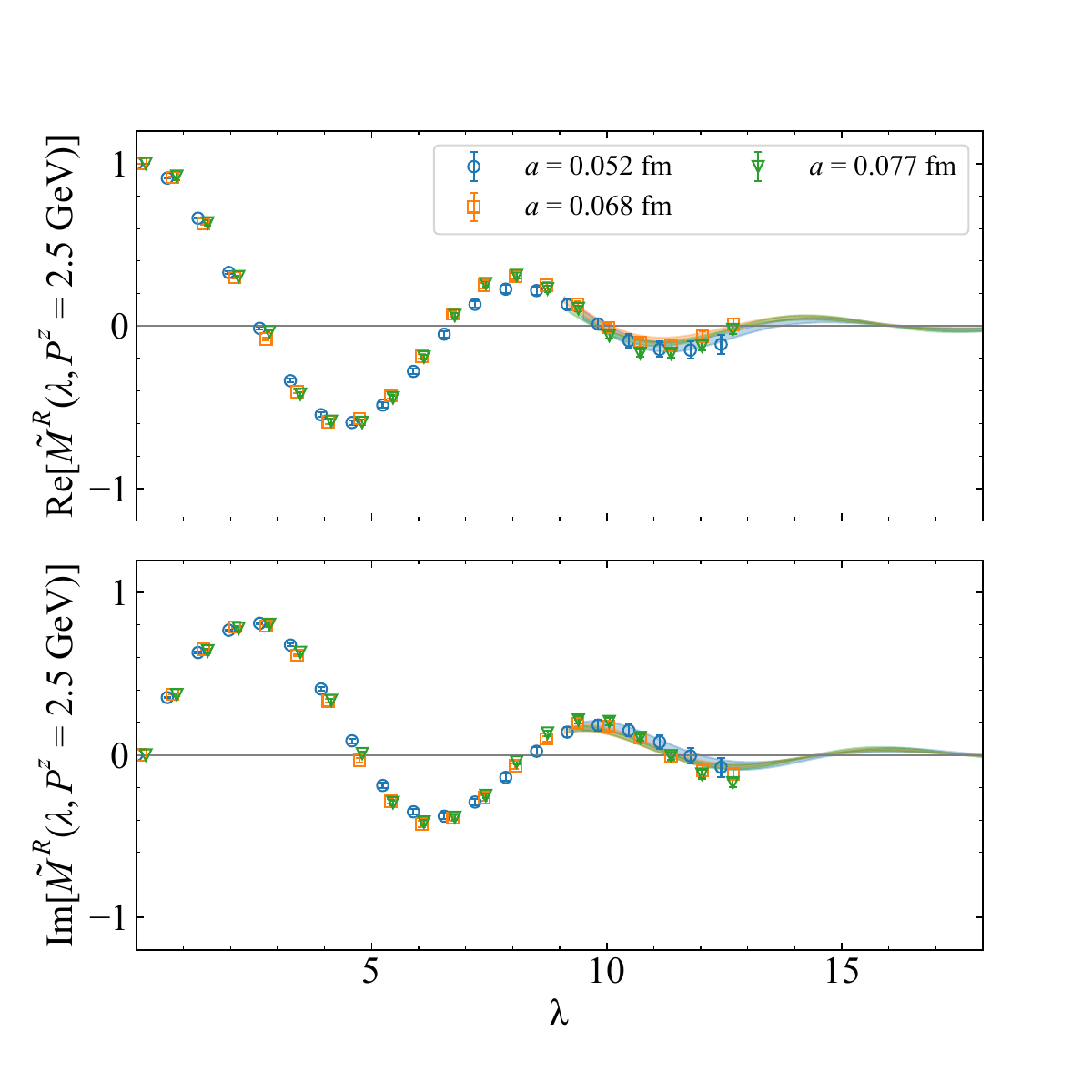}
  \includegraphics[width=0.45\linewidth]{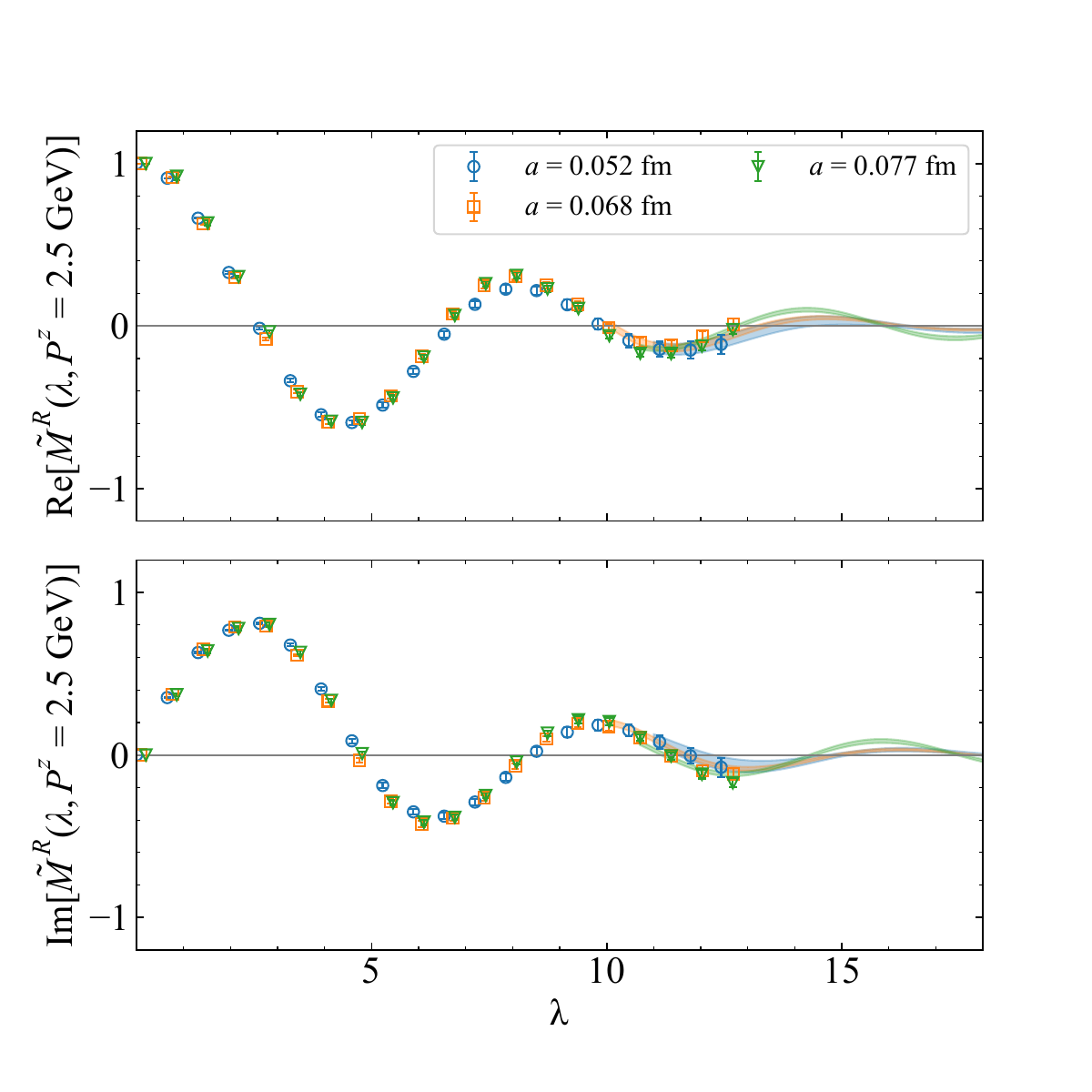}
  \caption{$\lambda$-extrapolation results (colored bands) at $P^z\simeq2.5\,{\rm GeV}$ with different lattice spacings(upper: real part; lower: imaginary part). The left and right panels correspond to shifting the starting point of the 
$\lambda$ extrapolation by one data point backward (smaller $\lambda$) and forward (larger $\lambda$), respectively.}
  \label{fig:more_lambda_extrapolation_2}
\end{figure*}

\begin{figure*}[http]
  \centering
  \includegraphics[width=0.45\linewidth]{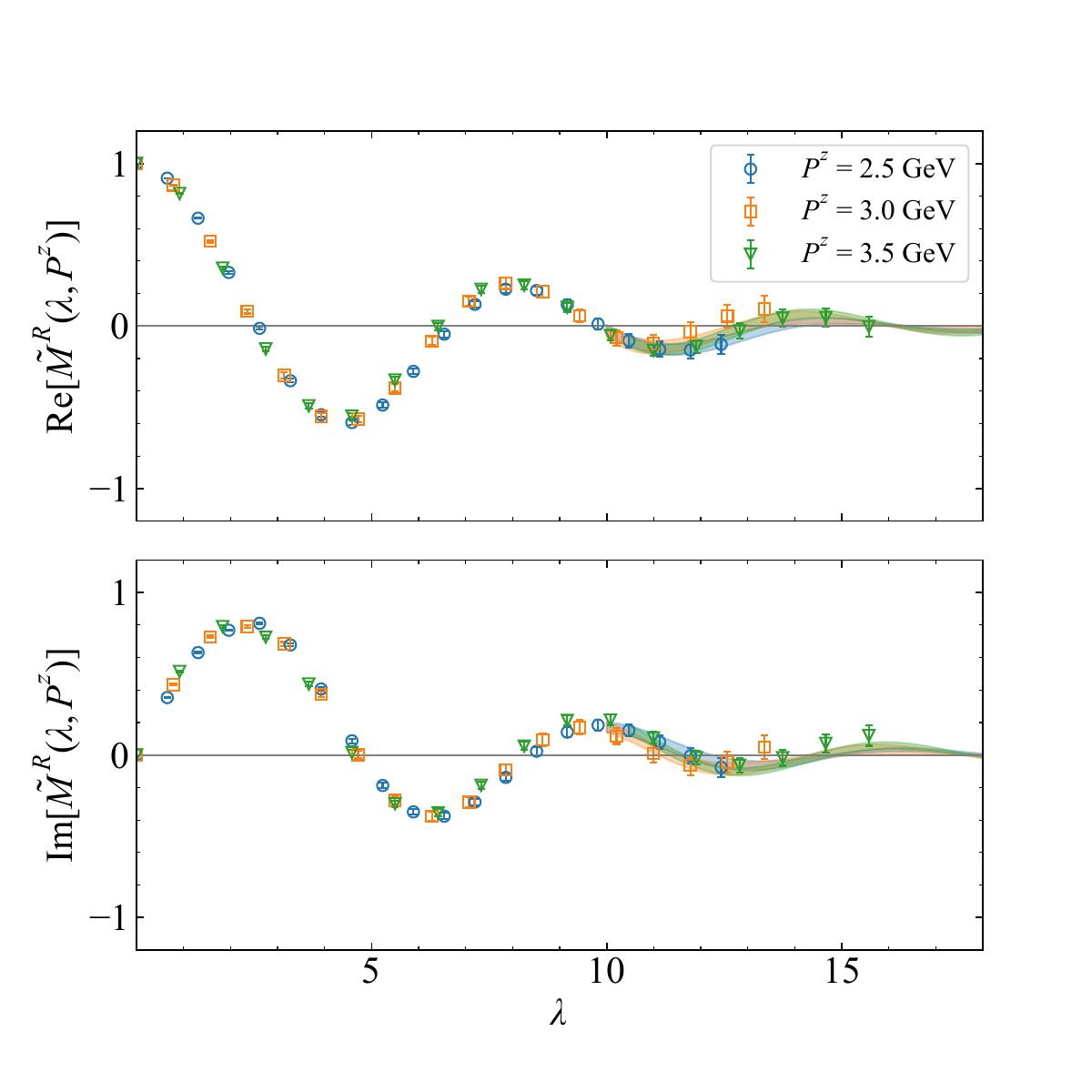}
  \includegraphics[width=0.45\linewidth]{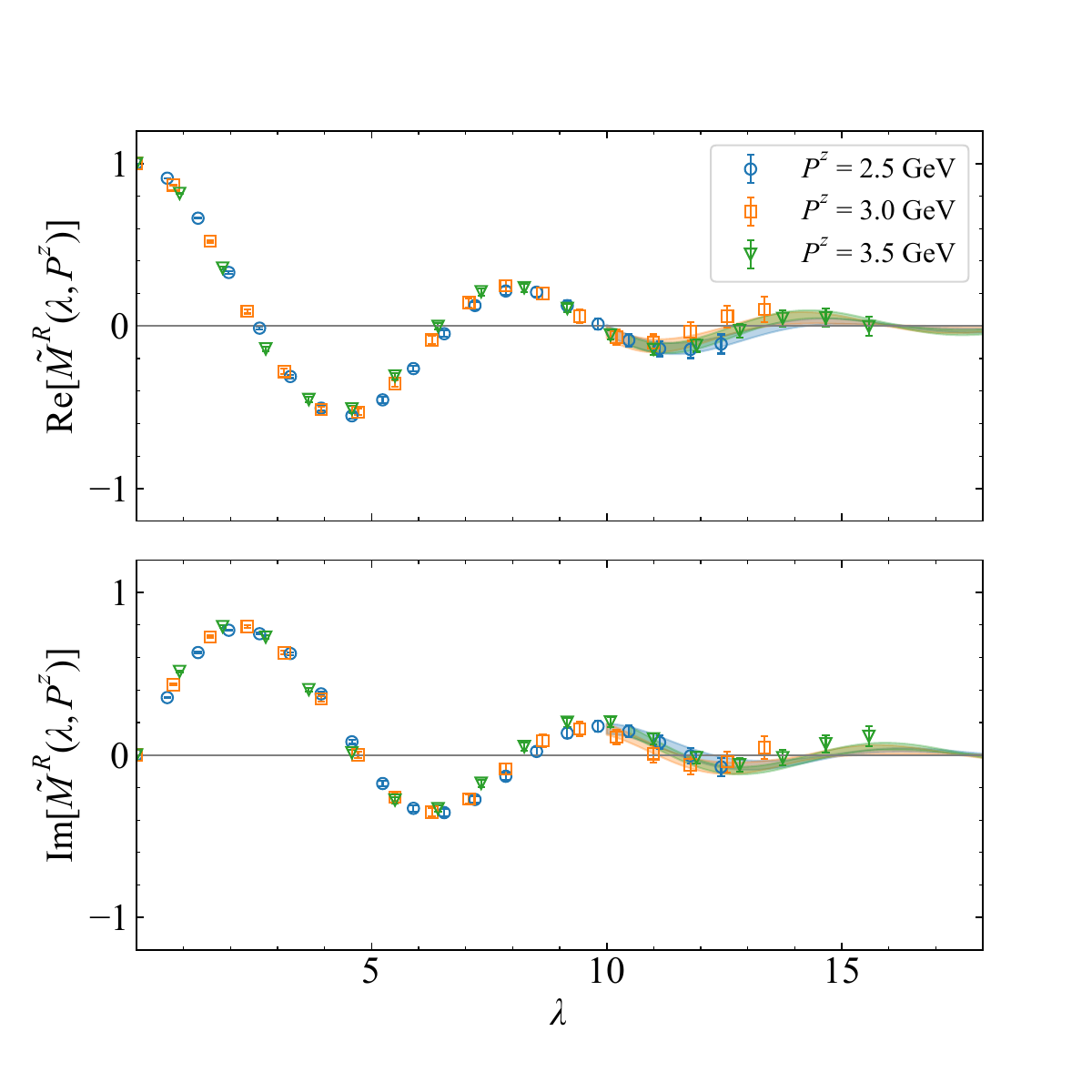}
  \caption{$\lambda$-extrapolation results (colored bands) at $P^z\simeq2.5$, $3.0$ and $3.5~{\rm GeV}$ on H48P32 ensemble(upper: real part; lower: imaginary part). The left and right panels correspond to $z_s=0.16$ fm and $z_s=0.26$ fm, respectively.}
  \label{fig:more_lambda_extrapolation_3}
\end{figure*}

\begin{figure*}[http]
  \centering
  \includegraphics[width=0.45\linewidth]{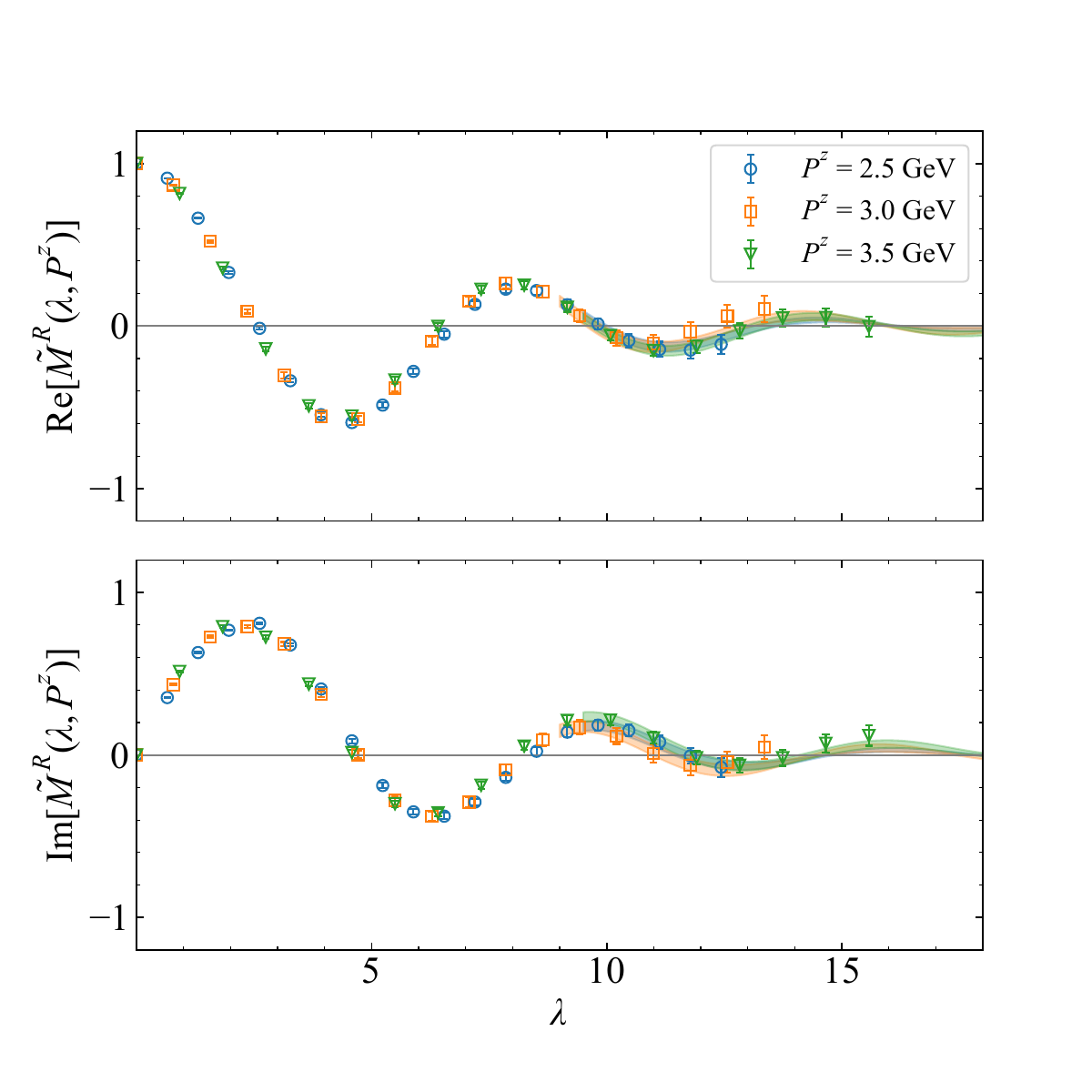}
  \includegraphics[width=0.45\linewidth]{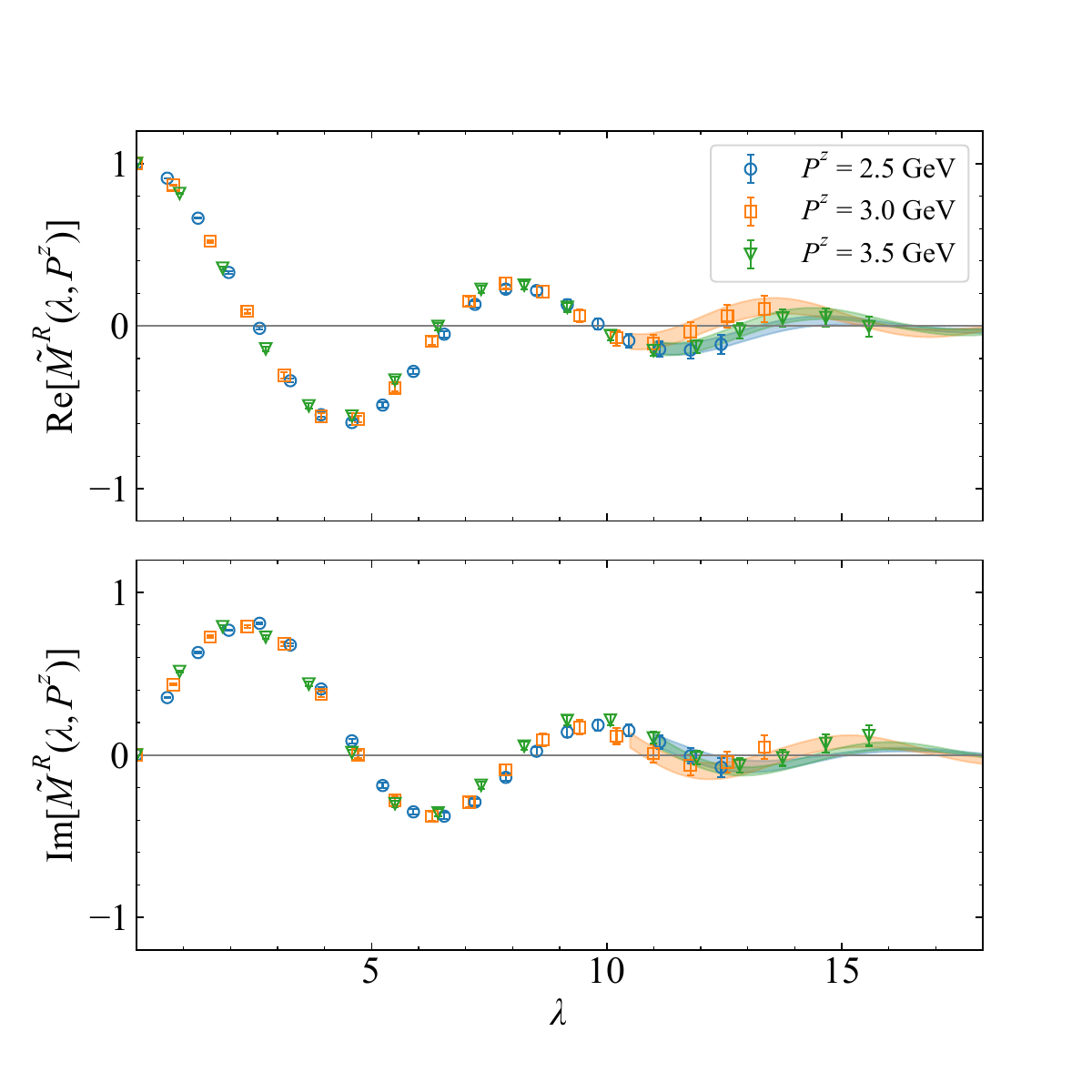}
  \caption{$\lambda$-extrapolation results (colored bands) at $P^z\simeq2.5$, $3.0$ and $3.5~{\rm GeV}$ on H48P32 ensemble(upper: real part; lower: imaginary part). The left and right panels correspond to shifting the starting point of the $\lambda$ extrapolation by one data point backward (smaller $\lambda$) and forward (larger $\lambda$), respectively.}
  \label{fig:more_lambda_extrapolation_4}
\end{figure*}

\section{More Results for quasi-DA in momentum space}
\label{ax:more_quasi-DA}

For completeness, we also present the quasi-DAs in momentum space, obtained from the Fourier transform of the $\lambda$-extrapolated coordinate-space matrix elements. Figure~\ref{fig:quasi-DA_in_momentum_space} provides an intuitive illustration of their dependence on the hadron boost momentum $P^z$ and of the discretization effects at fixed $P^z$ across different lattice spacings. These momentum-space quasi-DAs serve as a useful diagnostic for the subsequent matching and complement the coordinate-space analysis in the main text.

\begin{figure*}[http]
\centering
\includegraphics[width=0.45\linewidth]{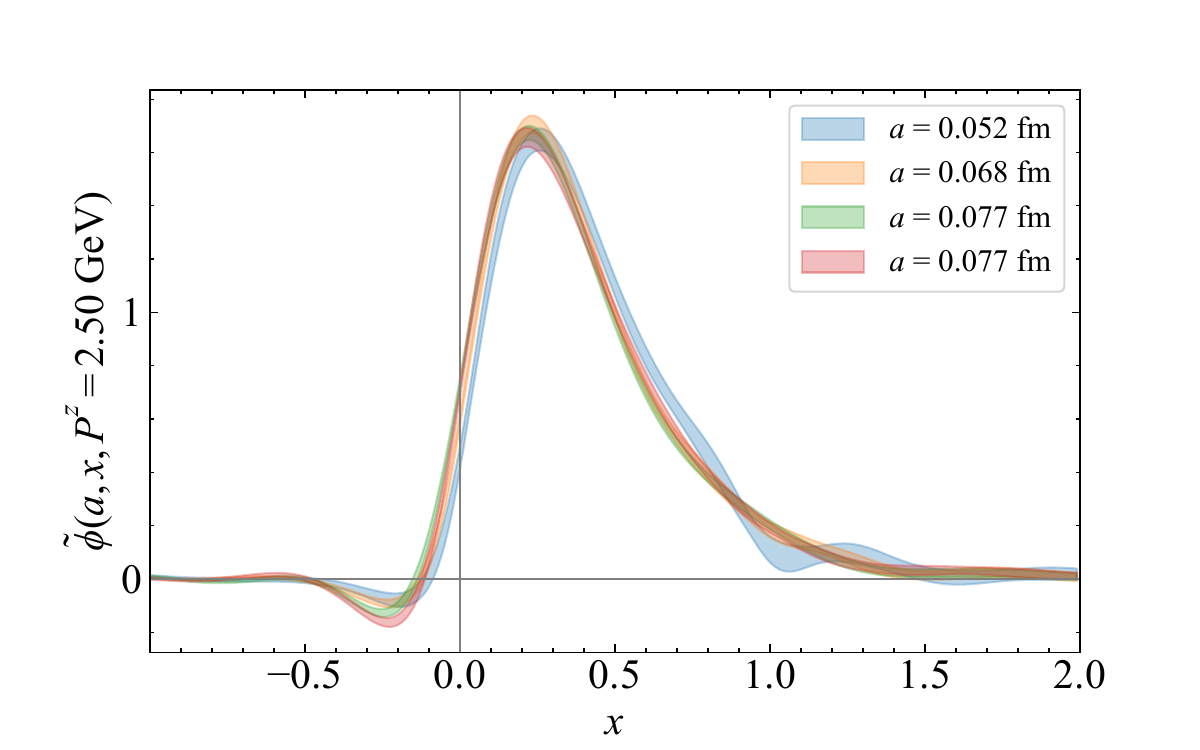}
\includegraphics[width=0.45\linewidth]{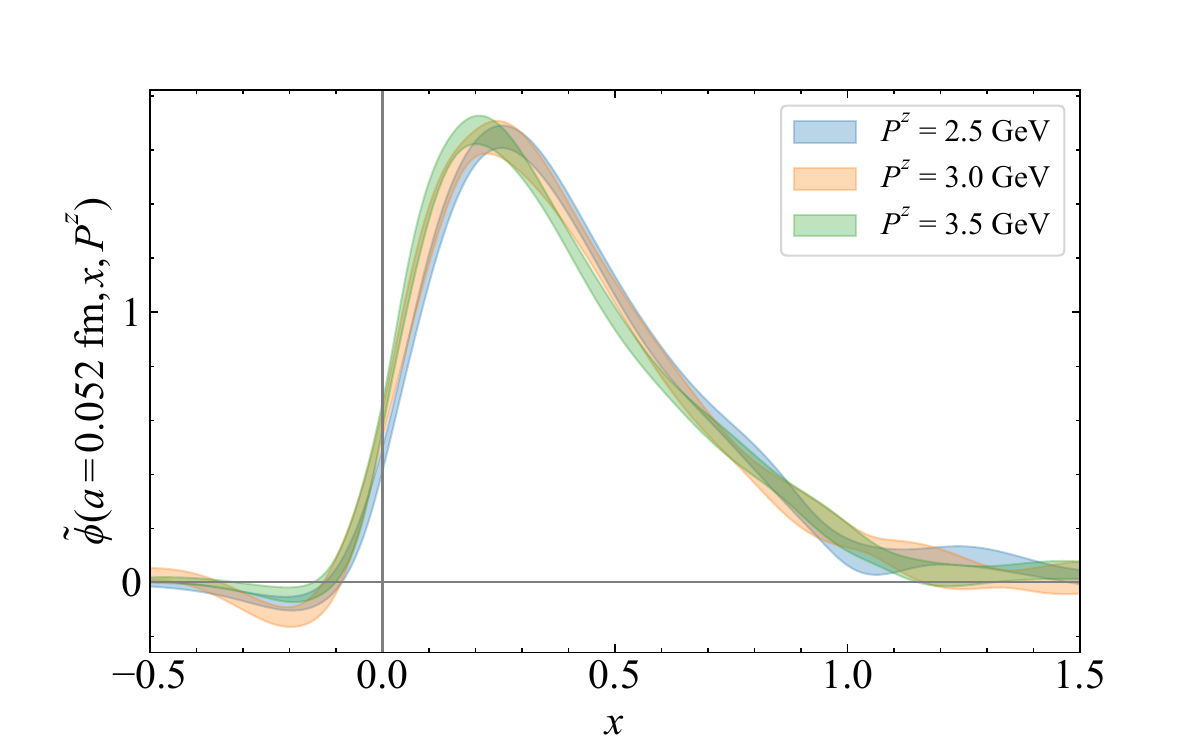}
\caption{Quasi-DA in momentum space at $\it\{a,P^z\}$ = \{0.1053 fm,2.5 GeV\},\{0.0775 fm,2.5 GeV\},\{0.0683 fm,2.5 GeV\},\{0.0519 fm,2.5 GeV\},\{0.0519 fm,3.0 GeV\},\{0.0519 fm,3.5 GeV\}, respectively. The left and right panels correspond to different lattice spacings and different momenta, respectively.}
\label{fig:quasi-DA_in_momentum_space}
\end{figure*}

\end{widetext}
\end{appendix}

\end{document}